\documentclass[12pt,preprint]{aastex}

\slugcomment{Submitted to {\it The Astrophysical Journal}}
\shorttitle{Multi-Colour Photometry for Pulsating sdB Stars}
\shortauthors{Randall et al.}

\newcommand{\gta}{\lower 0.5ex\hbox{$ \buildrel>\over\sim\ $}}
\newcommand{\lta}{\lower 0.5ex\hbox{$ \buildrel<\over\sim\ $}}
\newcommand{\Teff}{T_{\rm eff}}
\newcommand{\teff}{T_{\rm eff}}
\newcommand{\Msun}{M_{\rm \odot}}

\begin{document}

\title{The Potential of Multi-Colour Photometry for Pulsating Subdwarf B Stars\footnote{Predicted colour-amplitude ratios for a series of representative EC 14026 and PG 1716 stars are available upon request. Interested collaborators please contact S.K. Randall or G. Fontaine.} }

\author{S.K. Randall, G. Fontaine, P. Brassard, and P. Bergeron} 

\affil{D\'epartement de Physique, Universit\'e de Montr\'eal, C.P.
6128, Succ. Centre-Ville, Montr\'eal, Qu\'ebec, Canada H3C 3J7}

\email{randall@astro.umontreal.ca, fontaine@astro.umontreal.ca,
  brassard@astro.umontreal.ca, bergeron@astro.umontreal.ca}

\begin{abstract}

We investigate the potential of multi-colour photometry for partial mode 
identification in both long- and short-period variable subdwarf B stars. 
The technique presented is based on the fact that the frequency dependence 
 of an oscillation's amplitude and phase bears the signature of the
 mode's degree index $l$, among other things. Unknown contributing factors 
can be eliminated through the evaluation of the amplitude ratios and phase
 differences arising from the brightness variation in different wavebands, 
theoretically enabling the inference of the degree index from observations 
in two or more bandpasses. Employing a designated model atmosphere code,
we calculate the brightness variation expected across the visible disk
during a pulsation cycle in terms of temperature, radius, and surface
gravity perturbations to the emergent flux for representative EC 14026
and PG 1716 star models. Nonadiabatic effects are considered in detail
and found to be significant from nonadiabatic pulsation calculations
applied to our state-of-the-art models of subdwarf B stars. Our results
indicate that the brightness variations observed in subdwarf B stars are
caused primarily by changes in temperature and radius, with surface
gravity perturbations playing a small role. For PG 1716 stars, 
temperature effects dominate in the limit of long periods with the result 
that the oscillatory amplitudes and phases loose their period dependence 
and nonadiabatic effects become unimportant. Outside this regime however,
 their values are strongly influenced by both factors. We find that the 
phase shifts between brightness variations in different wavebands are 
generally small, but may lie above the experimental detection threshold 
 in certain cases. The prospect of mode discrimination seems much more 
promising on the basis of the corresponding amplitude ratios. While in
 EC 14026 stars the amplitude ratios predicted are very similar for 
modes with $l$ = 0, 1 or 2, they are well separated from those of modes 
with $l$ = 3, $l$ = 5, and $l$ = 4 or 6, each of which form a distinct group. 
 For the case of the PG 1716 stars it should be possible to discriminate
 between modes with $l$ = 1, 2 , 4 or 6 and those of degree indices
 $l$ = 3 and $l$ = 5. Identifying modes within a given group is
 challenging for both types of pulsator and requires multi-colour photometry
 of extremely high quality. Nevertheless, we demonstrate that it is feasible 
 using the example of the largest amplitude peak detected for the fast 
pulsator KPD 2109+4401 by Jeffery et al. (2004).

\end{abstract}

\keywords{stars: horizontal-branch --- stars: interiors --- stars:
  oscillations --- subdwarfs} 

\section{INTRODUCTION}

Subdwarf B (sdB) stars are evolved extreme horizontal branch stars with
effective temperatures in the range 20,000$-$42,000 K and low masses of
around 0.5 $\Msun$ (e.g., Saffer et al. 1994). They are believed to be
composed of a helium-burning core surrounded by a hydrogen envelope too
thin for them to ascend the asymptotic giant branch after core helium
exhaustion (Heber 1986; Dorman 1995). Instead, they evolve off and along the
horizontal branch and eventually end their lives as low-mass white
dwarfs (Bergeron et al. 1994). The discovery of pulsators among subdwarf
B stars (Kilkenny et al. 1997) has opened them up to asteroseismological
probing, an important tool for understanding the internal structure of these
interesting objects. Pulsating subdwarfs can be divided into two
categories: the rapidly pulsating EC 14026 stars and the slowly 
oscillating long-period variable subdwarf B stars (or PG 1716 stars for
short). 

The first EC 14026 stars were discovered to be pulsating
multi-periodically with typical periods in the range 100-200 s in 1997
(Kilkenny et al. 1997; Koen et al. 1997; Stobie et al. 1997; O'Donoghue
et al. 1997). At about the same time and completely independently,
Charpinet et al. (1996, 1997) predicted the existence of pressure mode
($p$-mode) instabilities in these stars caused by a classical kappa
mechanism associated with the iron opacity peak in the subdwarfs'
envelopes. This opacity peak in turn is dependent on a local
overabundance of iron, which is achieved by the competitive action of
gravitational settling and radiative levitation. As the relative
contributions of these two processes are determined by the surface
gravity and the effective temperature of the star, all pulsating
subdwarfs should lie in a designated instability strip on the 
log $g-\Teff$ diagram. To date, the number of known EC 14026
stars has risen to 33 (see Kilkenny 2002, Silvotti et al. 2002,  
Bill\`eres et al. 2002, and Fontaine et al. 2004 for a recent census),
all of which fall into the theoretically predicted instability strip,
clustering around $\log{g}\sim$5.75 and $T_{\rm eff}\sim$33,000. Since
theoretical and observed pulsational properties are in good agreement,
it is believed that a qualitative understanding of the pulsation
mechanism has been reached. Beyond this, detailed quantitative
interpretations of observed period spectra have been possible in a few
instances (Brassard et al. 2001; Charpinet et al. 2003, 2005), allowing
the identification of the modes excited through the so-called forward 
method. Being the ultimate goal in asteroseismology, this led to the
determination of the stars' fundamental parameters, including the mass
fraction of the thin hydrogen shell which cannot be deduced otherwise.
It goes without saying that independent observational tests of these
mode idendifications would be most welcome. In particular, the dense
 period spectra detected in certain EC 14026 pulsators force the
 controversial inclusion of modes with $l$ = 3 and 4 in the 
asteroseismological process, since there are not enough theoretical modes
 with $l$  = 0, 1, and 2 to account for the mode density observed.

Long-period variables constitute a newer, less extensively studied class
of subdwarf B pulsator. Their variability was first announced by Betsy
Green at the ``Asteroseismology across the H-R diagram'' conference held
in Porto in July 2002 (Green et al. 2003a; see also Green et al.
2003b). Slowly pulsating sdB stars are distinctly cooler than their short
period counterparts and show multi-periodic, low-amplitude ($\sim$ 1
milli-magnitude) luminosity variations with typical periods between 0.8
and 1.6 hours. These are about a factor of 30 longer than those of the
EC 14026 stars, and automatically imply high radial order gravity modes
($g$-modes). Early ideas on a possible excitation mechanism for this new
class of star included tidal excitation in a close binary system
(Fontaine et al. 2003a) and a mechanism involving a slow reviving of the
hydrogen shell (Green et al. 2003a), but proved unfruitful. A much more
promising mechanism capable of qualitatively explaining the observed
periods was brought forward more recently (Fontaine et al. 2003b) and
involves the same driving process that so successfully explains the
instabilities in the EC 14026 stars. This is analogous to the
case of the $\beta$ Cep/slowly pulsating B stars on the main sequence
(Dziembowski \& Pamyatnykh 1993; Dziembowski, Moskalik, \& Pamyatnykh
1993; Gautschy \& Saio 1993).

In their paper describing the excitation mechanism in the slow pulsators,
Fontaine et al. (2003b) built a series of representative models of
subdwarf B stars along the extreme horizontal branch and computed their
nonadiabatic pulsation properties in a wide range of periods for modes
with $l$ = 0 to $l$ = 8. This uncovered two distinct ``islands'' of
instability in the pulsational period vs $T_{\rm eff}$ diagram (as can
be seen in their Figure 4). The first of these is clustered around
$T_{\rm eff}\sim 31,000$ K and corresponds to the short period,
low-order $p$-modes typical of the EC 14026 stars, where modes with
degree values of $l$ = 0 and upward are driven. The second region of
instability, attributed to the long period pulsators, lies at lower
temperatures between 22,000 and 26,000 K and features $g$-modes with
long periods similar to those observed in the slow oscillators. However,
in the majority of models only modes with degree indices of $l$ = 3 and
higher can currently be excited, which would imply that the
long-period luminosity variations observed in the cooler subdwarfs
correspond to modes with $l$ = 3, 4, and 5. This is in conflict with
canonical wisdom, which suggests that modes with $l\ge 3$ should
generally not be observable due to cancellation effects when
integrating over the visible disk of the star. While Fontaine et
al. (2003b) argue that the low amplitudes of the variations observed
point to modes of relatively high degree indices, they concede that the
instability calculations are subject to a real blue-edge problem. In
particular, we now know of long-period pulsators with effective temperatures
as high as $\sim 28,000$ K, which according to current theory should
excite only modes with $l\ge 8$. As it does not seem feasible that these
could be observed even in the best of circumstances, it is clear that we face
major challenges as far as mode identification for the slow pulsators is
concerned. However, there is scope for improvement on the theoretical
side. Following a suggestion of Hideyuki Saio, we are currently
developing more realistic subdwarf B star models that incorporate the
presence of helium in the iron radiative levitation calculations, rather
than adopting the pure hydrogen background assumed by Fontaine et
al. (2003b). While this may well lead to a better description of the 
PG 1716 stars' blue edge, we nevertheless feel it would be
highly beneficial  if the degree indices of the periods observed could
be determined  using a method completely independent of the instability
calculations.  

The best way of achieving this is to exploit the wavelength dependence
of a mode's pulsational amplitude, which depends on $l$ as well as on
other parameters such as the viewing angle and the intrinsic amplitude
of the oscillation in question. By calculating the ratio of amplitudes
in different wavebands, the latter unknown quantities can be elimitated,
leaving (to a first approximation) a dependence only on $l$ and the
atmospheric parameters of the star. The theoretical colour-amplitude
ratios can thus be computed for a given target and compared to
pulsational amplitudes from multi-colour photometry in order to
determine the modes' degree indices. Likeways, phase shifts between 
oscillations in different wavebands may be exploited in certain 
cases. This method is by no means revolutionary and has
been applied to many different types of pulsating star, such as $\delta$
Scuti stars (Garrido, Garc\'ia-Lobo, \& Rodriguez 1990), $\beta$
Cepheids (Cugier, Dziembowski, \& Pamyatnykh 1994), ZZ Ceti white
dwarfs (Robinson et al. 1995; Fontaine et al. 1996), and $\gamma$
Doradus stars (Breger et al. 1997), to name just a few. Attempts to 
understand the observed period spectrum on the basis of multi-colour
photometry have also been made for EC 14026 stars, first by Koen
(1998) and more recently by Jeffery et al. (2004). The former study
gives a qualitative interpretation of the periods observed for KPD
2109+4401 based on the theory of Watson (1988), while the latter asserts 
to have provisionally identified the modes detected for the fast
oscillators KPD 2109+4401 and HS 0039+4302. In this case, the
identification is based on amplitude ratios computed by Ramachandran,
Jeffery, \& Townsend (2004) as well as on comparisons with the
pulsational properties of the evolutionary sequences published by
Charpinet et al. (2002).  

In the present study, we assess the potential of multi-colour
photometry for pulsating sdB stars in some detail. Using a grid of
 specially designed model atmospheres, we are able to compute vital 
quantities such as the emergent specific intensities and their derivatives
 to high accuracies. Moreover, we carry out full nonadiabatic pulsation
calculations in order to obtain eigenfunctions that are as realistic as
possible, an aspect that was not studied by Ramachandran et al. (2004). 
In the next section, we briefly review the theory of colour-amplitude
 variations in pulsating stars. We then describe our computations for
 subdwarf B stars, and present results for a representative EC 14026 
and PG 1716 model before examining the influence of the atmospheric 
parameters of the model in question. We end with a discussion of the
 practical applications of the tools developed.

\section{BASIC THEORY}

The theoretical foundations for the modelling of lightcurves of a star
undergoing non-radial pulsations in the linear regime were laid by the
pioneering calculations of Osaki (1971), Dziembowski (1977), Balona \&
Stobie (1979), and Buta \& Smith (1979). Based on the Baade-Wesselink
technique, their work exploited the wavelength dependence of a
pulsational mode's amplitude and phase and enabled the inference of its
degree index $l$ based on light and radial velocity observations. The
equations were reformulated by Stamford \& Watson (1981) and Watson
(1988) for use with multi-colour photometric data alone. Their approach
proved very convenient for practical purposes, and has been applied to
many types of non-radially pulsating stars by a host of different
authors. More recent versions, comparable in scope to the Watson method, 
have been presented by Cugier, Dziembowski, \& Pamyatnykh (1994),
Heynderickx, Waelkens, \& Smeyers (1994), Balona \& Evers (1999),
Cugier \& Daszy\'nska (2001), and Townsend (2002). The latest advancement 
was proposed by Dupret et al. (2003) and consists of including a detailed
 discussion of non-adiabatic effects in the optically-thin atmospheric
 layers. All of the above treatments model the lightcurves of non-radially
 pulsating stars in terms of perturbations to the photospheric pressure or
 surface gravity, the effective temperature and the stellar radius. For cases
 where temperature effects completely dominate the brightness variations $-$ in
pulsating white dwarfs for instance $-$ a simpler approach is possible, as
was first discussed by Robinson, Kepler and Nather (1982) and later
exploited for ZZ Ceti white dwarfs by Brassard, Fontaine, \& Wesemael 
(1995, hereafter BFW95). For our application to pulsating sdB stars
below, we mostly follow the presentations of Cugier and Daszy\'nska
(2001) and Townsend (2002), but adopt a notation closer to that of BFW95.    

The Lagrangian perturbations to the stellar radius $R$, the effective
temperature $\teff$, and surface gravity $g_s$ caused by a single
non-radial pulsation may be expressed as 
\begin{eqnarray}
\frac{\delta R}{R}(\theta,\phi,t) & = & \Re\left[\frac{\delta r}{r^0}Y_l^m(\theta,\phi)e^{i\omega t}\right],\\
\frac{\delta \teff}{\teff}(\theta,\phi,t) & = & \Re\left[\frac{\delta T}{T^0}Y_l^m(\theta,\phi)e^{i\omega t}\right], \\ 
\frac{\delta g_s}{g_s}(\theta,\phi,t) & = & \Re\left[\frac{\delta g}{g^0}Y_l^m(\theta,\phi)e^{i\omega t}\right],
\end{eqnarray}
where $\Re [...]$ denotes the real part of a complex quantity, $\omega$
is the complex angular eigenfrequency of the mode (we define
$\omega_{nlm}\equiv\Re[\omega]=2\pi/P_{nlm}$, where $P_{ nlm}$ is the
pulsation period of a mode of radial order $n$, harmonic degree $l$ and
azimuthal order $m$), $t$ is the time coordinate, $Y_l^m(\theta,\phi)$
is the usual spherical harmonic function describing the angular
dependence of the mode, and $\frac{\delta r}{r^0}$, $\frac{\delta
  T}{T^0}$ and $\frac{\delta g}{g^0}$ are the radial components of the
complex eigenfunctions. We define the spherical harmonics in terms of
the polar angle $\theta$ and the azimuthal angle $\phi$ using the
nomenclature of Jackson (1975) as 
\begin{equation}
Y_l^m(\theta,\phi)={\sqrt{\frac{2l+1}{4\pi}\frac{(l-m)!}{(l+m)!}}P_l^m(\cos{\theta})e^{im\phi}},
\end{equation}
where the associated Legendre functions $P_l^m$ can be generated by the
equation 
\begin{equation}
P_l^m(x)=\frac{(-1)^m}{2^ll!}(1-x^2)^{m/2}\frac{d^{l+m}}{dx^{l+m}}(x^2-1)^l.
\end{equation}
With this definition, we may introduce the real quantity
\begin{equation}
\bar Y_l^m(\theta)\equiv Y_l^m(\theta,\phi)e^{-im\phi},
\end{equation}
which will be useful below and is described in more detail by BFW95. 
The radial components of the complex eigenfunctions may be written as
\begin{eqnarray}\label{eq:reigenfunction}
\frac{\delta r}{r^0} =&\left | \frac{\delta r}{r^0}\right | e^{i\phi_r}& \equiv \epsilon_r e^{i\phi_r}, \\
\frac{\delta T}{T^0} =&\left | \frac{\delta T}{T^0}\right | e^{i\phi_T}& \equiv \epsilon_T e^{i\phi_T}, \\
\label{eq:geigenfunction}
\frac{\delta g}{g^0} =&\left | \frac{\delta g}{g^0}\right | e^{i\phi_g}& \equiv \epsilon_g e^{i\phi_g},
\end{eqnarray}
where we have introduced the dimensionless amplitudes (moduli)
$\epsilon_r$, $\epsilon_T$ and $\epsilon_g$ of the complex radial
eigenfunctions, as well as their phases $\phi_r$, $\phi_T$ and $\phi_g$
with respect to some arbitrary value, which by convention is set to
$\phi_r$ = 0. Note that the upperscript `` $^0$ '' indicates the
unperturbed value of the variable of interest in the atmospheric layers.  

In the Watson approach, the brightness variation of a non-radially
pulsating star is calculated by assuming a dependence of the local flux
on the local instantaneous effective temperature and surface gravity,
and then integrating over the observed disk taking into account the
geometry of the mode and the radial distortions involved. As implemented
by Townsend (2002), the first order perturbation to the emergent
Eddington flux of a star undergoing a single non-radial pulsation can
thus be written in terms of the radial components of the complex
eigenfunctions detailed in equations 7$-$9 as 
{\setlength\arraycolsep{2pt}
\begin{eqnarray} \label{eq:townsend}
\frac{H_\nu^1}{H_\nu^0} & = & \Re\displaystyle\bigg[\displaystyle\bigg(\left\{(2+l)(1-l)\frac{\delta r}{r^0}\frac{I_{l\nu}^0}{I_{0\nu}^0}\right\}+\left\{\frac{\delta T}{T^0}\frac{1}{I_{0\nu}^0}\frac{\partial I_{l\nu}^0}{\partial\ln T^0}\right\}+{} \nonumber\\
& & {}+\left\{\frac{\delta g}{g^0}\frac{1}{I_{0\nu}^0}\frac{\partial I_{l\nu}^0}{\partial \ln g^0}\right\} \displaystyle\bigg) Y_l^m(\theta_0,\phi_0)e^{i\omega t}\displaystyle\bigg].
\end{eqnarray}}
Keeping with the notation of BFW95, $H_\nu^0$ is the unperturbed
emergent monochromatic Eddington flux, $H_\nu^1$ is its first-order
perturbation, and $I_{l\nu}^0$ is the angle-integrated unperturbed
emergent monochromatic specific intensity 
\begin{equation}\label{eq:ilnu}
I_{l\nu}^0=\int_0^1 I_\nu^0(\mu)P_l(\mu)\mu d\mu ,
\end{equation} 
with a weight function given by a Legendre polynomial $P_l(\mu)$. The
angles $(\theta_0,\phi_0)$ are the angular coordinates of the observer
in the spherical coordinate system of the star. The quantity
$I_{0\nu}^0$ refers to the specific case of $I_{l\nu}^0$ where the degree
index $l=0$. It is also understood that the derivative of $I_{l\nu}^0$
with respect to the effective temperature $T^0$ (surface gravity $g^0$)
is taken at constant surface gravity (effective temperature). 

In order to simplify equation (\ref{eq:townsend}) and to enable the
eventual elimination of the intrinsic amplitude of the
oscillation by taking flux ratios obtained at different wavelengths, it
is necessary to link the radial components of the three
eigenfunctions. By convention, we express the surface gravity and
temperature perturbations in terms of the radius
eigenfunction. Regarding the former, Cugier and Daszy\'nska (2001; see
also Dupret et al. 2003) have argued that one may approximate 
\begin{equation}\label{eq:delg}
\frac{\delta g}{g^0}\simeq -(2+\sigma_{nlm}^2)\frac{\delta r}{r^0}\equiv -D_{nlm} \frac{\delta r}{r^0}
\end{equation}
in the limit where the radial gradient of the amplitude of the pressure
perturbation ($|\delta P/P^0|$) is small. The real dimensionless
quantity $\sigma_{nlm}$ is given by 
\begin{equation}
\sigma_{nlm}\equiv \omega_{nlm}\sqrt{\frac{R}{g_s}},
\end{equation}
where $R$ is the stellar radius and $g_s$ is the surface gravity. Given
the definition of $\delta r/r^0$ and $\delta g/g^0$ in equations
(\ref{eq:reigenfunction}) and (\ref{eq:geigenfunction}), the
relationship implies that the radius and surface gravity variations
occur in phase, i.e., $\phi_g=\phi_r$ (=0, by convention). It must be
mentioned that equation (\ref{eq:delg}) differs from the more standard
expression employed in the Watson approach (see, e.g., equation (11) of
Cugier et al. (1994)), which suggests 
\begin{equation}\label{eq:watson}
\frac{\delta g}{g^0}\simeq -\left(4+\sigma_{nlm}^2-\frac{l(l+1)}{\sigma_{nlm}^2}\right)\frac{\delta r}{r^0}\equiv -C_{nlm}\frac{\delta r}{r^0}.
\end{equation}

In the original theory outlined by Stamford and Watson (1981) and Watson
(1988), the coefficient $C_{nlm}$ is multiplied by an additional term of
order unity denoted by $P^*(\equiv
\partial\log{g^0}/\partial\log{P^0}|_{\tau=1})$ when used in the context
of equation (\ref{eq:watson}). There has been some debate as to the
proper value to adopt for $P^*$, with some authors (e.g., Cugier et
al. 1994, Balona \& Evers 1999, Townsend 2002) arguing that it should
be taken strictly at unity, while most others employ grids of model
atmospheres to compute precise values. Regardless of whether or not the
$P^*$ quantity is included in equation (\ref{eq:watson}), the expression
remains considerably different from that recommended by Cugier and
Daszy\'nska (2001; our equation (\ref{eq:delg})), particularly when
considering long period modes. The latter is deemed more physical, as it
is more consistent with the outer boundary conditions employed during
pulsation calculations for a stellar model. We thus adopt the result of
Cugier and Daszy\'nska (2001) after verifying that we do indeed find
$|\delta P/P^0|$ to be mostly flat in the atmospheric layers of our subdwarf B
star models (as discussed below). 

The radial component of the temperature eigenfunction for its part may
be related to the radius perturbation by the expression 
\begin{equation}\label{eq:deltr}
\frac{\delta T}{T^0}=\frac{\left|\frac{\delta T}{T^0}\right|}{\left|\frac{\delta r}{r^0}\right |}e^{i(\phi_T-\phi_r)}\frac{\delta r}{r^0}\equiv\frac{\epsilon_T}{\epsilon_r}e^{i\psi_T}\frac{\delta r}{r^0},
\end{equation}
where we have introduced $\psi_T$ as the phase lag between the
temperature and the radius perturbation. In the adiabatic approximation,
we set $\psi_T=\pi$, since maximum temperature occurs at
minimum radius for $p$-modes. In the non-adiabatic case, the phase lag
can be evaluated using 
\begin{equation}\label{eq:psit}
\psi_T=\tan^{-1}\left(\frac{\Im\left[\frac{\delta T}{T^0}\right]}{\Re\left[\frac{\delta T}{T^0}\right]}\right)-\tan^{-1}\left(\frac{\Im\left[\frac{\delta r}{r^0}\right]}{\Re\left[\frac{\delta r}{r^0}\right]}\right),
\end{equation}
where $\Im [...]$ indicates the imaginary part, and $\Re [...]$ denotes
the real part of a complex quantity. We will apply this relation to
detailed non-adiabatic calculations with the aim of modelling $\psi_T$
as a function of depth, period, and degree index $l$ below.

In the standard Watson approach, it is customary to introduce a second
dimensionless real parameter $R$, which measures the departure from
adiabacity of the amplitude factor in equation (\ref{eq:deltr}), and can
hence be defined as 
\begin{equation}\label{eq:r}
R\equiv \frac{\epsilon_T/\epsilon_r}{\left|\left(\frac{\delta T}{T^0}\right)_{ad}\right|/\left|\left(\frac{\delta r}{r^0}\right)_{ad}\right|}.
\end{equation}   
It is obvious that, in the adiabatic limit, $R$=1. As for the
non-adiabatic case, many authors seem to believe that physically
acceptable values of $R$ are confined to the range $0\leq R\leq 1$. We
beg to differ, since we see no physical reason for $R$ not to exceed 1
and, indeed, find no justification for the constraint in the
literature. As far as we can see, it arose primarily from the fact that
observations of certain types of pulsating stars (such as the $\beta$
Cephei, $\delta$ Scuti and short-period Cepheids mentioned by Stamford
\& Watson 1981) indicated $0.25\leq R\leq 1$. While similar values
have been recovered computationally for other classes of pulsator (see,
e.g., Townsend 2002 describing slowly pulsating B stars), predictions
for white dwarfs have yielded $R$ values greater than 1 in some cases
(Robinson, Kepler, \& Nather, 1982). This is also what our pulsation
calculations indicate for the slowly pulsating subdwarf B stars, as will
be decribed in the next section.   

The quantities appearing in the denominator of equation (\ref{eq:r}) can
be described by recalling a well-known expression relating the moduli of
the temperature perturbation and the radius perturbation under the
assumption of adiabacity and the Cowling approximation, namely 
\begin{equation}
\left|\left(\frac{\delta T}{T^0}\right)_{ad}\right|=-\nabla_{ad}C_{nlm}\left|\left(\frac{\delta r}{r^0}\right)_{ad}\right|,
\end{equation}
where $\nabla_{ad} (=1-\Gamma_2^{-1})$ is the usual adiabatic temperature
gradient and $C_{nlm}$ is the same coefficient as introduced in equation
(\ref{eq:watson}). Here the eigenfunctions $(\delta T/T^0)_{ad}$ and $(\delta
r/r^0)_{ad}$ are real quantities, and the minus
sign corresponds to a phase shift of $\psi_T=\pi$. Using this relation
together with the definition of $R$, we can reformulate equation
(\ref{eq:deltr}) to yield 
\begin{equation}\label{eq:delt}
\frac{\delta T}{T^0}=R\nabla_{ad}C_{nlm}e^{i\psi_T}\frac{\delta r}{r^0}.
\end{equation}
Note that, in practice, the most direct way of computing the
relationship between $\delta T/T^0$ and $\delta r/r^0$ is to use
equation (\ref{eq:deltr}), calculating their relative phase $\psi_T$
from equation (\ref{eq:psit}), and evaluating their amplitude ratio
using 
\begin{equation}
\frac{\epsilon_T}{\epsilon_r}=\frac{\left\{\left(\Re\left[\frac{\delta T}{T^0}\right]\right)^2+\left(\Im\left[\frac{\delta T}{T^0}\right]\right)^2\right\}^{1/2}}{\left\{\left(\Re\left[\frac{\delta r}{r^0}\right]\right)^2+\left(\Im\left[\frac{\delta r}{r^0}\right]\right)^2\right\}^{1/2}}. 
\end{equation}
In addition, we then explicitly evaluate the adiabacity parameter $R$ using
 equation (\ref{eq:delt}). This will prove useful in 
numerical experiments aimed at assessing the impact of non-adiabatic
effects on the predicted pulsational amplitudes and phases in different
filters, as $\psi_T$ and $R$ can be input directly and set to the
adiabatic values for instance. 

Returning to equation (\ref{eq:townsend}), the expression describing the
perturbation to the emergent Eddington flux, we now seek to reformulate
the intensity terms and link them to both the BFW95 and the standard
Watson (1988) notation. It can be shown that the specific intensity term
appearing in the temperature perturbation can be related to the
monochromatic quantity $A_{l\nu}$ defined by BFW95 by the expression 
\begin{equation}\label{eq:tint1}
\frac{1}{I_{0\nu}^0}\frac{\partial I_{l\nu}^0}{\partial\ln T^0}=\frac{T^0 A_{l\nu}}{H_{\nu}^0},
\end{equation}
where, according to equation (17) of BFW95,
\begin{equation}
A_{l\nu}=\frac{1}{2}\int_0^1\frac{\partial I_{\nu}^0}{\partial T^0}P_l(\mu)\mu d\mu.
\end{equation}
By following the steps described in Appendix B of BFW95, we can re-write
equation (\ref{eq:tint1}) using the notion of the limb-darkening law
$h_{\nu}(\mu)$ as 
\begin{equation}\label{eq:tint}
\frac{1}{I_{0\nu}^0}\frac{\partial I_{l\nu}^0}{\partial\ln T^0}=\alpha_{T\nu}b_{l\nu}+\frac{\partial b_{l\nu}}{\partial \ln T^0},
\end{equation}
where 
\begin{equation}
\alpha_{T\nu}\equiv\frac{\partial \ln H_{\nu}^0}{\partial \ln T^0},
\end{equation}
and
\begin{equation}\label{eq:blnu}
b_{l\nu}\equiv \frac{\int_0^1 h_{\nu}(\mu)P_l(\mu)\mu d\mu}{\int_0^1 h_{\nu}(\mu)\mu d\mu}.
\end{equation}
The two last quantities, the logarithmic derivative of the emergent flux
with respect to the effective temperature, and the weighted
monochromatic limb darkening integral, are familiar notions in the
Watson model. It can be seen that, when expressing the intensity term of
the temperature perturbation, one has the choice of using a single
expression such as the left-hand side of equation (\ref{eq:tint})
employed by BFW95 or Townsend (2002), or the splitted terms on the right
hand side of the equation favoured in the more traditional
implementations of the Watson model. We choose to adopt the latter here
in order to facilitate comparisons with the Watson approach.  

Clearly, the specific intensity component of the gravity perturbation in
equation (\ref{eq:townsend}) can be expressed in an equivalent way,
yielding 
\begin{equation}\label{eq:gint}
\frac{1}{I_{0\nu}^0}\frac{\partial I_{l\nu}^0}{\partial\ln g^0}=\alpha_{g\nu}b_{l\nu}+\frac{\partial b_{l\nu}}{\partial\ln g^0},
\end{equation}
with
\begin{equation}
\alpha_{g\nu}\equiv \frac{\partial\ln H_{\nu}^0}{\partial\ln g^0}.
\end{equation}
Finally, we use the fact that 
\begin{equation}
I_{\nu}^0(\mu)=I_{\nu}^0(0)h_{\nu}(\mu)
\end{equation}
together with the definition of $I_{l\nu}^0$ (equation (\ref{eq:ilnu}))
and equation (\ref{eq:blnu}) to reformulate the specific intensity
coefficient in the radius term of equation (\ref{eq:townsend}) as 
\begin{equation}\label{eq:rint}
\frac{I_{l\nu}^0}{I_{0\nu}^0}=b_{l\nu}.
\end{equation}

We can now re-write the perturbation to the relative instantaneous
emergent monochromatic Eddington flux of equation (\ref{eq:townsend}),
using equations (\ref{eq:delg}), (\ref{eq:delt}), (\ref{eq:tint}),
(\ref{eq:gint}) and (\ref{eq:rint}), as 
{\setlength\arraycolsep{2pt}
\begin{eqnarray}\label{eq:eddmono}
\frac{H_{\nu}^1}{H_{\nu}^ 0} & = & \epsilon_r \bar Y_l^m(i)
\displaystyle\bigg[\left\{(2+l)(1-l)b_{l\nu}-D_{nlm}b_{l\nu}\alpha_{g\nu}-D_{nlm}\frac{\partial b_{l\nu}}{\partial \ln g^0}\right\}{} \nonumber\\
& & {}\times\cos(m\phi_0+\omega_{nlm}t)+{} \nonumber\\
& & {}+\left\{R\nabla_{ad}C_{nlm}b_{l\nu}\alpha_{T\nu}+R\nabla_{ad}C_{nlm}\frac{\partial b_{l\nu}}{\partial\ln T^0}\right\}{} \nonumber\\
& & {}\times\cos(m\phi_0+\omega_{nlm}t+\psi_T)\displaystyle\bigg],
\end{eqnarray}}
where we used the fact that $\theta_0=i$, the inclination angle, in
$\bar Y_l^m(i)$, the real function giving the viewing aspect (see BFW95
for details). 

In practice, the last equation will be applied to broadband, rather than
monochromatic photometry, so it is necessary to express it in terms of
frequency-integrated quantities. We thus introduce 
\begin{eqnarray}
b_{lx} & \equiv & \frac{\int_0^\infty W_{\nu}^xb_{l\nu}d\nu}{\int_0^\infty W_{\nu}^xd\nu}, \\
\alpha_{Tx} & \equiv & \frac{\int_0^\infty W_{\nu}^x\alpha_{T\nu}d\nu}{\int_0^\infty W_{\nu}^xd\nu}, \\
\alpha_{gx} & \equiv & \frac{\int_0^\infty W_\nu^x\alpha_{g\nu}
  d\nu}{\int_0^\infty W_\nu^x d\nu}, 
\end{eqnarray}      
where $W_\nu^x$ represents the transmission function for filter $x$,
convolved, in principle, with the response of the telescope/detector
combination and the atmospheric extinction curve at a given site. 

We can now regroup the various components of equation (\ref{eq:eddmono})
into five terms analogous to those employed in the traditional Watson
model. We thus introduce the following short-hand notation: 
\begin{eqnarray}
T_1 & \equiv & R\nabla_{ad}C_{nlm}b_{lx}\alpha_{Tx}, \\
T_2 & \equiv & R\nabla_{ad}C_{nlm}\frac{\partial b_{lx}}{\partial \ln T^0}, \\
T_3 & \equiv & (2+l)(1-l)b_{lx}, \\
T_4 & \equiv & -D_{nlm}b_{lx}\alpha_{gx}, \\
T_5 & \equiv & -D_{nlm}\frac{\partial b_{lx}}{\partial \ln g^0}.
\end{eqnarray}
We finally follow Koen (1998) and define
\begin{eqnarray}
\gamma_1 & \equiv & T_1+T_2, \\
\gamma_2 & \equiv & T_3+T_4+T_5,
\end{eqnarray}
which represent the effects on the brightness variation during a
pulsation cycle due to effective temperature ($\gamma_1$) and
radius/gravity perturbations ($\gamma_2$) respectively. Using a
trigonometric identity we may then obtain our final expression for the
relative flux in a photometric bandpass $x$ arising from the excitation
of a single pulsation mode 
\begin{equation}\label{eq:amplitude}
\frac{H_x^1}{H_x^0}=\epsilon_r\bar Y_l^m(i)A_{nlm}^x\cos(m\phi_0+\omega_{nlm}t+\phi_{nlm}^x),
\end{equation}
with the wavelength-dependent amplitude given by
\begin{equation}
A_{nlm}^x=(\gamma_1^2+\gamma_2^2+2\gamma_1\gamma_2\cos{\psi_T})^{1/2},
\end{equation}
and the wavelength-dependent phase given by
\begin{equation}\label{eq:phase}
\phi_{nlm}^x=\tan^{-1}\left(\frac{\gamma_1\sin{\psi_T}}{\gamma_1\cos{\psi_T}+\gamma_2}\right).
\end{equation}
Strictly speaking, this expression is valid only for non-rotating stars
where the amplitude and phase depend on the radial order $n$ and the
degree index $l$, but not the azimuthal order $m$. However, Cugier \&
Daszy\'nska (2001) have argued that it may also be applied to slowly rotating
stars, loosely defined as those with a spin parameter
$S=2\Omega/\omega_{nlm}<0.5$ (where $\Omega$ is the rotation
frequency). In this case, the $\omega_{nlm}t$ term in the cosine
function of equation (\ref{eq:amplitude}) is replaced by
$(\omega_{nlm}-m\Omega)t$. The amplitude and phase also become sensitive
to the azimuthal index $m$, but only through the period dependence of an  
$m$ component in a rotationally split $(2l+1)$ multiplet. For fast
rotators with higher values of $S$, things become more complicated and a
rigorous treatment such as the one developed by Townsend (2003) becomes
necessary.    

In practice, equation (\ref{eq:amplitude}) is not used directly since it
depends on the unknown factor $\epsilon_r\bar Y_l^m(i)$. Instead, we take
advantage of the wavelength independence of the latter and eliminate it
by calculating the amplitude ratios $A_{nlm}^x/A_{nlm}^y$ and phase
differences $\phi_{nlm}^x-\phi_{nlm}^y$ arising from the lightcurves for
two different bandpasses $x$ and $y$. Mode discrimination can then
proceed by exploiting the fact that the wavelength dependence of an
oscillation's amplitude and phase may depend strongly on its degree index
$l$, and that observational amplitude ratios and phase differences are
readily obtainable from multi-colour photometry. 

We note that if non-adiabatic effects are neglected in the calculations
of the theoretical quantities, then $\psi_T=\pi$, the phase $\phi_{nlm}$
calculated from equation (\ref{eq:phase}) will always be zero, and no
phase shifts will be predicted between different bandpasses. However, a
small or negligible observed phase shift does not necessarily imply the
absence of non-adiabatic effects, and one may well encounter situations
where the expected phase shifts remain quite small despite important
deviations of the adiabacity parameter $R$ from its adiabatic value of
$R=1$. This is the case for our model of a typical EC 14026 star
discussed below. On the other hand, the observed phase shifts will be
expected to be negligible if effective temperature perturbations
completely dominate the brightness variation and radius/surface gravity
effects can safely be ignored ($\gamma_2=0$ in the limiting case). In
that instance, equation (\ref{eq:phase}) immediately infers
$\phi_{nlm}^x=\psi_T$ irrespective of the bandpass in
question. Moreover, the period dependence of the $T_1$ and $T_2$ terms
through their common factor $RC_{nlm}$ cancels out in the calculation of the 
amplitude ratios, with the result that the latter bear the signature
only of the degree index $l$ and not the period dependent radial order
$n$. By the same logic, the amplitude ratios are no longer affected by
non-adiabatic effects, a point on which we concur with Ramachandran 
et al. (2004). In retrospect, this alleviates the worry expressed by
Robinson et al. (1982) about the legitimacy of 
using the adiabatic relationship between $\delta T/T^0$ and $\delta
r/r^0$ in their discussion of colour variations in pulsating white
dwarfs (see also BFW95). When considering the pulsations of white dwarfs
in the linear regime, the brightness variations are completely dominated
by temperature effects, which implies an absence of phase shifts between
the light curves of different colours, as well as amplitude ratios
insensitive to non-adiabatic effects and the period of the mode in
question. The latter then depend only on the degree index $l$, a
situation also encountered for high-order $g$-modes in our typical PG
1716 star model discussed below. 

\section{MODEL ATMOSPHERES AND MONOCHROMATIC QUANTITIES}

The quantities required in the theoretical framework discussed above 
can broadly be divided into three groups: those that can be inferred 
observationally, those that must be computed on the basis of full
stellar models, and finally those that are derived from model atmospheres. 
In this section we focus on the latter group, which includes the 
monochromatic quantities  $\alpha_{T\nu}$, $\alpha_{g\nu}$, $b_{l\nu}$,
$b_{l\nu,T} \equiv {\partial b_{l\nu}}/{\partial \ln T^0}$, and 
$b_{l\nu,g} \equiv {\partial b_{l\nu}}/{\partial \ln g^0}$. Since their
computation involves not only the standard specific intensities,
 but also the corresponding derivatives with respect to effective temperature
 and surface gravity across the visible disk, we needed to modify our model 
atmosphere code for subdwarf B stars in order to carry out the task 
efficiently and accurately. We constructed a grid of model atmospheres
 defined at 9 gravity points (log $g$ = 4.8 to 6.4 in steps of 0.2 dex) and 11 
temperature points ($\Teff$ = 20,000 to 40,000 K in steps of 2000 K) 
representative of the distribution of sdB stars in the $\log{g}-\teff$ plane. 
 Detailed interpolation within this grid enables the calculation of the 
 desired quantities for any given $\log{g}-\teff$ combination. The 
 model atmospheres are computed under the assumption of LTE and
 uniform composition specified by log $N(He)/N(H)$ =$-$2.0, a typical
 value for sdB's. Metals were not included, since subdwarf B stars are known to
 be chemically peculiar, and metal abundances vary from one target to
 the next. While it could prove interesting to incorporate representative 
 metal abundances and thus assess the importance of metals in this kind 
of calculation in the future, our H/He LTE model grid is quite sufficient 
 for the purposes of this study. 

As mentioned above, the colour-amplitude technique is usually applied to
 broadband photometry, but we find it instructive to first examine the
behavior of the key monochromatic quantities. We begin with the unperturbed
 emergent Eddington flux, which is shown in the top panel of
Figure 1 in the optical domain for a model with $\Teff$ = 33,000 K and
 log $g$ = 5.75. These are also the atmospheric parameters we adopt for
 our representative EC 14026 star model in the next section,
 where more details are provided. As is typical for the observed optical
 spectra of sdB stars, the Eddington flux is characterized by the presence of
broad hydrogen Balmer lines and several weak and narrow helium lines. 
For comparison, the middle panel illustrates the predicted first-order
perturbation to the emergent Eddington flux assuming a nonradial
pulsation with a period of 150 s (a typical low-order $p$-mode in an EC
14026 star) and six values of the degree index from $l$ = 0 (top curve)
to $l$ = 5 (bottom curve). Note that here, $H_{\nu}^1$ is divided by the
unknown factor $\epsilon_r\bar Y_l^m(i)$. It is evident that the pulsational 
amplitude rapidly decreases with increasing $l$, which is a direct
manifestation of the well-known geometric cancellation effects
associated with an increasing number of nodal lines crisscrossing the
visible disk. For values of $l$ $>$ 2, the decrease of the amplitude with
increasing $l$ at a given frequency is no longer monotonic, but also depends
 on the limb-darkening law of the model atmosphere in question in quite a 
complex way. The amplitude  
of the $l$ = 4 curve for instance is higher than that of the $l$ = 3 curve 
in the optical domain shown, and the latter dips below the $l$ = 5 curve 
above $\sim$ 4000 \AA. By dividing the curves by the unperturbed flux
$H_{\nu}^0$, one obtains the $relative$ monochromatic amplitude of the
assumed mode (again to within the unknown parameter $\epsilon_r\bar
Y_l^m(i)$), as indicated in the bottom panel of the figure. It is the 
 latter quantity that forms the basis of the colour-amplitude technique,
 which relies on comparing the relative amplitude at two wavelengths 
in order to eliminate the unknown factor. 

The result of such an operation is illustrated in Figure 2, where we have 
divided the relative monochromatic amplitude curves for each $l$ by 
the corresponding relative amplitude at an (arbitrary) frequency point
 in the continuum with $\lambda$ = 3650 $\rm\AA$. In this particular 
example, there is little difference between curves with $l$ = 0, 1, 
and 2, however those with $l$ = 3, 4, and 5 bear a stronger and more 
dictinct signature of their degree index. Along with the amplitude 
ratios, the phase differences between the brightness variations at different 
wavelengths may also be used to infer the degree $l$ of a pulsation
 mode. This is shown in Figure 3, where we plot the monochromatic
 phase difference with respect to the spectral point at 3650 $\rm\AA$
 for the same assumed pulsation mode with a period of 150 s and $l$ 
values from 0 to 5 as indicated. In our example, the phase shifts remain
 relatively small (less than a few degrees) over the optical domain. 
This implies that, in practice, mode discrimation will be difficult 
to achieve on the basis of phase differences alone. 

 The amplitude ratios and phase shifts discussed in the following
 sections are simply frequency-integrated counterparts to the
 monochromatic curves pictured in the last two figures. In this
 context it should be noted that the spikes associated with the central cores
 of the absorption lines, and in particularly those related to the narrow
 helium lines, do not significantly contribute to the bandpass integrated 
quantities. We would like also to recall that the behaviour of the amplitude
 ratios and phase shifts with wavelength depends not only on the degree
 index $l$, but also on the period of the mode and, of course, on the
 atmospheric parameters of the model in question. 

\section{NONADIABATIC EFFECTS IN REPRESENTATIVE MODELS OF PULSATING SDB
  STARS}

We now evaluate the quantities $R$ and $\psi_T$ through the use of full
nonadiabatic pulsation calculations. To do this, we use the same 
numerical tools employed earlier by Charpinet et al. (1997; see also
Fontaine et al. 1998, Charpinet et al. 2001, and Fontaine et al. 2003b)
to construct their second-generation stellar models. These are characterized
 by four free parameters, the effective temperature $\Teff$, the surface
gravity log $g$, the fractional mass contained in the H-rich envelope
$M(H)/M_*$, and the total mass $M_*$. The models feature an opacity profile
that largely depends on the nonuniform distribution of iron as a
function of depth. This distribution results from the competition
between gravitational settling and radiative levitation and has been
shown to be responsible for the excitation of low-order $p$-modes in
 models of EC 14026 stars as well as high-order $g$-modes in models of
 PG 1716 stars (Fontaine et al. 2003b) through the $\kappa$-mechanism. 
Note that, according to the same authors, standard models of
sdB stars with uniform metallicity are unable to excite pulsation modes.

In this section we will focus on a representative model of an EC 14016 star
and a PG 1716 pulsator respectively. According to Figure
1 of Fontaine et al. (2004), which summarizes the location of subdwarf B 
stars on the H-R diagram, a typical EC 14026 star has log $g$  $\simeq$ 5.75
 and $\Teff$ $\simeq$ 33,000 K, which implies that it is significantly denser
 and hotter than its typical PG 1716 counterpart at log $g$ $\simeq$ 5.40 
and $\Teff$ $\simeq$ 27,000 K. We adopt these values, and list some of  
the models' other characteristics, including the two defining variables
 $M(H)/M_*$ and $M_*$ in Table 1. We explicitly give the value of the total
radius as well as that of the adiabatic temperature gradient
averaged over the atmospheric layers (see below), as both of these
quantities enter into our equations. For completeness, we also
provide the value of the quantity $P^*$ evaluated from our model
atmosphere grid at the appropriate values of ($\Teff$, log $g$), despite the 
fact that we do not use that variable in our calculations, since we choose 
to relate the surface gravity and radius perturbation via the expression 
developed by Cugier $\&$ Daszy\'nska (2001). 

Taking into account the range of periods observed in typical EC 14026
stars, we compute all modes with periods in the range 80$-$300 s and
with values of $l$ from 0 to 5 for our representative short-period pulsator
 model. For the case of our PG 1716 star model, we calculate modes with
 periods in the range 2000$-$6000 s and with values of $l$ = 1, 2, and 3.
 The first results are illustrated in Figures 4 and 5, where we show
the modulus of the radial component of each mode's nonadiabatic pressure
eigenfunction as a function of depth in the outermost layers for the
 EC 14026 and PG 1716 model respectively. We define the ``atmospheric 
layers of interest'' as those lying between the optical depths $\tau$ = 0.1
 and $\tau$ = 10.0. It can be seen that about half of the modes
 considered in the EC 14026 model show small gradients across these
 layers as is required in the Cugier \& Daszy\'nska (2001) expression
 connecting the surface gravity perturbation to the total radius 
perturbation. The other modes, which systematically correspond to those with
 shorter periods, do not strictly pass this test but, lacking a more 
generally applicable relation, there is little we can do to remedy 
this shortcoming. The situation is better for the PG 1716 model, where 
the vast majority of the modes considered show very small
$|\delta P/P|$ gradients across the atmospheric layers, with the exception
of the few modes with the longest periods. As it turns out, we will
find out below that surface gravity perturbations contribute very little
to the brightness variations in pulsating sdB stars, which is lucky in 
the context of our present ``problems'', as any inaccuracies in the surface 
gravity perturbations will have little impact on the final results.

Figures 6a and 6b show the behaviour of the adiabaticity parameter $R$
 in the atmospheric layers of our EC 14026 model for all modes of interest.
 Figures 7a and 7b illustrate similar results for the phase lag $\psi_T$. 
The figures give the distinct impression that both $R$ and $\psi_T$ depend
 primarily on the period and very little on the degree index $l$. This can, 
in fact, be confirmed quantitatively by computing an ``average'' atmospheric
 value of both quantities for each mode, and plotting them as
functions of the mode's period as shown in Figure 8. The averaging
process used is a simple unweighted integration over all atmospheric
layers between $\tau$ = 0.1 and $\tau$ =10.0. We believe that this
approach is slightly more rigorous than simply taking the local values at the
photosphere itself ($\tau$ = 2/3). Figure 8 highlights the 
 almost perfect one-to-one relationship that exists between the value of
 $<R>$ and  the period of the mode for the range of interest. A similar
 situation  
 is encountered for the averaged quantity $<\psi_T>$. In both cases there
 is very little, if any, dependence on the degree index $l$, allowing us to
 model $<R>$ and $<\psi_T>$ as functions of the period $P$ in a simple and
 accurate way (we drop the subscript ``$nlm$'' in what follows for a more 
concise notation). While this is not necessary when examining a specific
 equilibrium model for which individual values can be obtained for each
 mode (e.g., the small circles in Figure 8), the procedure will prove useful
 in our applications below, where we want to treat the pulsation period 
as a free continuous variable rather than a discrete eigenvalue.
 Values of $R$ and $\psi_T$ sufficiently accurate for our needs were
 obtained by $\chi^2$-fitting the data points in Figure 8 to cubic curves
 as illustrated. The cubic solutions are given by 
\begin{equation}
<R> =
0.3459-3.031\times10^{-4}P+1.784\times10^{-5}P^2-3.040\times10^{-8}P^3,
\end{equation}
and
\begin{equation}
<\psi_T> = \pi + 
1.1010-1.415\times10^{-2}P+3.702\times10^{-5}P^2-2.122\times10^{-8}P^3,
\end{equation}
which are formally valid in the period range 80$-$300 s and for values
of the degree index in the range $l$ = 0$-$5. Figure 8 reaffirms the
well-known fact that non-adiabatic effects are never negligible in
stellar atmospheres, but it is the particular dependence of $<R>$ (and
$<\psi_T>$) on the period that is of central interest here and that
cannot be ignored in the computations. In future studies, 
the accuracy of the estimates for $R$ and $\psi_T$ could be further 
improved by examining the nonadiabatic pulsation equations in the presence
 of optically thin layers more closely, for example by following the 
theory of Dupret (2001) and Dupret et al. (2003).

The situation is slightly more complicated for the high-order $g$-modes in
our representative PG 1716 model, where $R$ and $\psi_T$
depend on both the period $P$ and the index $l$ as is illustrated in
Figures 9 and 10. The presence of a few  trapped modes (trapped
above the H-rich envelope/He core interface) renders things somewhat
more challenging in terms of defining an ``average'' behavior for a
pulsation mode. Furthermore, while $<\psi_T>$ depends monotonically on
the period for a given $l$ (at least in the period range of interest),
this is not the case for $<R>$, making it ever more tedious to 
acceptably fit the data points. As illustrated in Figure 11, we were
finally able to find fits to $<R>$ and $<\psi_T>$ sufficiently accurate
  for our present needs by using the following analytic relationships
\begin{equation}
<R>l^{0.13} =
0.5817+6.399\times10^{-4}(Pl^{0.38})-1.164\times10^{-7}(Pl^{0.38})^2+6.082\times10^{-12}(Pl^{0.38})^3,  
\end{equation}
and
\begin{equation}
<\psi_T> = \pi -2.30e^{(Pl^{0.38}/1800)}.
\end{equation}

Figure 11 clearly indicates that nonadiabatic effects are important in
the atmospheres of PG 1716 stars, and should not be neglected in the 
computations. As mentioned in the theory section, the adiabaticity
 parameter $<R>$ may become larger than 1 for the long period $g$-modes
characteristic of these objects. However, as we will see below, temperature
variations increasingly dominate the brightness variations in the
 limit of long periods, causing the $observable$ effects of a departure
 from adiabaticity to disappear.

\section{APPLICATION TO UBVRI PHOTOMETRY}

As a representative application, we consider multi-colour photometry in
the standard Johnson-Cousins system $UBVRI$. For each reference model
(specified by a value of $\Teff$ and a value of log $g$ for the model
atmosphere calculations), we computed the
frequency-integrated quantities $\alpha_{Tx}$, $\alpha_{gx}$, $b_{lx}$, 
$b_{lx,T} \equiv {\partial b_{lx}}/{\partial \ln T^0}$, and 
$b_{lx,g} \equiv {\partial b_{lx}}/{\partial \ln g^0}$, where the
subscript ``$x$'' denotes one of the wavebands. In our modelling of the
effective response curves $W_{\nu}^x$, we convolved the standard
transmission functions of the $UBVRI$ filters with the extinction curve
of Kitt Peak National Observatory (representative of sites with
2000$-$2500 m altitudes). We also assumed a grey response for the
instrument/telescope combination. Our values for the
frequency-integrated quantities are given in Table 2 (Table 3) for our
reference EC 14026 (PG 1716) star model. For the former model, we list the 
values for modes with degree indices in the range $l$ = 0$-$5,
while for the latter they are provided for modes with $l$ between 1 
and 6 since the case $l$ = 0 is of no interest for the long-period 
variables (as a rule, the $p$-branch period spectra of sdB stars do not
reach into the range of the long periods observed in PG 1716 stars). 

We draw attention to the fact that in both Table 2 and Table 3
 $\alpha_{gx} << \alpha_{Tx}$ and $b_{lx,g} << b_{lx,T}$ with only
 very few exceptions. This means that the $T_4$ and $T_5$ terms 
will generally be very small compared to the $T_1$ and $T_2$ terms 
in the equations discussed in Section 2, implying that in sdB stars
 the contributions to the brightness variations due to surface gravity
 perturbations are small compared to the effects of effective
temperature perturbations ($T_1$ and $T_2$) and radius changes
($T_3$). Thus, the exact expression used to relate the surface gravity
 perturbation to the radius perturbation (see our discussion of the 
Cugier \& Daszy\'nska 2001 method versus the more traditional one in 
Section 2) is of minor importance for sdB stars. 

Employing the data listed in Tables 1, 2, and 3 as well as our polynomial 
models for $R$ and $\psi_T$ given by equations (44) through (47), we
computed amplitudes and phases for a number of modes according to our
equations (42) and (43). The period was treated as a free parameter, which 
enabled a detailed examination of its influence on the final results. 
Keeping with tradition, the amplitudes ratios and phases shifts were 
 calculated with respect to the bluest filter, in this case the $U$ bandpass.
 The results for our representative EC 14026 star model and those
for our PG 1716 model are discussed separately in the following subsections.

\subsection{Results for our representative EC 14026 star model}

We first show, in Figure 12, the ratio $|\gamma_1/\gamma_2|$ as a
function of effective wavelength in the five filters considered for three
 representative periods spanning the range of those observed in a typical 
EC 14026 pulsator. All the modes illustrated are associated with degree 
indices $l$ = 0$-$5 and correspond to low-order $p$-modes. Excepting the
modes with $l$ = 1, for which $T_3$ is always equal to 0 (see our
equation (36)), as well as the $l$ = 3 modes in the $I$ bandpass, for
which the contributions of the effective temperature perturbations are
particularly small, the figure infers that both effects
(i.e., temperature and radius changes) contribute to the brightness
 variations in a non-negligible way. Unlike for white dwarfs, 
it is thus not appropriate to assume that temperature effects completely 
dominate in EC 14026 stars. While we already mentioned that the $T_4$
 and $T_5$ terms usually remain quite small, it is the $T_3$ term, whose
 numerical value dominates $\gamma_2$, that becomes appreciable 
compared to $\gamma_1$ (= $T_1 + T_2$).  

In order to assess the importance of non-adiabatic effects, we repeated
the calculation of $|\gamma_1/\gamma_2|$, this time imposing the adiabatic
 values $R$ = 1 and $\psi_T$ = $\pi$ for all modes considered. The results
 are represented by the dotted lines in Figure 12. Keeping in mind the 
logarithmic scale of the ordinate axis, it is obvious that non-adiabatic
 effects are significant in the composition of brightness variations in 
EC 14026 pulsators.

Figure 13 shows the expected phase shifts for the same set of modes. It can 
 be seen that, in the optical domain, the phase shifts depend on the period
 quite sensitively, while remaining relatively small and reaching a maximum
 of a few degrees at the most in the period range of interest. In the 
adiabatic approximation we expect no phase shifts at all. This means that, 
in practice, it will be very difficult to exploit the (weak) $l$ dependence
of such phase shifts. The situation seems more promising for the
 amplitude ratios illustrated in Figure 14. However,
while the modes with $l$ = 3, 4, and 5 bear distinct signatures of their 
degree index, the capacity to discriminate between modes with $l$ = 0, 1,
 and 2 on the basis of optical photometry appears much more limited. 
Observational amplitude measurements for such modes will have to be 
unusually precise if they are to be used as mode discriminators in EC
14026 stars. We would like to point out that non-adiabatic effects on the
amplitude ratios are generally non-negligible, as can be deduced by
comparing the continuous lines (non-adiabatic results) to the dotted
 lines (adiabatic values) in Figure 14. It may also be worth  
mentioning that the results depicted in the middle panels
of Figures 13 and 14 (the case with $P$ = 150 s) are consistent with
 the monochromatic results illustrated in Figures 2 and 3, as is to be
 expected.

In Figures 15a and 15b (separated for visualization purposes), we plot
the expected amplitude ratios for modes with $P$ = 100 s (dotted lines),
150 s (dashed lines), and 200 s (long-dashed lines), with the aim of
 emphasizing their strong period dependence. In order to compare the
 results to corresponding period-independent values, we also include
 amplitude ratios (solid lines) obtained by postulating that effective
 temperature 
 perturbations completely dominate the brightness variations (i.e.
by setting $\gamma_2$ = 0 in the calculations). As suggested above,
 it becomes evident that this not a good working assumption for the EC 14026 
stars.

Finally, in Figure 16, we show the results of a numerical experiment in
which we explicitly adopt $T_3$ = 0 in order to assess the relative
impact of the $T_4$ and $T_5$ terms on the amplitude ratios. The
resulting period-dependent ratios are indicated by the dotted lines and 
contrast with the continuous lines corresponding to the period-independent
 limiting case in which $\gamma_2$ = 0. While we previously argued that the
 contributions of the $T_4$ and $T_5$ terms are generally be quite small
 in pulsating sdB stars, the figure reveals that non-negligible effects 
remain for certain cases, especially in the $B$ band. We thus recommend 
that these terms be kept by in the calculations, particularly since they 
are readily computable.

\subsection{Results for our representative PG 1716 star model}

The computations for our reference PG 1716 model uncover two distinct 
regimes as far as the relative importance of the temperature perturbation 
is concerned. At the lower end of the period range of interest, contributions
 from both the $\gamma_1$ and $\gamma_2$ terms are significant, whereas the 
brightness variations are dominated by temperature changes for the longer
 periods involved. This can be seen from Figures 17a and 17b, 
where we show the ratio $|\gamma_1/\gamma_2|$ as a function of effective
wavelength in a format similar to that of Figure 12. In these and the 
following figures, we illustrate our results for six different periods between
 $P$= 2400 s and $P$ = 6000 s. Comparing Figures 17a and 17b clearly reveals
that the  ratio $|\gamma_1/\gamma_2|$ monotonically increases with period 
for all filters and degree indices $l$. This is a direct consequence of the 
 the $\gamma_1$ term increasing in absolute value with period via the
period-dependent coefficient $C_{nlm}$ for the long periods considered,
 while $T_3$ remains constant for a given filter and degree index. We recall
 that modes with $l$ = 1 represent a special case, since $T_3$ = 0 in that
 instance.  

This transition from one regime to the other strongly influences
 the phase shifts and amplitude ratios calculated. The former are depicted 
in Figures 18a and 18b for the same set of modes as considered above. 
For the shorter periods we predict significant phases shifts, which reach
more than 10 degrees for the 2400 s period. Unlike the much smaller phase 
shifts expected for the EC 14026 stars, this is well within achievable
measurement accuracy. At the same time, we find the magnitude of the
 expected phase shifts to drop very rapidly with increasing period in
 PG 1716 stars (note, in particular, the change of the ordinate scale
 between Fig, 18a and Fig. 18b), to the point where they would not be 
detected with any confidence for the longer period modes. 

The expected amplitude ratios are shown in Figures 19a and 19b. Once again, 
the transition from the period-dependent regime, where $\gamma_1$ and
$\gamma_2$ are of similar importance, to the period-independent case
 dominated by effective temperature perturbations is striking. The plots
 also emphasize the diminishing importance of nonadiabatic effects with 
increasing period, as can be seen from the convergence of the
 dotted (adiabatic) and continuous (non-adiabatic) lines. Thus, in our 
PG 1716 model nonadiabatic effects range from extremely significant 
for $P$ = 2400 s to practically negligible for $P$ = 6000 s.  

In Figures 20a and 20b (separated for visualization purposes), we compare
the expected amplitude ratios for modes with $P$ = 3000 s (dotted lines),
4000 s (dashed lines), 5000 s (long-dashed lines), 6000 s (dot-dashed
lines), 7000 s (dot-long-dashed lines), and 8000 s (dashed-long-dashed
lines), again with the aim of highlighting the period dependence of 
the results. The convergence of the ratios to the period-independent 
values (solid lines) in the limiting case of long periods is recovered 
nicely. As before, the period-independent ratios are obtained by 
assuming that effective temperature perturbations completely dominate 
the brightness variations (i.e., by setting $\gamma_2$ = 0 in the 
calculations). 

Analogously to the case presented in Figure 16 for our EC 14026
model, we carry out supplementary calculations where we explicitly set
 $T_3$ = 0 in order to assess the relative impact of the $T_4$ and
 $T_5$ terms on the amplitude ratios. The results of this
numerical experiment are shown in Figure 21. As in Figure 16, the resulting
period-dependent ratios are indicated by dotted lines and are compared
to the solid lines representing the period-independent limiting case
for which $\gamma_2$ = 0. In contrast to the previous plot, the contributions
 of the $T_4$ and $T_5$ terms are essentially negligible in PG 1716 stars, 
apart for in the very shortest periods. For the three longest periods (not 
illustrated) the two curves are not distinguishable.
 
Having discussed the signature of the period and the degree index, 
as well as that of the different terms contributing to the brightness 
variations on bandpass-integrated wavelength, we now focus specifically on 
results expected from the $U-I$ filter combination (which corresponds to the 
longest wavelength base in the system used here). Figure 22 shows the $I$ to
 $U$ amplitude ratio as a function of the degree index $l$ on the basis of
calculations extending to $l$ = 8. From bottom to top, the various curves
 correspond to periods with $P$ = 2400, 3000, 4000, 5000, 6000, 7000 and
8000 s respectively. The top curve represents an infinite period, to which
 the finite period curves converge in the limiting case. While evident from 
previous figures, the peculiar behaviour of the $l$=3 and 5 modes is 
particularly striking here. In practice, these are the modes most likely to 
be identified from multi-colour photometry, since the amplitude ratios 
of the remaining modes may not differ enough to the extent where they
 can be resolved observationally. 

Finally, Figure 23 and 24 respectively show the $I/U$ amplitude ratio
 and the $I-U$ phase shift expected for our PG 1716 model as a function 
of period for modes with $l$=1 (black), 2 (grey), 3 (blue), 4 (cyan),
 5 (green) and 6 (red). Note that here, the periods illustrated were not
 considered as free parameters, but are the solution of the eigenvalue
 problem. Once again, the convergence of values to the 
period-independent value in the limiting case of long periods is evident.
 Interestingly, in Figure 23 the results for the very shortest periods 
break with the systematic decrease in amplitude ratio with period, and 
slightly increase instead.  These same periods are associated with
 particularly large (potentially detectable) phase shifts in Figure 24,
 however it should be noted that these are shorter than the periods so
 far observed in PG 1716 stars and of primarily theoretical interest.  

\subsection{Results for other models: Dependence on $\Teff$ and log $g$}

In this section, we go beyond the two representative subdwarf B star models 
explored above and discuss the dependence of amplitude ratios and phase
 differences on the atmospheric parameters $\log{g}$ and $\teff$. To this 
end, we constructed a sequence of subdwarf B star models parallel to the
 zero-age extreme horizontal branch, the vital parameters of which are
 detailed in Table 4. Note that the surface gravity increases with effective 
temperature, while the thickness of the hydrogen-rich outer layer decreases.
 The division between the PG 1716 and EC 14026 regime is based on 
spectroscopically determined values of $\log{g}$ and $\teff$ for the two 
types of pulsator, such as those illustrated in Figure 1 of Fontaine et al. 
(2004). In addition to the fundamental parameters, Table 4 also gives the 
approximate period range in which modes are believed to be excited for each 
model. For the EC 14026 stars, these correspond to modes 
that are found to be unstable from our non-adiabatic pulsation calculations,
 since the latter have been shown to predict the observed period ranges very 
accurately. Note that we have uniformly decreased the iron abundance in our 
models by a factor of 3 compared to the original ``second-generation'' models, 
 as suggested by Fontaine et al. (2003b). In the case of the PG 1716
 stars the situation is more complicated due to the discrepancies that 
still exist between modelled
 and observed instability regions (see our discussion in the Introduction). 
 We can however constrain the unstable period ranges for the PG 1716 models 
in our sequence on the basis of the frequencies extracted from long-period 
variable subdwarfs observed. To date, a quantitative analysis of the period 
spectrum has been possible for three targets spanning the range of PG 1716 
stars in effective temperature: PG 1627+017 at $\teff\sim$ 23,000 K (Randall
 et al. 2004a), PG 1338+481 at $\teff\sim$ 26,000 K (Randall et al. 2004b) and 
PG 0101+039 at $\teff\sim$ 28,000 K (Randall et al. 2005). The study of these
 objects revealed that the width and mode density of the excited period range, 
as well as the numerical values of the periods themselves, decrease 
systematically with increasing temperature, enabling us to infer approximate 
instability bands for all models in our sequence by interpolation of the 
observed values. It is these estimated period ranges, rather than precisely 
modelled quantities, that are given in Table 4.

For each model in the sequence, we computed the necessary model atmosphere
 parameters in the same bandpasses as used for the representative models for 
modes with $l=0-6$ according to their specific $\log{g}-\teff$ combination. 
 The stellar radius was likewise determined for every model on the basis of 
its mass and surface gravity, and its characteristic periods were evaluated
from pulsation theory in the range of interest. On the other hand, 
the remaining three quantities derived from the full stellar models and 
non-adiabatic pulsation calculations were approximated to the values 
determined for the appropriate representative model. Thus the values of
 $<\nabla_{ad}>$ were taken from Table 1, while $<R>$ and $<\psi_T>$ were
 computed  using equations (44) and (45) or expressions (46) and (47) for
 the EC 14026 and PG 1716 models respectively. The resulting accuracy of 
$<R>$ and $<\psi_T>$ is deemed sufficient for the illustrative purposes 
sought here, however it is understood that for quantitative analyses of 
observed multi-colour photometry the fitting process outlined in section 4 
will have to be repeated for the designated object.

Figures 25 and 26 respectively show the phase shifts and amplitude ratios 
predicted for our sequence of subdwarf B star models from the $I-U$ bandpass 
combination. They are illustrated for modes with $l$ = 0 (yellow, EC 14026 
stars only), $l$ = 1 (black), $l$ = 2 (grey), $l$ = 3 (blue), $l$ = 4 
(cyan), $l$ = 5 (green) and $l$ = 6 (red). Points of different colour are
 shifted slightly in effective temperature for a given model in order to
 facilitate 
viewing and the PG 1716 and EC 14026 domains are separated by the vertical 
dotted line. From Figure 25 we find that the phase shifts predicted are 
small (less than 5 degrees) for the majority of modes in both EC 14026 and 
PG 1716 stars, and would most likely lie below the detection threshold of
 multi-colour observations. Larger phase shifts of up to $\sim$ 10
 degrees are expected for the very shortest periods in the hotter PG 1716 
 variables, as well as for the $l$ = 3 modes at the higher end of the period 
range excited in the hotter EC 14026 stars. While it seems plausible that 
the latter could be detected observationally, the potential for mode 
discrimination in subdwarf B stars clearly lies with the amplitude ratios 
depicted in Figure 26. In the case of the PG 1716 stars, the values for 
modes with $l$ = 1, 2, 4 or 6 are very similar, but can easily be
 distinguished from those of $l$ = 3 or $l$ = 5 modes. Even though the three
 regimes slowly approach each other with increasing temperature, they remain
 well separated irrespective of the model in question. In the case of the
 EC 14026 stars, we identify four distinct groups of modes with $l$ = 0, 1 or
 2, $l$ = 3, $l$ = 5 and $l$ = 4 or 6. Again, the influence of the model
 parameters is relatively small compared to the separation between the
 domains and becomes vital only when attempting to discriminate between
 the modes of a given group. 

It should be kept in mind that the results depicted in Figures 25 and 26 
 depend not only on the atmospheric parameters of the models in question, 
but also on the period range excited in each case. As such, they represent 
the phase shifts and amplitude ratios predicted in real subdwarf B stars 
and give a good indication of what would be expected from multi-colour 
photometry. However, in terms of a purely theoretical exploration we find 
it instructive to examine the impact of both the surface gravity and the
effective temperature on the amplitude ratios individually. To this end, 
we repeated the computation process detailed above for both EC 14026 and 
PG 1716 stars, keeping the effective temperature (surface gravity) constant
 at the representative value from Table 1, and varying the surface gravity 
(effective temperature) within the ranges found in the sequences of Table 
4. We imposed a representative period of $P$ = 150 s for the EC 14026 star 
domain and $P$ = 4500 s for the PG 1716 star regime. The variation of the 
amplitude ratio with effective temperature is illustrated in Figures 27a
 and 27b for short- and long-period variable subdwarf B stars respectively 
(see the figure captions for more detail). For both types of oscillator,
the curves are reminiscent of the amplitude ratios found for our sequence of 
subdwarf B stars presented in Figure 26, and change gradually and 
monotonically with temperature. The situation is more interesting for the 
variation with surface gravity, depicted in Figure 28a (28b) for EC 14026 
(PG 1716) models. In the case of the fast oscillators, the amplitude ratios 
for most of the modes move in the opposite direction compared to Figure 26, 
and the closely spaced $l$ = 0, 1, 2 graphs seem to diverge from the
 (constant) $l$ = 1 value as the surface gravity increases. For the PG 1716
 models, curves representing modes of different degree indices overlap in
 several  instances, giving rise to a behaviour similar to that encountered
 as a function of period (see Figure 23).

\section{FEASIBILITY OF APPLICATION TO MULTI-COLOUR DATA}

In this section we demonstrate the potential of the method developed 
by applying it to published multi-colour photometry. To date, the only
 sets of multi-colour observations that exist for pulsating subdwarf B
 stars are those of Koen (1998), Falter et al. (2003), Jeffery et al. (2004),
 and Oreiro et al. (2005). Among these, the Jeffery et al. (2004) data
 for the fast pulsator KPD 2109+4401 yielded the most accurate amplitude
 estimates. Based on three nights of Sloan filter u'g'r' photometry with
 ULTRACAM (Dhillon et al., in preparation) mounted on the 4.2-m William
 Herschel Telescope, the authors were able to extract seven periodicities
 in the 180$-$200 s range. For the purposes of our brief feasibility study,
 we attempt to determine the degree index of the highest amplitude mode
 at 182.4 s only.
 
The first step towards finding the theoretical amplitudes in different 
wavebands is the calculation of the monochromatic model atmosphere 
quantities $\alpha_{T\nu}$, $\alpha_{g\nu}$, $b_{l\nu}$, $b_{l\nu ,T}$ and 
$b_{l\nu,g}$ described in Section 3. Since these depend quite sensitively on
 the atmospheric parameters of the target, they need to be computed for
 KPD 2109+4401 specifically. We adopt $T_{\rm eff}$=31,380 K and $\log{g}$=5.65
 as derived from our model atmosphere fit to the hydrogen Balmer and helium
 lines present in the time-averaged high-resolution MMT spectrum obtained
 by Betsy Green (private communication). These values are part of an ongoing 
program designed to provide homogeneous estimates of the atmospheric 
parameters of a large sample of subdwarf B stars, and we refer the interested 
reader to the forthcoming Paper by Green, Fontaine, $\&$ Chayer 
(in preparation) for more information. The monochromatic quantities are then 
integrated over the effective u'g'r' wavebands, computed by convolving the 
Sloan bandpasses with the quantum efficiency curves of the ULTRACAM CCD 
chips (Vik Dhillon, private communication) and the atmospheric transparency 
curve of a representative observing site at 2000$-$3000 m altitude (in
this case Kitt Peak National  
Observatory). Next, we derive the adiabacity parameters $R$ and $\Psi_T$ 
(Section 4) for a period of 182.4 s from the non-adiabatic eigenfunctions
 of an envelope model characterised by the atmospheric parameters given
 above as well as representative values of $M_{\ast}$=0.48 $M_{\odot}$ and
 $\log{q(H)}$=$-$4.0. We finally calculate the pulsational amplitudes expected
 from the u', g', and r' photometry for degree indices from $l$=0 to $l$=5
 using the equations given in Section 2.
 
The predicted multi-colour amplitudes are fit to those observed using a 
$\chi^2$ minimisation routine following Fontaine et al. (1986). Compared
to the standard normalisation 
 of all amplitudes to one particular waveband this is a more objective way
 of determining the quality of a match, since the data from all bandpasses 
are weighted evenly. For every degree index $l$ the theoretical amplitudes
 $a_{theo}$ in each of the three bandpasses i are multiplied by a scale
 factor $f_l$, chosen in such a way as to minimise
\begin{equation}
\chi^2=\sum_{i=1}^3\displaystyle\bigg(\frac{f_l a_{theo}^i-a_{obs}^i}{\sigma^{i}}\displaystyle\bigg)^2,
\end{equation}
where $a_{obs}^i$ is the amplitude observed in a given waveband and $\sigma^i$
 is the error on the measurement. The results of this operation for the 
182.4-s mode of KPD 2109+4401 detected by Jeffery et al. (2004) are 
illustrated in Figure 29, and the corresponding $\chi^2$ and quality-of-fit 
(Q) values are listed in Table 5. It is immediately obvious that the data
 are matched well by the predictions for an $l$=0 mode, the theoretical
 values falling within the (very small) error bars in all bandpasses. The 
next best fit is that of the $l$=1 mode, however the associated $\chi^2$ 
residuals are a factor of 30 larger than those for $l$=0. In fact, if we 
adopt the canonical notion that a fit cannot be considered convincing unless
 the quality-of-fit $Q>0.001$ (see, e.g., Press et al. 1986), the
 theoretical $l$=0 mode is the $only$  
one that can reproduce the observed amplitudes in a satisfactory manner. 
While this implies an unambiguous identification of the mode's degree index,
 it should be kept in mind that the estimated $\chi^2$ and $Q$ values are 
 sensitive to the formal uncertainties on the observed amplitudes, which may
 well have been underestimated. Assuming, in an extreme case scenario,
 that the true errors are twice as large as those calculated by Jeffery et al. 
(2004), both $l$=0 and $l$=1 would provide acceptable fits. Nevertheless, 
the match for $l$=0 remains far superior, and we believe that our 
identification of the degree index is sound.  

We would like to point out that this is the first partial mode identification
 in a subdwarf B star on the basis of its amplitude-wavelength dependence 
alone. While the net result -- the mode has a degree index of $l$=0 -- is the
 same as that of Jeffery et al. (2004), we were not forced to invoke additional
 constraints to distinguish modes with $l$=0, 1, and 2. The greater accuracy
 achieved is undoubtedly due to the fact that we were able to compute a
 detailed model atmosphere characteristic of KPD 2109+4401 specifically,
 and could incorporate non-adiabatic effects from realistic envelope models.

\section{CONCLUSION}

We have modelled the brightness variations expected across the visible disk 
during a pulsation cycle for both short- and long-period variable subdwarf
B stars taking into account the effects of temperature, radius, and surface 
gravity perturbations. The quantities related to the emergent intensity and 
its derivatives with respect to effective temperature and surface gravity 
were computed with the aid of a full model atmosphere code specifically 
modified for this purpose. Employing full model atmospheres has led to a
degree of self-consistency not often achieved in this kind of
calculation for other types of pulsating stars. For instance, in the case of
pulsators near the main sequence most researchers use the
Kurucz (1993) model atmospheres (or their equivalent) to compute
$\alpha_{Tx}$ and $\alpha_{gx}$. However, since these data do not treat
 limb darkening, the same authors are forced to turn elsewhere, in
particular to the popular tables of Wade \& Rucinski (1985), for limb
darkening coefficients that allow them to approximately estimate
 the quantities $b_{lx}$, $b_{lx,T}$, and $b_{lx,g}$. Notwithstanding
the fact that sdB stars are not main sequence stars, our approach
alleviates the uncertainties associated with this mixed procedure. 

In contrast to previous studies concerning the potential of multi-colour
 photometry for mode identification in sdB stars, we were able to model 
 non-adiabatic effects in some detail. Applying full non-adiabatic 
pulsation calculations to representative EC 14026 and PG 1716 star models
 showed that these are by no means negligible. In EC 14026 stars, the two
 parameters measuring the departure from adiabacity of the eigenfunctions
 in the atmospheric layers of interest ($<R>$ and $<\psi_T>$) vary
 systematically with period, but are independent of  the degree index $l$.
 This is fortunate as it implies that $<R>$ and $<\psi_T>$ can be
 accurately computed for observed periodicities without
 assuming any prior knowledge of $l$. We should mention here that, while 
we use the fits to $<R>$ and $<\psi_T>$ obtained from our respresentative 
model throughout this explorative study, the latter are sensitive to the
 atmospheric parameters of the model in question. As such, they must be
 computed individually according to the specifications of each target
 if a quantitative interpretation of the observational data is to be achieved. 
For the PG 1716 stars, the situation is more complicated as the departure 
from adiabacity depends on $l$ as well as on the period. The values of $<R>$
 and $<\psi_T>$ attributed to oscillations observed will therefore 
constitute only rough estimates, introducing inaccuracies into the amplitude 
ratios and phase shifts computed. Fortunately, measurements of the
 longer periods excited in these stars are insensitive to non-adiabatic
 effects since they are dominated by temperature perturbations and
 $<R>,<\psi_T>$ cancel out in the amplitude ratios and phase shifts. It is 
of interest to note that our computations return $R>1$ for the majority 
of modes believed to be excited in long-period variable subdwarf B stars. 
Although this is in conflict with the prevailing sentiment that $R$ must lie 
in the range $0<R<1$, we find no physical justification for this and believe 
our results to be accurate.

According to our computations, the brightness variations observed in subdwarf 
B stars are caused primarily by temperature and radius perturbations, 
 the contribution of surface gravity changes being small in EC 14026 stars 
and negligible in PG 1716 stars. For the latter, temperature effects alone
 dominate the flux changes in the limit of long periods. In this 
regime, non-adiabatic effects loose their influence on the amplitude 
ratios and phase shifts, and the period of the mode is no longer an issue. 
The adiabatic approximation is thus valid in this particular case, which
 immediately implies that oscillations should occur in phase at all
 wavelengths. Note however that conversely a lack of observed phase shifts 
does not automatically justify use of the adiabatic approximation. Even
 outside the temperature dominated regime, phase shifts are generally
 predicted to be small, although they may reach up to $\sim$ 10 
degrees for the shortest periods in PG 1716 stars and certain $l$=3
 modes in EC 14026 stars. Whereas this may be large enough for an 
observational detection, it is clear that mode discrimination will occur 
primarily on the basis of the amplitude ratios. We note that for main
sequence $g$-mode oscillators (e.g., Aerts et al. 2004 or De Cat et al. 2005)
as well as for white dwarfs (e.g., Robinson et al. 1982) the measured
phase shifts are negligible.

In the case of the EC 14026 stars, it should be relatively straightforward to 
distinguish modes with $l$ = 0, 1, or 2 from those with $l$ = 3, $l$ = 4 or 6
 and $l$ = 5. This could well prove invaluable as a consistency check for the 
``forward approach'' in asteroseismology, which has been used to claim 
mode identification in a number of short period variables through 
the inference of modes with $l$ = 0, 1, 2 and 3/4 (e.g., Charpinet et
al. 2005). It is interesting to note, in this connection, that the
visibility of the $l$ = 3 modes in the optical domain is less than that
of the $l$ = 4 modes (see Fig. 1), very much like the situation in main
sequence $p$-mode pulsators (e.g., Heyndericks et al. 1994), and this
should be taken into account in future asteroseismological exercises of
the sort. Discrimination between modes with $l$ = 0, 1, or 2 is
much more challenging and requires multi-colour photometry of
unprecedented quality as well as accurate spectroscopic estimates of the
atmospheric parameters for the target observed. Nevertheless, we have
 demonstrated that it is feasible using the highest amplitude
 mode detected for KPD 2109+4401 (Jeffery et al. 2004) as an example. 
It is not yet clear whether similar results can be achieved on the basis of 
other published datasets, or even for the lower-amplitude modes detected 
in KPD 2109+4401. This can only be answered through detailed quantitative 
analyses, which we plan to carry out in the near future. It will be
 particularly interesting to see whether we can confirm the tentative
 mode identifications reported by Jeffery et al. (2004) for the remaining 
modes detected for KPD 2109+4401 and HS 0039+4302. Regardless of the outcome 
of this project, we feel confident that colour-amplitude ratios
 and phase shifts could be measured to sufficient accuracy from future
 observations provided that the targets are bright enough and light curves
 with well resolved frequency peaks in the Fourier domain are obtained.
 Several consecutive nights on a 4 m-class telescope, or even a single night
 on an 8 m telescope for a well-chosen ``simple'' pulsator such as
 PG 1219+534 (see Charpinet et al. 2005) would likely be adequate.
 Alternatively, simultaneous ground- and space-based observations would be
 useful insofar as the frequency baseline could be extented to the UV, where
 the signature of the degree index on the pulsational amplitude is greater
 than in the visible domain. Given the observing time on the appropriate
 instrument, it would be interesting to monitor a target for which an
 asteroseismological analysis has already been completed, and compare
 the degree indices inferred from the two independent methods.
 
For the PG 1716 stars, discrimination on the basis of colour-amplitude 
ratios seems feasible between modes with degree indices $l$ = 1, 2, 4, 6
 and those with $l$ = 3 and $l$ = 5. To date, the only substantial set of 
multi-colour photometry for a long-period variable subdwarf consists of the 
 $\sim$ 250 hours of simultaneous (Johnson-Cousins) U/R data obtained for 
PG 1338+481. While detailed results will be presented elsewhere, a 
preliminary analysis of the photometry indicates amplitude ratios consistent 
with those predicted for $l$ = 1, 2, 4 or 6 rather than $l$ = 3 or $l$ =
5.\footnote{We point out in this context that main sequence $g$-mode
  pulsators with good empirical mode identification all have $l$ = 1.}  
Compared to the study of EC 14026 stars, that of the PG 1716 stars is still 
in its infancy, which is partly due to deficiencies in the models, and partly 
a result of the considerable observational challenges presented by the low 
amplitudes and long periods of the pulsations. Unambiguous mode identification 
in these objects will likely be possible only by using a combination of 
the ``forward approach'' employed for the EC 14026 stars, and inference of 
the degree index from multi-colour photometry, which we have developed 
the tools for. In the immediate future we hope that restricting, if not 
identifying, the degree index will clarify the current discrepancies between 
predicted and observed instabilities and pave the way for a more mature 
understanding of these exciting objects.
 
This work was supported in part by the Natural Sciences and Engineering
Research Council of Canada and by the Fonds de recherche sur la nature
et les technologies (Qu\'ebec). G.F. also acknowledges the contribution
of the Canada Research Chair Program.


\clearpage
\begin{deluxetable}{ccc}
\tablecaption{Basic Properties of our Representative sdB Models}
\tablewidth{0pt}
\tablehead{
\colhead{} & \colhead{EC 14026} & \colhead{PG 1716} \\}
\startdata
$\teff$ (K) & 33,000 & 27,000 \\
$\log{g}$ & 5.75 & 5.40 \\
$M_*/M_{\odot}$ & 0.48 & 0.48 \\
$\log{M(H)/M_{\ast}}$ & $-$4.0 & $-$2.5 \\
$R_*/R_{\odot}$ & 0.1528 & 0.2286 \\
$<\nabla_{ad}>$ & 0.345 & 0.350 \\
($P*$) & (1.86) & (1.90) \\
\enddata
\end{deluxetable}

\clearpage
\begin{deluxetable}{cccccc}
\tablecaption{Model atmosphere parameters for our reference EC14026 star model}
\tablewidth{0pt}
\tablehead{
\colhead{Filter} & \colhead{$\alpha_{Tx}$} & \colhead{$\alpha_{gx}$} & \colhead{$b_{lx}$} & \colhead{$b_{lx,T}$} & \colhead{$b_{lx,g}$}}
\startdata
\hline
$l$=0 & {}& {}& {}& {}& {}\\
\hline
 U &  2.92408e+00 &$-$8.79918e$-$03 & 1.00000e+00 & 0.00000e+00 & 0.00000e+00\\
 B &  2.39047e+00 &$-$1.03963e$-$02 & 1.00000e+00 & 0.00000e+00 & 0.00000e+00\\
 V &  2.22439e+00 &$-$3.58973e$-$03 & 1.00000e+00 & 0.00000e+00 & 0.00000e+00\\
 R &  2.20732e+00 &$-$4.44731e$-$03 & 1.00000e+00 & 0.00000e+00 & 0.00000e+00\\
 I &  2.16038e+00 &$-$4.66709e$-$03 & 1.00000e+00 & 0.00000e+00 & 0.00000e+00\\
\hline
$l$=1 & {}& {}& {}& {}& {}\\
\hline
 U &  2.92408e+00 &$-$8.79918e$-$03 & 6.82603e$-$01 &$-$8.27838e$-$02 & 1.50712e$-$04\\
 B &  2.39047e+00 &$-$1.03963e$-$02 & 6.80962e$-$01 &$-$7.07727e$-$02 &$-$4.30909e$-$04\\
 V &  2.22439e+00 &$-$3.58973e$-$03 & 6.79500e$-$01 &$-$6.48752e$-$02 & 4.72654e$-$05\\
 R &  2.20732e+00 &$-$4.44731e$-$03 & 6.77752e$-$01 &$-$5.72272e$-$02 & 3.63862e$-$05\\
 I &  2.16038e+00 &$-$4.66709e$-$03 & 6.75976e$-$01 &$-$4.98751e$-$02 & 1.40290e$-$04\\
\hline
$l$=2 & {}& {}& {}& {}& {}\\
\hline
 U &  2.92408e+00 &$-$8.79918e$-$03 & 2.77831e$-$01 &$-$1.45151e$-$01 & 2.72343e$-$04\\
 B &  2.39047e+00 &$-$1.03963e$-$02 & 2.74875e$-$01 &$-$1.23325e$-$01 &$-$7.48759e$-$04\\
 V &  2.22439e+00 &$-$3.58973e$-$03 & 2.72346e$-$01 &$-$1.12924e$-$01 & 7.83689e$-$05\\
 R &  2.20732e+00 &$-$4.44731e$-$03 & 2.69307e$-$01 &$-$9.93220e$-$02 & 5.76610e$-$05\\
 I &  2.16038e+00 &$-$4.66709e$-$03 & 2.66224e$-$01 &$-$8.62079e$-$02 & 2.41525e$-$04\\
\hline
$l$=3 & {}& {}& {}& {}& {}\\
\hline
 U &  2.92408e+00 &$-$8.79918e$-$03 & 2.21585e$-$02 &$-$1.13544e$-$01 & 2.28955e$-$04\\
 B &  2.39047e+00 &$-$1.03963e$-$02 & 1.96901e$-$02 &$-$9.49294e$-$02 &$-$5.71154e$-$04\\
 V &  2.22439e+00 &$-$3.58973e$-$03 & 1.77843e$-$02 &$-$8.66973e$-$02 & 5.26364e$-$05\\
 R &  2.20732e+00 &$-$4.44731e$-$03 & 1.54585e$-$02 &$-$7.57038e$-$02 & 3.31297e$-$05\\
 I &  2.16038e+00 &$-$4.66709e$-$03 & 1.31074e$-$02 &$-$6.50246e$-$02 & 1.79893e$-$04\\
\hline
$l$=4 & {}& {}& {}& {}& {}\\
\hline
 U &  2.92408e+00 &$-$8.79918e$-$03 &$-$3.37428e$-$02 &$-$2.92580e$-$02 & 8.03034e$-$05\\
 B &  2.39047e+00 &$-$1.03963e$-$02 &$-$3.45850e$-$02 &$-$2.23291e$-$02 &$-$1.25024e$-$04\\
 V &  2.22439e+00 &$-$3.58973e$-$03 &$-$3.49776e$-$02 &$-$2.01295e$-$02 & 2.41962e$-$06\\
 R &  2.20732e+00 &$-$4.44731e$-$03 &$-$3.55043e$-$02 &$-$1.68782e$-$02 &$-$7.22134e$-$06\\
 I &  2.16038e+00 &$-$4.66709e$-$03 &$-$3.60262e$-$02 &$-$1.36059e$-$02 & 3.36702e$-$05\\
\hline
$l$=5 & {}& {}& {}& {}& {}\\
\hline
 U &  2.92408e+00 &$-$8.79918e$-$03 & 2.94649e$-$03 & 1.84351e$-$02 &$-$1.87196e$-$05\\
 B &  2.39047e+00 &$-$1.03963e$-$02 & 3.17349e$-$03 & 1.73638e$-$02 & 1.16289e$-$04\\
 V &  2.22439e+00 &$-$3.58973e$-$03 & 3.57060e$-$03 & 1.60135e$-$02 &$-$1.73842e$-$05\\
 R &  2.20732e+00 &$-$4.44731e$-$03 & 4.01634e$-$03 & 1.44965e$-$02 &$-$1.84547e$-$05\\
 I &  2.16038e+00 &$-$4.66709e$-$03 & 4.47573e$-$03 & 1.31376e$-$02 &$-$4.11307e$-$05\\
\enddata
\end{deluxetable}

\clearpage
\begin{deluxetable}{cccccc}
\tablecaption{Model atmosphere parameters for our reference PG1716 star model}
\tablewidth{0pt}
\tablehead{
\colhead{Filter} & \colhead{$\alpha_{Tx}$} & \colhead{$\alpha_{gx}$} & \colhead{$b_{lx}$} & \colhead{$b_{lx,T}$} & \colhead{$b_{lx,g}$}}
\startdata
\hline
$l$=1 & {}& {}& {}& {}& {}\\
\hline
 U &  2.55827e+00 & $-$9.44029e$-$03 & 6.86440e$-$01 & $-$1.35877e$-$02 & $-$2.13902e$-$04 \\
 B &  2.02976e+00 & $-$9.88741e$-$03 & 6.84945e$-$01 & $-$1.93784e$-$02 & $-$9.19858e$-$04 \\
 V &  1.81454e+00 & $-$4.38065e$-$04 & 6.82739e$-$01 & $-$1.60495e$-$02 & $-$1.91847e$-$04 \\
 R &  1.71773e+00 & $-$1.30255e$-$03 & 6.80274e$-$01 & $-$9.93252e$-$03 & $-$1.67575e$-$04 \\
 I &  1.59319e+00 & $-$2.00633e$-$03 & 6.77747e$-$01 & $-$3.46771e$-$03 & $-$3.34334e$-$05 \\
\hline
  $l$=2 & {}& {}& {}& {}& {}\\
\hline   
 U &  2.55827e+00 &$-$9.44029e$-$03 & 2.84671e$-$01 &$-$2.52137e$-$02 &$-$3.57770e$-$04 \\
 B &  2.02976e+00 &$-$9.88741e$-$03 & 2.81883e$-$01 &$-$3.41905e$-$02 &$-$1.61503e$-$03 \\
 V &  1.81454e+00 &$-$4.38065e$-$04 & 2.78062e$-$01 &$-$2.86214e$-$02 &$-$3.41634e$-$04 \\
 R &  1.71773e+00 &$-$1.30255e$-$03 & 2.73758e$-$01 &$-$1.81031e$-$02 &$-$2.98321e$-$04 \\
 I &  1.59319e+00 &$-$2.00633e$-$03 & 2.69344e$-$01 &$-$6.92282e$-$03 &$-$5.62310e$-$05 \\
\hline
  $l$=3 & {}& {}& {}& {}& {}\\
\hline
 U &  2.55827e+00 &$-$9.44029e$-$03 & 2.77314e$-$02 &$-$2.24244e$-$02 &$-$2.45991e$-$04 \\
 B &  2.02976e+00 &$-$9.88741e$-$03 & 2.52149e$-$02 &$-$2.70928e$-$02 &$-$1.26624e$-$03 \\
 V &  1.81454e+00 &$-$4.38065e$-$04 & 2.23207e$-$02 &$-$2.32581e$-$02 &$-$2.77800e$-$04 \\
 R &  1.71773e+00 &$-$1.30255e$-$03 & 1.89973e$-$02 &$-$1.54544e$-$02 &$-$2.42536e$-$04 \\
 I &  1.59319e+00 &$-$2.00633e$-$03 & 1.55810e$-$02 &$-$7.03719e$-$03 &$-$3.96414e$-$05 \\
\hline
 $l$=4 & {}& {}& {}& {}& {}\\
\hline
 U &  2.55827e+00 &$-$9.44029e$-$03 &$-$3.20092e$-$02 &$-$9.31317e$-$03 &$-$1.84146e$-$05 \\
 B &  2.02976e+00 &$-$9.88741e$-$03 &$-$3.31155e$-$02 &$-$7.28935e$-$03 &$-$3.27026e$-$04 \\
 V &  1.81454e+00 &$-$4.38065e$-$04 &$-$3.37299e$-$02 &$-$6.98455e$-$03 &$-$8.61390e$-$05 \\
 R &  1.71773e+00 &$-$1.30255e$-$03 &$-$3.45198e$-$02 &$-$5.57709e$-$03 &$-$7.54679e$-$05 \\
 I &  1.59319e+00 &$-$2.00633e$-$03 &$-$3.53407e$-$02 &$-$3.89855e$-$03 &$-$5.21899e$-$06 \\
\hline
$l$=5 & {}& {}& {}& {}& {}\\  
\hline   
 U &  2.55827e+00 &$-$9.44029e$-$03 & 2.29813e$-$03 & 7.35094e$-$04 & 7.80329e$-$05 \\
 B &  2.02976e+00 &$-$9.88741e$-$03 & 2.29739e$-$03 & 4.39081e$-$03 & 2.10197e$-$04 \\
 V &  1.81454e+00 &$-$4.38065e$-$04 & 2.88750e$-$03 & 3.20848e$-$03 & 3.19444e$-$05 \\
 R &  1.71773e+00 &$-$1.30255e$-$03 & 3.49885e$-$03 & 1.36516e$-$03 & 2.70668e$-$05 \\
 I &  1.59319e+00 &$-$2.00633e$-$03 & 4.12129e$-$03 &$-$5.06197e$-$04 & 9.33007e$-$06 \\
\hline
 $l$=6 & {}& {}& {}& {}& {}\\
\hline    
 U &  2.55827e+00 &$-$9.44029e$-$03 & 2.84314e$-$02 & 1.14630e$-$03 & 7.21713e$-$06 \\
 B &  2.02976e+00 &$-$9.88741e$-$03 & 2.85101e$-$02 & 1.43976e$-$03 & 5.62154e$-$05 \\
 V &  1.81454e+00 &$-$4.38065e$-$04 & 2.86265e$-$02 & 1.36594e$-$03 & 9.79649e$-$06 \\
 R &  1.71773e+00 &$-$1.30255e$-$03 & 2.87730e$-$02 & 9.96041e$-$04 & 7.43661e$-$06 \\
 I &  1.59319e+00 &$-$2.00633e$-$03 & 2.89270e$-$02 & 5.57062e$-$04 &$-$1.04396e$-$06 \\
\enddata
\end{deluxetable}

\clearpage
\begin{deluxetable}{ccccc}
\tablecaption{Equilibrium Models for a sequence of subdwarf B stars with $M_{\ast}/M_{\odot}$=0.48}
\tablewidth{0pt}
\tablehead{
\colhead{No.} & \colhead{$T_{\rm eff}$(K)} & \colhead{$\log{g}$} & \colhead{$\log{q(H)}$} & \colhead{$\Delta P$ (s)} \\}
\startdata
PG 1716 stars \\
\hline
1 & 22,000 & 5.13 & $-$1.64 & 4500$-$9500 \\
2 & 23,000 & 5.19 & $-$1.81 & 4000$-$8000 \\
3 & 24,000 & 5.25 & $-$1.97 & 3500$-$6500 \\
4 & 25,000 & 5.31 & $-$2.13 & 3000$-$5700 \\
5 & 26,000 & 5.37 & $-$2.30 & 2500$-$5000 \\
6 & 27,000 & 5.44 & $-$2.46 & 2300$-$4000 \\
7 & 28,000 & 5.50 & $-$2.62 & 2000$-$3500 \\
\hline
EC 14026 stars \\
\hline
8 & 29,000 & 5.56 & $-$2.79 & 220$-$450 \\
9 & 30,000 & 5.62 & $-$2.95 & 190$-$450 \\
10 & 31,000 & 5.68 & $-$3.12 & 170$-$360 \\
11 & 32,000 & 5.75 & $-$3.28 & 150$-$260 \\
12 & 33,000 & 5.81 & $-$3.44 & 110$-$200 \\
13 & 34,000 & 5.87 & $-$3.61 & 100$-$150 \\
14 & 35,000 & 5.93 & $-$3.77 & 95$-$110 \\
\enddata
\end{deluxetable}
\clearpage

\clearpage
\begin{deluxetable}{ccc}
\tablecaption{Fit of predicted u'g'r' amplitudes to those observed for the 182.4-s mode of KPD 2109+4401 by Jeffery et al. (2004)}
\tablewidth{0pt}
\tablehead{
\colhead{Degree index $l$} & \colhead{$\chi^2$} & \colhead{Q} }
\startdata
0 & 0.736 & 0.692 \\
1 & 22.6  & 1.24$\times 10^{-5}$ \\
2 & 96.0  & 1.43$\times 10^{-24}$ \\
3 & 1470  & 0.00 \\
4 & 1150  & 0.00 \\
5 & 4860 & 0.00 \\
\enddata
\end{deluxetable}
\clearpage

\clearpage

\clearpage
\centerline{\bf{FIGURE CAPTIONS}}

\noindent
Fig. 1 --- Behavior of some key monochromatic quantities for our
representative EC 14026 star model. The latter has log $g$ = 5.75 and
$\Teff$ = 33,000 K. Top panel: Unperturbed emergent Eddington flux in
the optical domain. Middle panel: First order perturbation to the flux
caused by a nonradial pulsation with a period of 150 s ($p$-mode); at
3500 $\rm\AA$, from top to bottom, each curve is characterized by a
value of the degree index $l$ = 0, 1, 2, 4, 3, and 5. Bottom panel:
Similar to the middle panel, but, this time, illustrating the $relative$
amplitude of the perturbation.

\noindent
Fig. 2 --- Logarithm of monochromatic amplitude ratios with respect to
an arbitrary spectral point at 3650 $\rm\AA$. This again refers to our
representative EC 14026 star model and for a mode with a period of 150 s
and degree index $l$ = 0 (cyan), 1 (red), 2 (blue), 3 (green), 4
(magenta), and 5 (black).

\noindent
Fig. 3 --- Similar to Figure 2, but for monochromatic phase differences.

\noindent
Fig. 4 --- Behaviour of the radial component of the pressure eigenfunction 
in the atmospheric layers of interest for our representative EC 14026
star model on the basis of modes with degree indices from $l$ = 0 to 5
for all periods in the range 80$-$300 s. The vertical dotted lines
correspond, from right to left, to Rosseland optical depths of $\tau$ =
0.1, 1.0, and 10.0, respectively.

\noindent
Fig. 5 --- Similar to Figure 4, but for our representative PG 1716 model 
and considering modes with $l$= 1, 2 and 3 in the period range 2000$-$6000 s.

\noindent
Fig. 6 --- $R$ values for our representative EC 14026 model in the
 atmospheric layers of interest for periods in the range 80$-$300 s.
 The curves for different degree indices are illustrated separately as 
indicated. 

\noindent
Fig. 7 --- Similar to Figure 6, but for the quantity $\psi_T-\pi$.

\noindent
Fig. 8 --- Cubic fit to computed $<R>$ (continuous line) and $<\psi_T>$
(dashed line) values as a function of period for our representative EC
14026 star model.

\noindent
Fig. 9 --- $R$ values for our representative PG 1716 model in the 
atmospheric layers of interest for periods in the range 2000$-$6000 s. 
The curves for different degree indices are illustrated separately as 
indicated. 

\noindent
Fig. 10 --- Similar to Figure 9, but for the quantity $\psi_T-\pi$.

\noindent 
Fig. 11 --- Cubic fit to computed $<R>$ (continuous line) and $<\psi_T>$
(dashed line) values as a function of period and of degree index $l$ for
our representative PG 1716 star model.

\noindent
Fig. 12 --- Relative importance of the temperature terms ($\gamma_1$) 
compared to the radius and surface gravity terms ($\gamma_2$) for our 
representative EC 14026 star model in the $UBVRI$ bandpasses. 
Continuous lines indicate the 
results obtained by fitting $<R>$ and $<\psi_T>$ according to equations 
(44) and (45) while dotted curves represent results obtained by 
setting the two parameters to their adiabatic values of $<R>=1$ and 
$<\psi_T>=\pi$. Three typical periods (low-order $p$-modes) of 100, 150,
and 200 s, as well as six values of the degree index, $l$ = 0$-$5, are
considered. The effective wavelengths of the various bandpasses are 3650
\AA ($U$), 4400 \AA ($B$), 5500 \AA ($V$), 6500 \AA ($R$), and 8000 \AA
($I$).  

\noindent
Fig. 13 --- Phase shifts relative to the $U$ filter calculated for 
our EC 14026 star model for the $UBVRI$ bandpasses. Continuous 
lines represent full non-adiabatic results while the dashed curve indicates 
the adiabatic values (which are always equal to zero).

\noindent
Fig. 14 --- Amplitude ratios relative to the $U$ filter calculated 
for our EC 14026 star model for the $UBVRI$ bandpasses. 
Continuous lines represent full non-adiabatic results while the dashed
 curves indicates the adiabatic values obtained by forcing $<R>=1$ and 
$<\psi_T>=\pi$. 

\noindent
Fig. 15 --- Amplitude ratios relative to the $U$ filter calculated 
for our EC 14026 star model for the $UBVRI$ bandpasses. The $l$ indices 
of the modes are indicated. Dotted, dashed and long-dashed 
lines refer to periods with 100, 150 and 200 s respectively. They are 
compared to the amplitude ratios obtained in the limit where $\gamma_2$ = 0
 (continuous curves).

\noindent
Fig. 16 --- Results of a numerical experiment in which the amplitude ratios 
relative to the $U$ band were obtained by adopting $T_3=0$ (dotted lines) 
and by setting $\gamma_2=0$ (continuous lines).

\noindent
Fig. 17 ---  Similar to Figure 12, but referring to our representative 
PG 1716 model. Six periods from 2400 s to 6000 s as well as six values
of $l$ from 1 to 6 are considered. For the continuous lines, $<R>$ and
$<\psi_T>$ were derived  from equations (46) and (47) while adiabatic
values were imposed for the dotted lines.

\noindent
Fig. 18 --- Similar to Figure 13, but referring to our representative 
PG 1716 model. Note the change of ordinate scale between panel a) and
panel b). 

\noindent
Fig. 19 --- Similar to Figure 14, but referring to our representative 
PG 1716 model.

\noindent
Fig. 20 --- Similar to Figure 15, but referring to our representative 
PG 1716 model. Amplitude ratios are illustrated for representative periods 
of $P$ = 3000 s (dotted lines), 4000 s (dashed lines), 5000 s (long-dashed 
lines), 6000 s (dot-dashed lines), 7000 s (dot-long-dashed lines) and 
8000 s (dashed-long-dashed lines). The continuous line refers to the 
case of $\gamma_2$ = 0.

\noindent
Fig. 21 --- Similar to Figure 16, but referring to our representative 
PG 1716 model.

\noindent
Fig. 22 --- $I$ to $U$ amplitude ratio as a function of degree index 
$l$ on the basis of modes extending up to $l$ = 8. From bottom to top, 
the various curves correspond to periods with $P$ = 2400, 3000, 4000, 
5000, 6000, 7000, 8000 and an infinite period respectively.

\noindent
Fig. 23 --- Variation of the $I/U$ amplitude ratio with period for our 
representative PG 1716 model. The values for different modes are 
indicated by points in black ($l$ = 1), grey ($l$ = 2), blue ($l$ = 3), 
cyan ($l$ = 4), green ($l$ = 5) and red ($l$ = 6).

\noindent
Fig. 24 --- Similar to Figure 23, but for phase differences between 
oscillations in the $I$ and $U$ bandpasses.

\noindent
Fig. 25 --- Phase differences between oscillations in the $I$ and $U$ 
bandpasses for the sequence of subdwarf B star models listed in Table 4. 
The dotted vertical line divides the PG 1716 star models on the left from 
the EC 14026 models on the right. We illustrate modes with $l$ = 0 
(yellow, EC 14026 models only), $l$ = 1 (black), $l$ = 2 (grey), 
$l$ = 3 (blue), $l$ = 4 (cyan), $l$ = 5 (green) and $l$ = 6 (red).

\noindent
Fig. 26 --- Similar to Figure 25, but for $I/U$ amplitude ratios. 

\noindent
Fig. 27 --- Behaviour of the $I/U$ amplitude ratio with effective 
temperature of a model. (a) EC 14026 regime: the surface gravity was kept
 constant at $\log{g}=5.75$, while $\teff$ was varied from 29,000 K $-$ 
35,000 K in increments of 1000 K. From top to bottom, the curves refer to 
modes with degree indices $l$ = 5, 6, 4, 2, 1, 0 and 3. (b) PG 1716 regime: 
we adopted the representative value of $\log{g}=5.40$ and varied the 
effective temperaure from 22,000 K $-$ 28,000 K, again in steps of 1000 K.
 From top to bottom, the modes in question correspond to $l$ = 5, 4, 1, 6, 
2 and 3.

\noindent
Fig. 28 --- Behaviour of the $I/U$ amplitude ratio with surface gravity. 
(a) EC 14026 regime: the effective temperature was kept constant at 
$\teff=33,000$ K, while $\log{g}$ was varied from 5.5 $-$ 5.9 in increments 
of 0.1 dex. From top to bottom on the right hand side, the curves refer 
to modes with degree indices $l$ = 5, 6, 4, 2, 1, 0 and 3. (b) PG 1716 
regime: we adopted the representative value of $\teff=27,000$ K and varied 
the surface gravity from 5.1 to 5.5, again in steps of 0.1 dex. From top to 
bottom on the right hand side, the modes in question correspond to 
$l$ = 5, 4, 1, 6, 2 and 3.   

\noindent
Fig. 29 --- Fit to the u', g' and r' pulsational amplitudes observed for the 
182.4 s mode of KPD 2109+4401 by Jeffery et al. (2004). The predicted 
amplitude-wavelength behaviours of modes with $l$=0 to $l$=5 have been 
fit to the observed values using a least-squares procedure. Only the $l$=0 
curve provides an acceptable fit.  

\clearpage
\begin{figure}[p]
\plotone{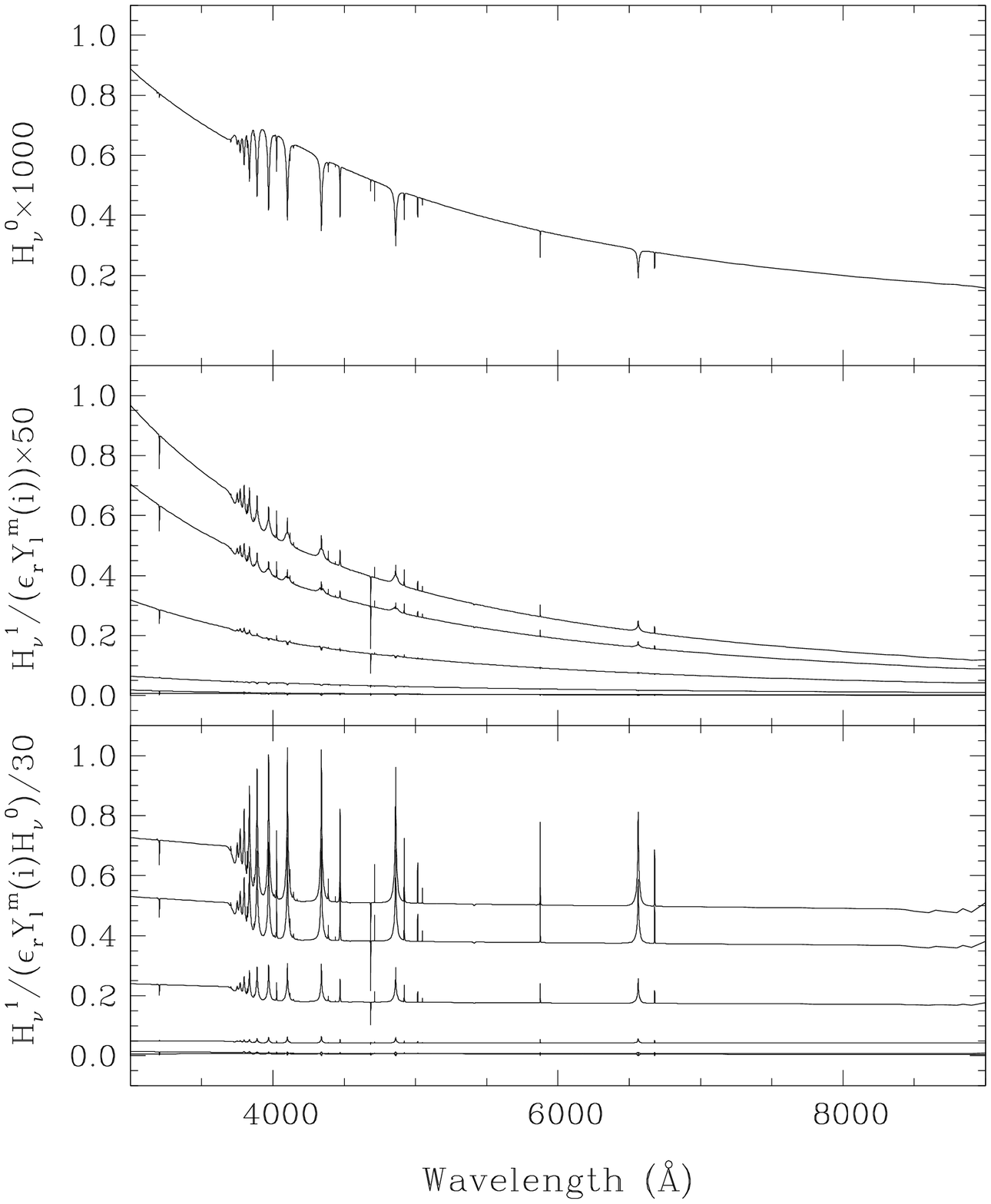}
\begin{flushright}
Figure 1
\end{flushright}
\end{figure}

\clearpage
\begin{figure}[p]
\plotone{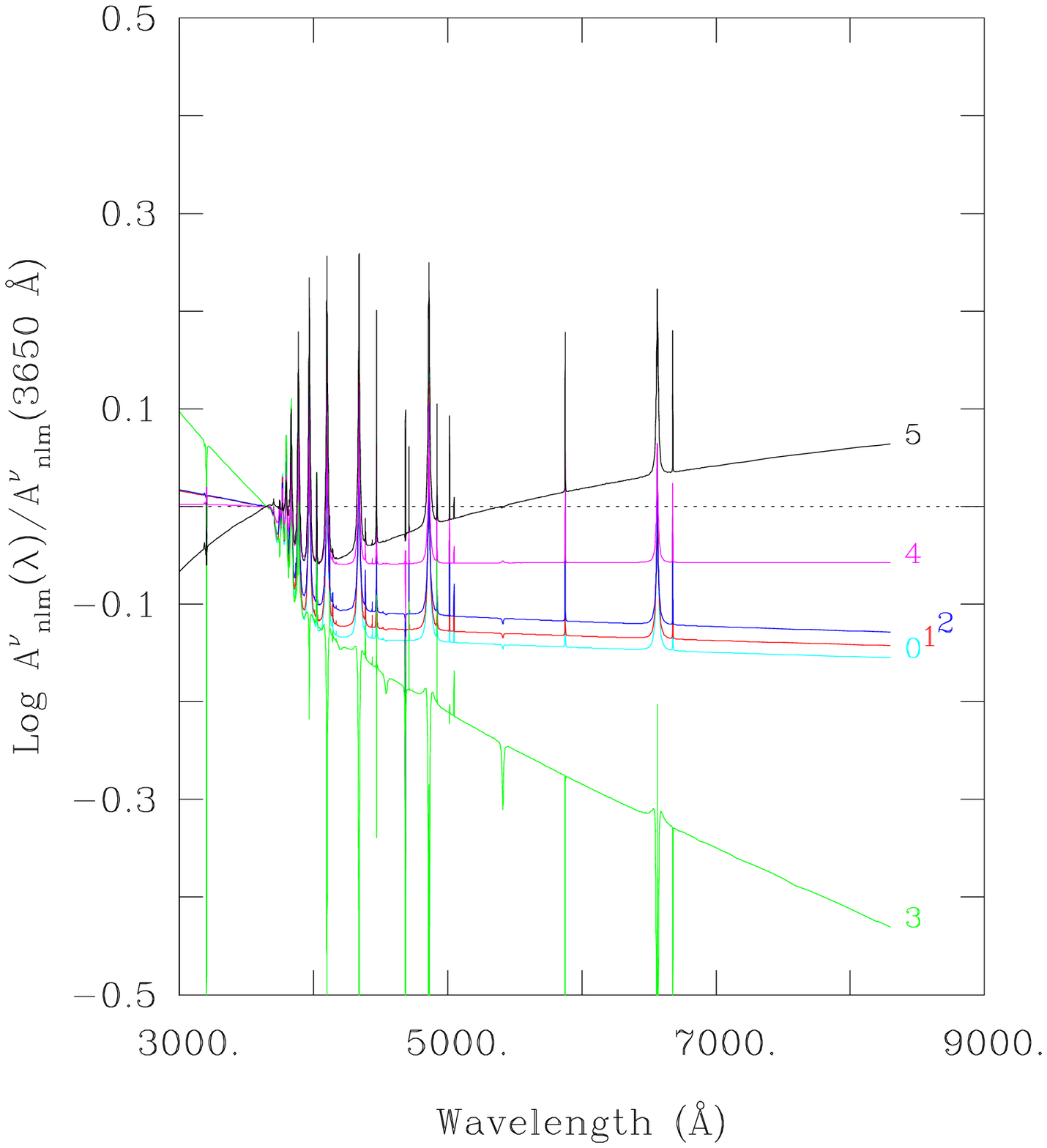}
\begin{flushright}
Figure 2
\end{flushright}
\end{figure}

\clearpage
\begin{figure}[p]
\plotone{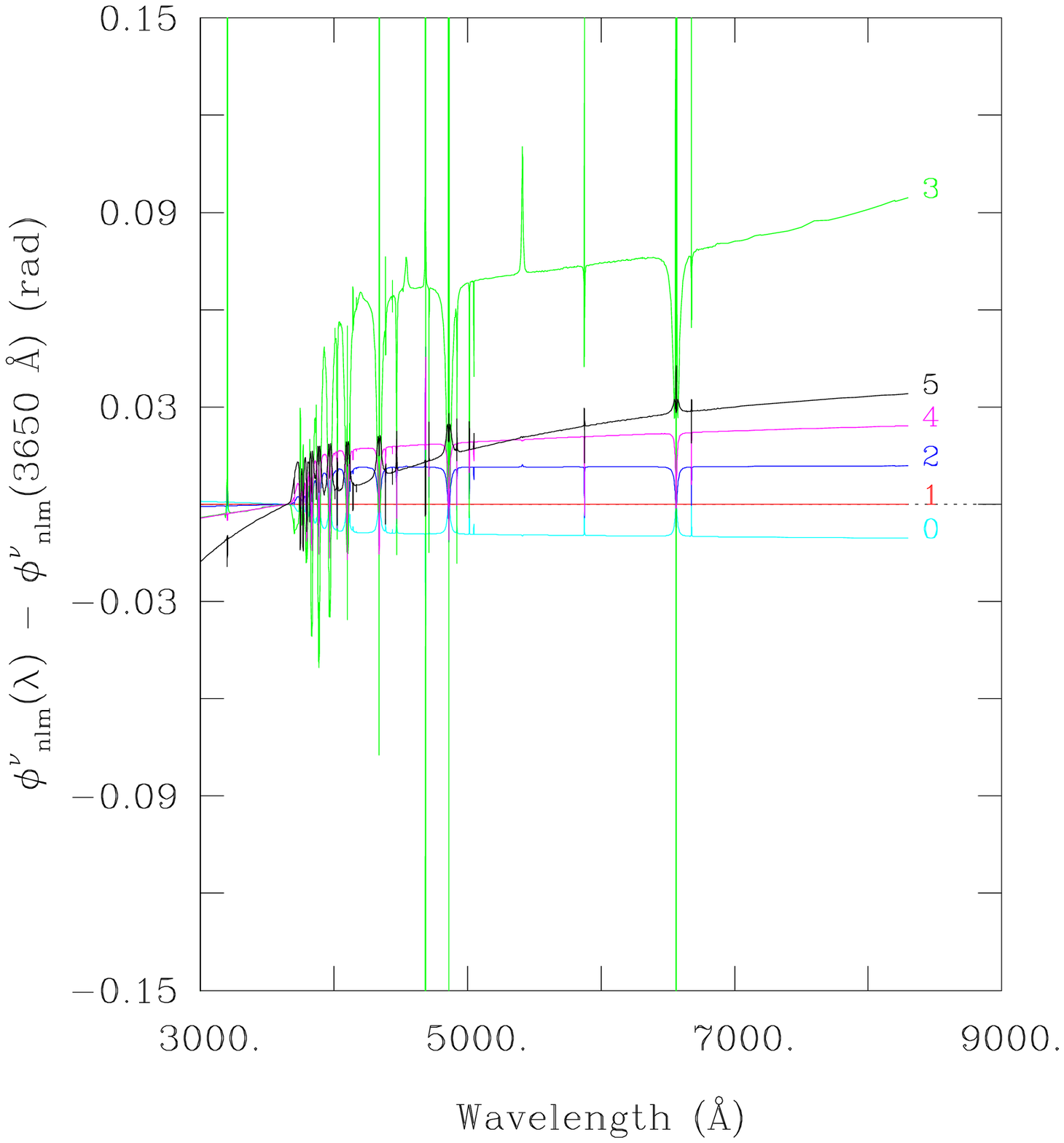}
\begin{flushright}
Figure 3
\end{flushright}
\end{figure}

\clearpage
\begin{figure}[p]
\plotone{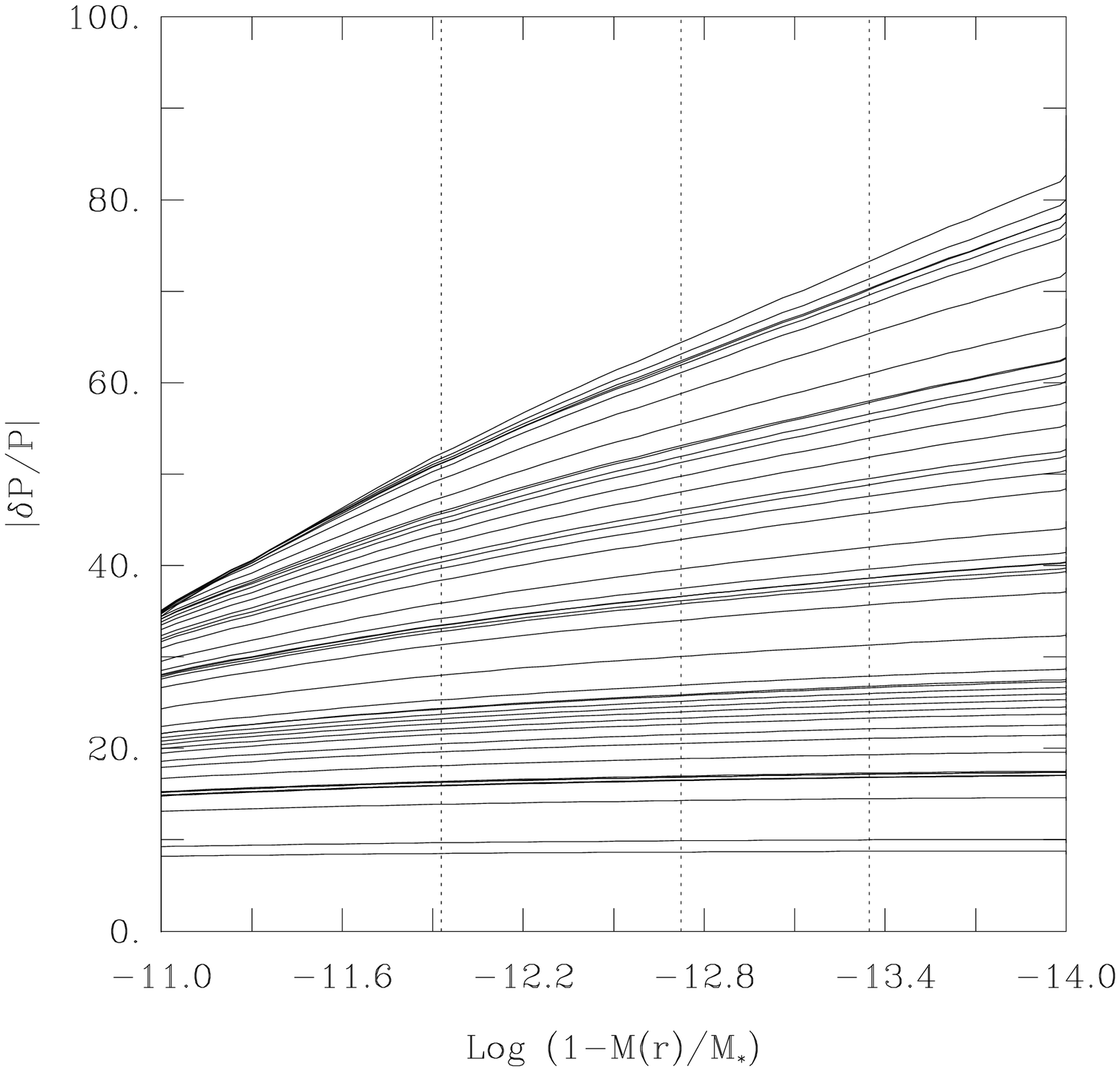}
\begin{flushright}
Figure 4
\end{flushright}
\end{figure}

\clearpage
\begin{figure}[p]
\plotone{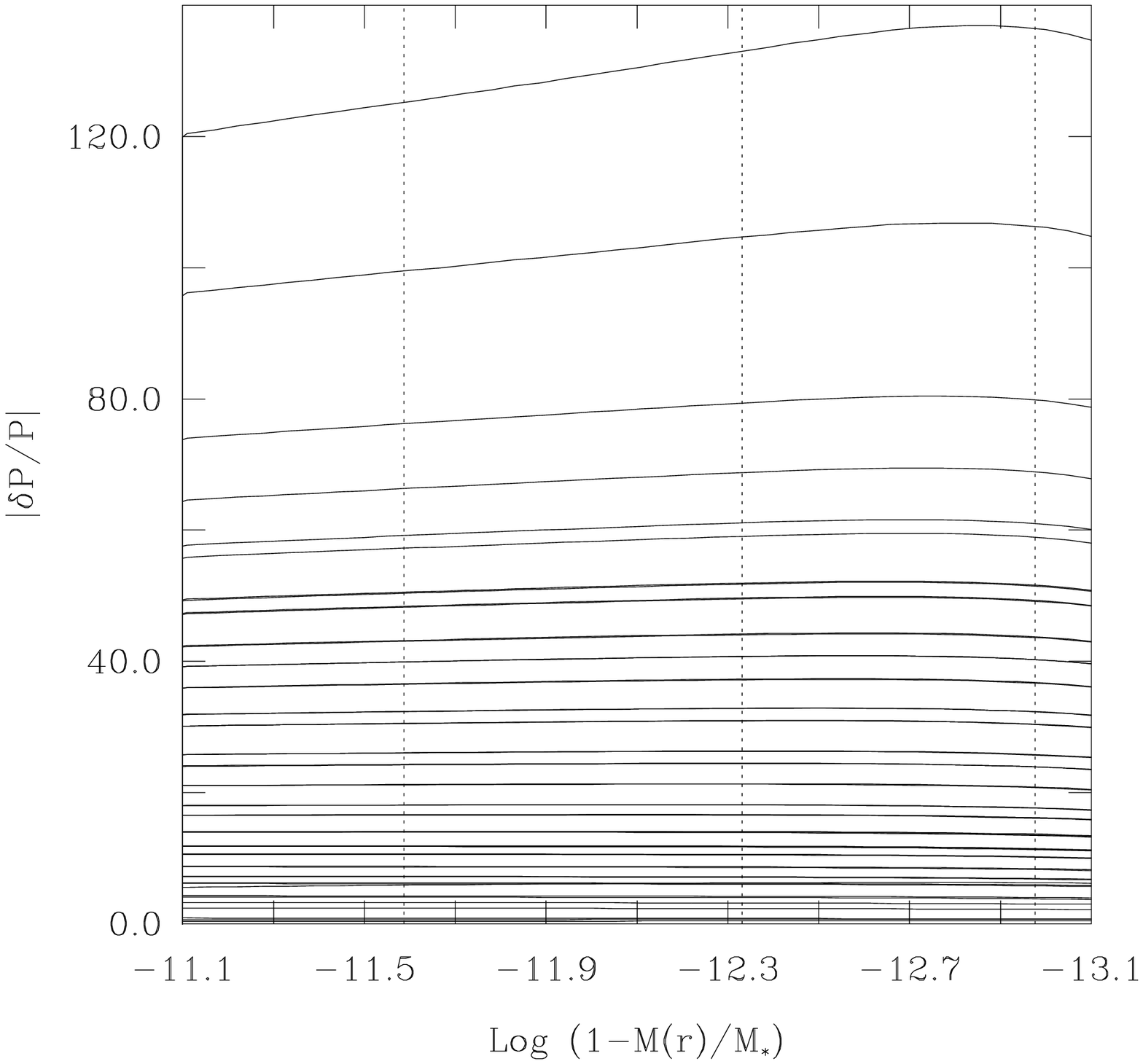}
\begin{flushright}
Figure 5
\end{flushright}
\end{figure}

\clearpage
\begin{figure}[p]
\plotone{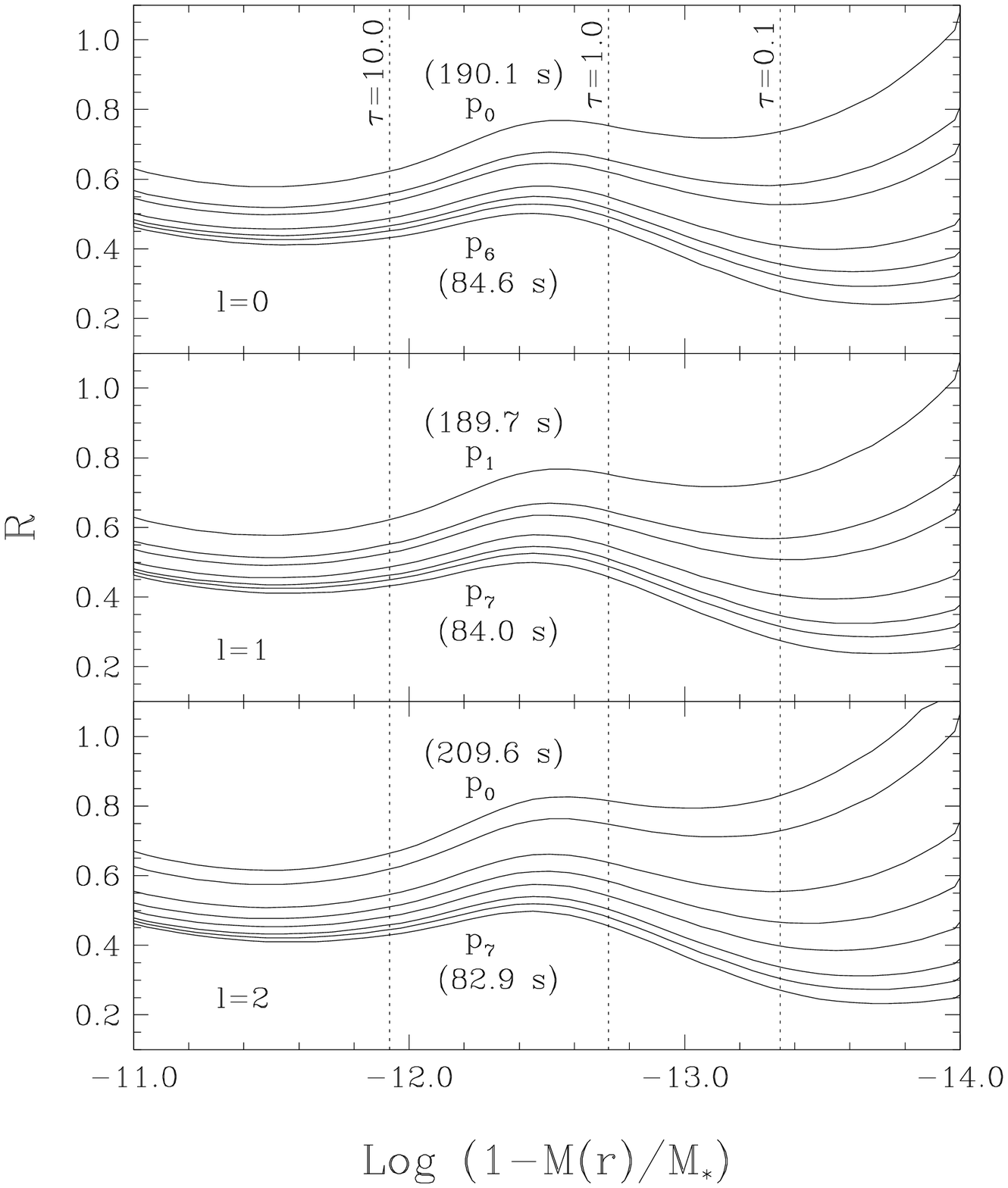}
\begin{flushright}
Figure 6a
\end{flushright}
\end{figure}

\clearpage
\begin{figure}[p]
\plotone{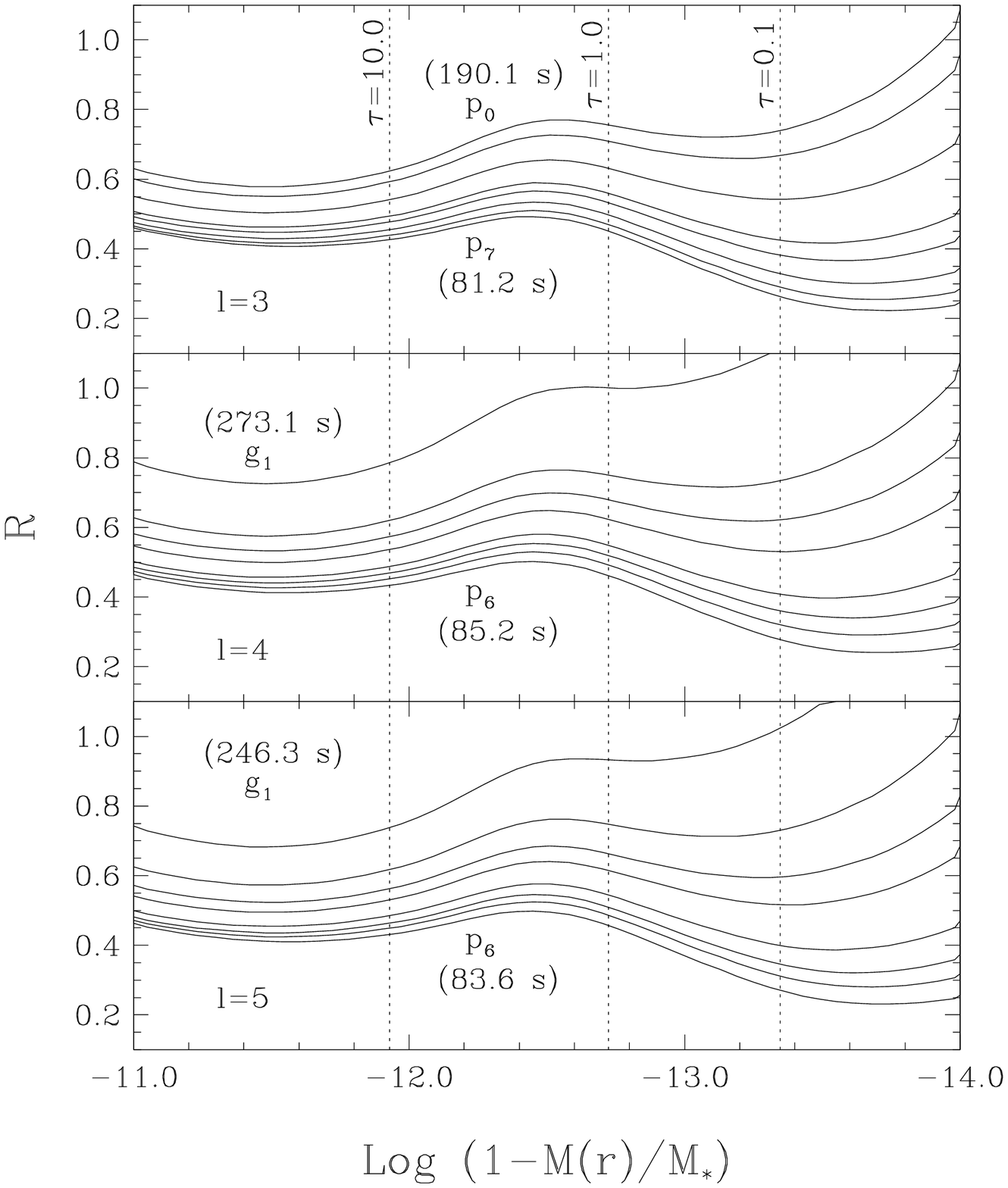}
\begin{flushright}
Figure 6b
\end{flushright}
\end{figure}

\clearpage
\begin{figure}[p]
\plotone{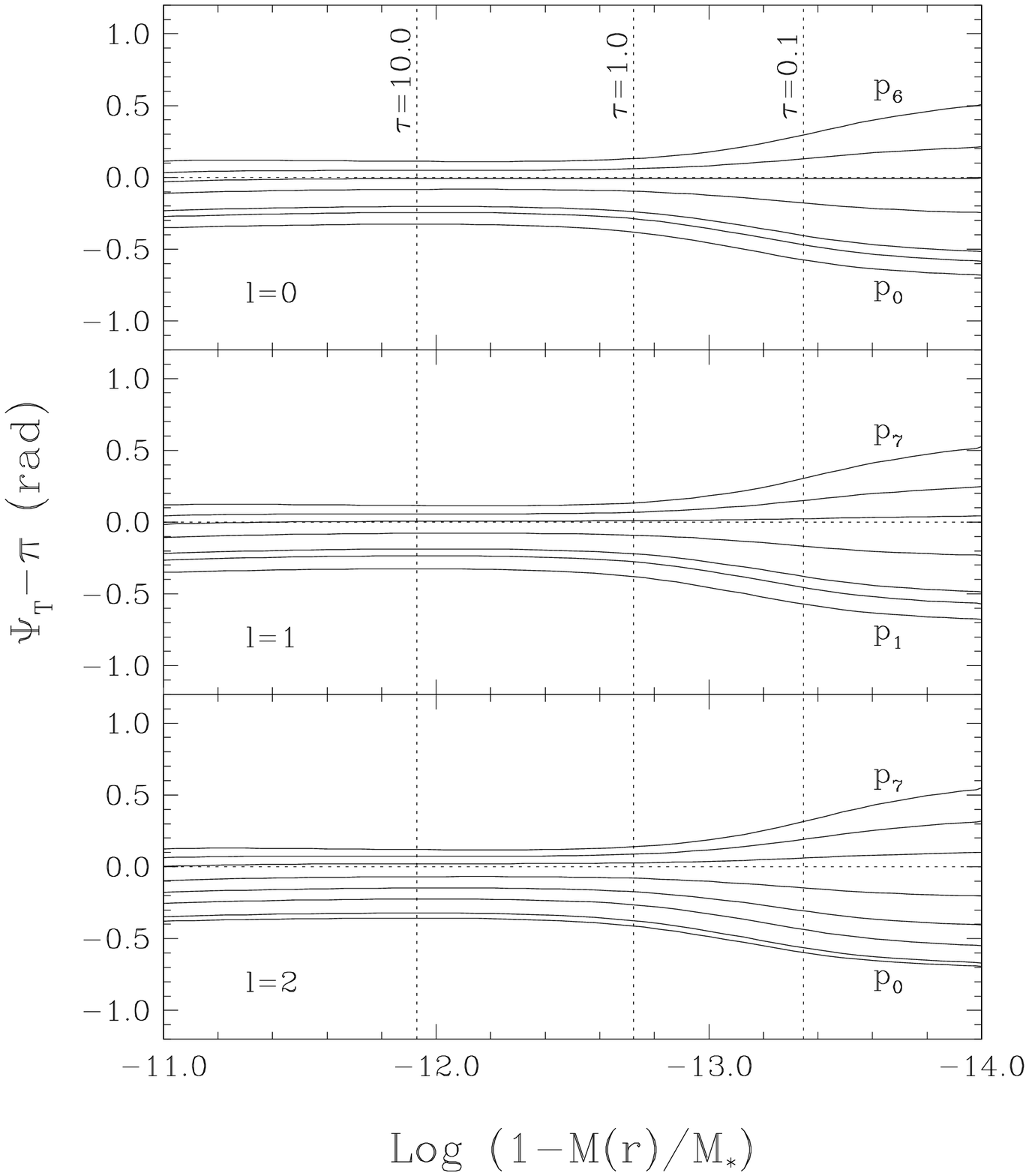}
\begin{flushright}
Figure 7a
\end{flushright}
\end{figure}

\clearpage
\begin{figure}[p]
\plotone{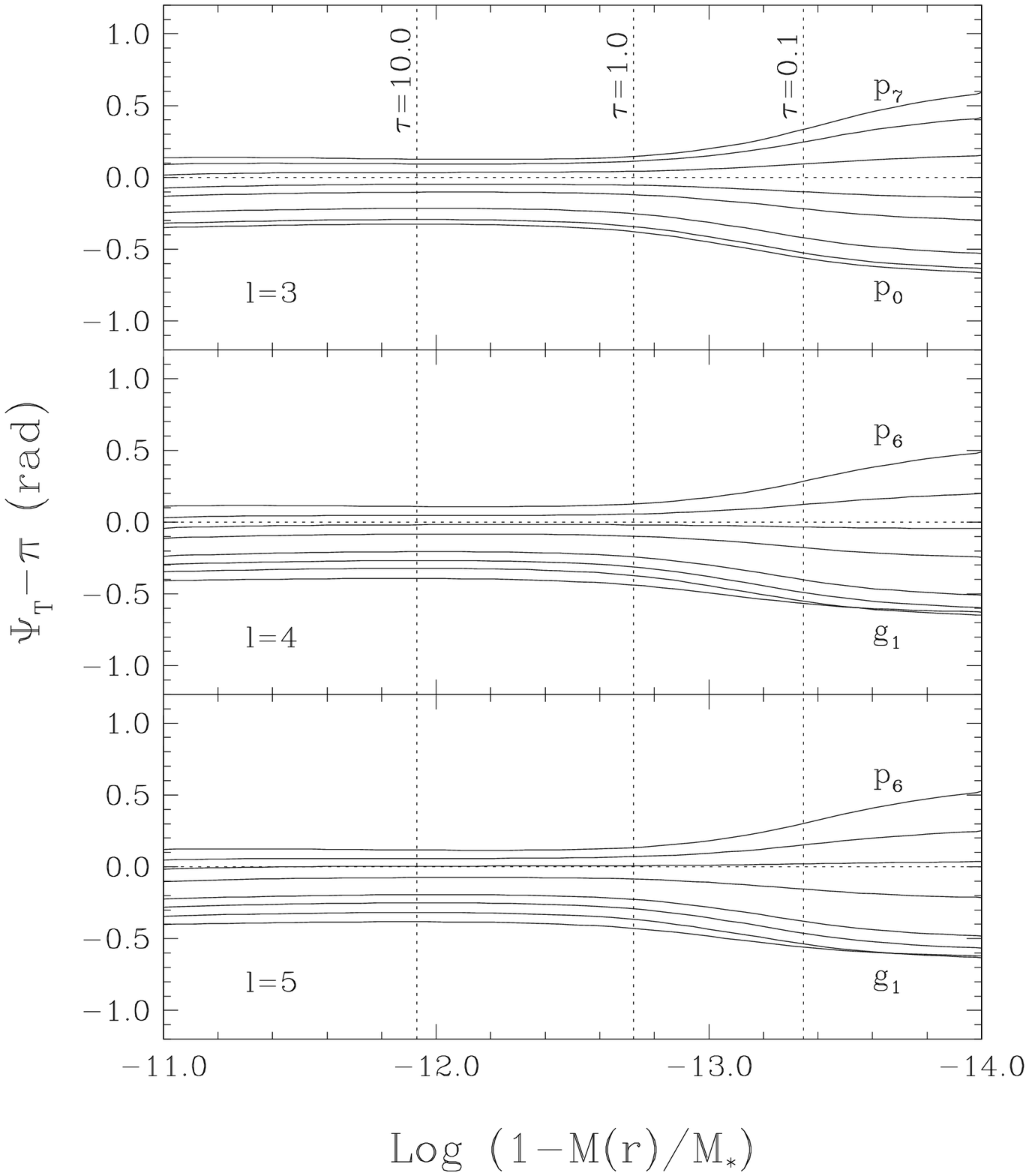}
\begin{flushright}
Figure 7b
\end{flushright}
\end{figure}

\clearpage
\begin{figure}[p]
\plotone{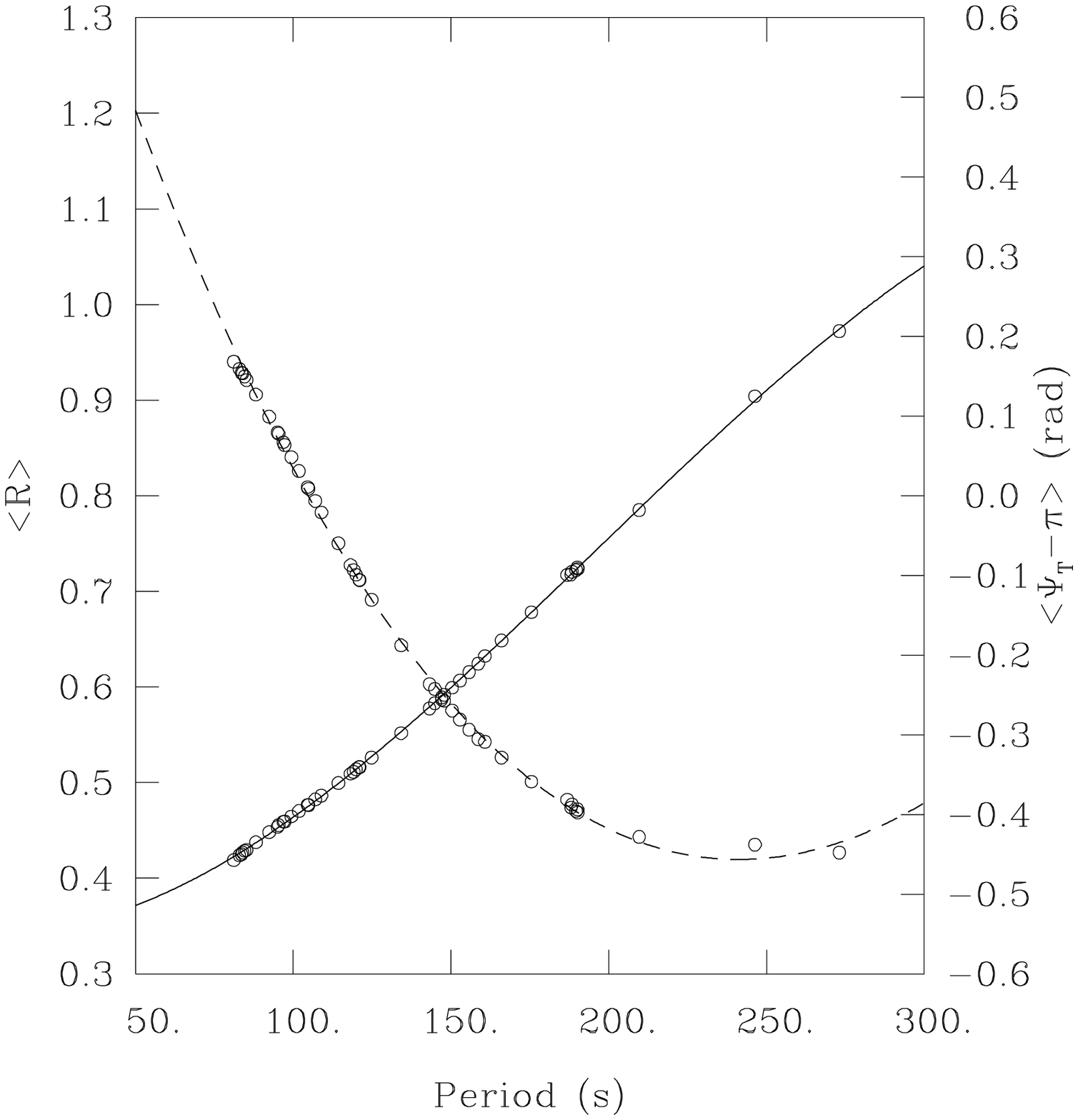}
\begin{flushright}
Figure 8
\end{flushright}
\end{figure}

\clearpage
\begin{figure}[p]
\plotone{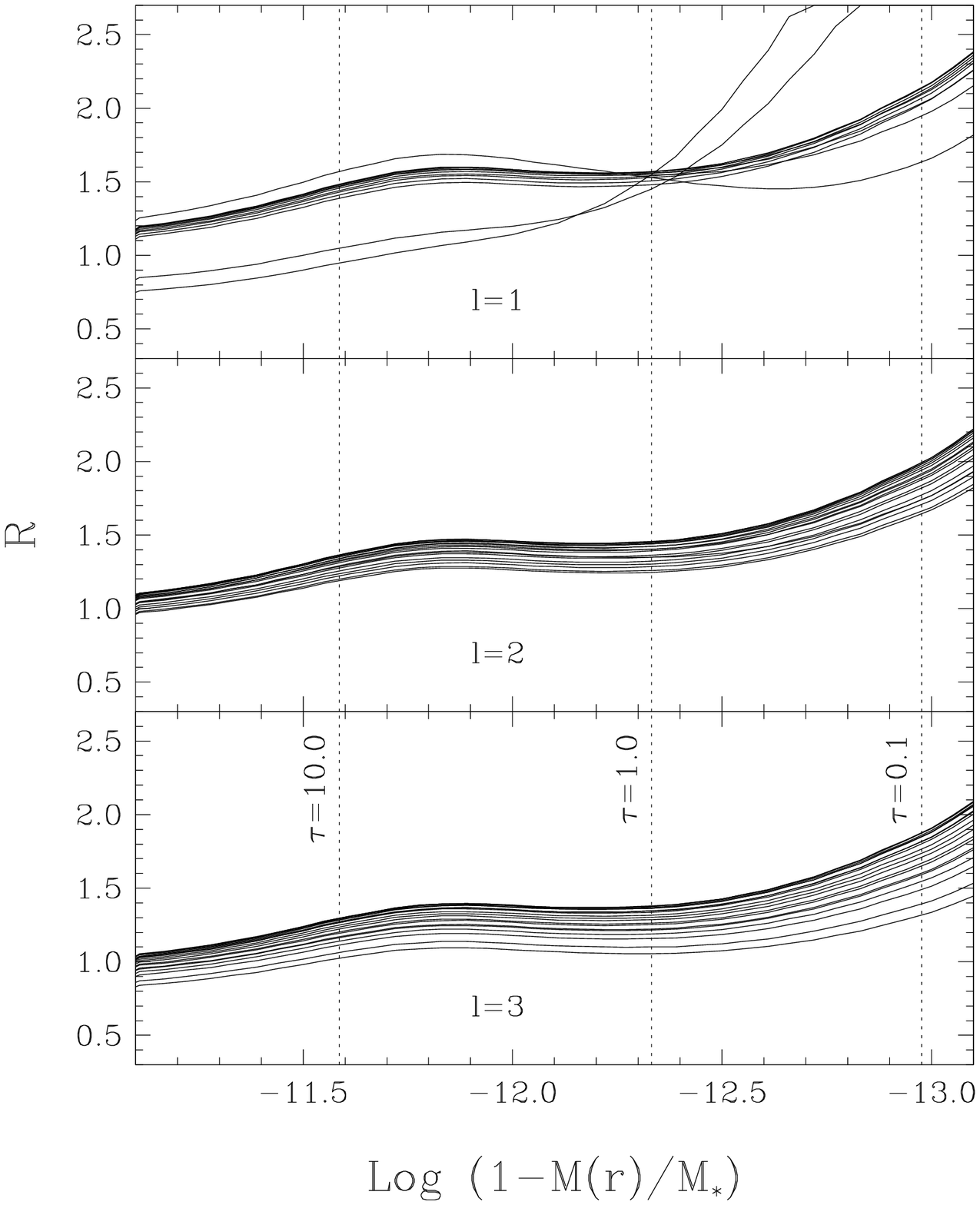}
\begin{flushright}
Figure 9
\end{flushright}
\end{figure}

\clearpage
\begin{figure}[p]
\plotone{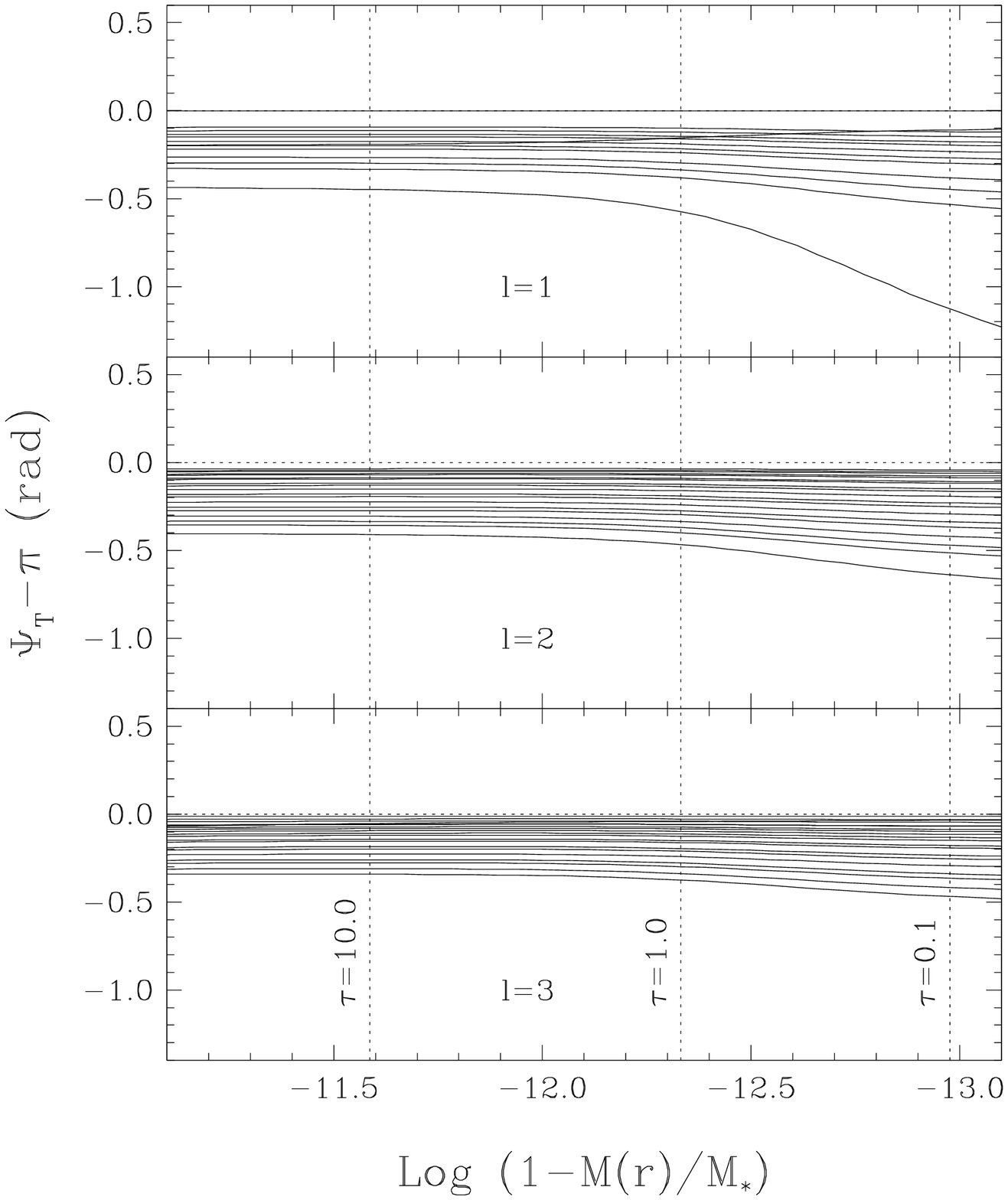}
\begin{flushright}
Figure 10
\end{flushright}
\end{figure}

\clearpage
\begin{figure}[p]
\plotone{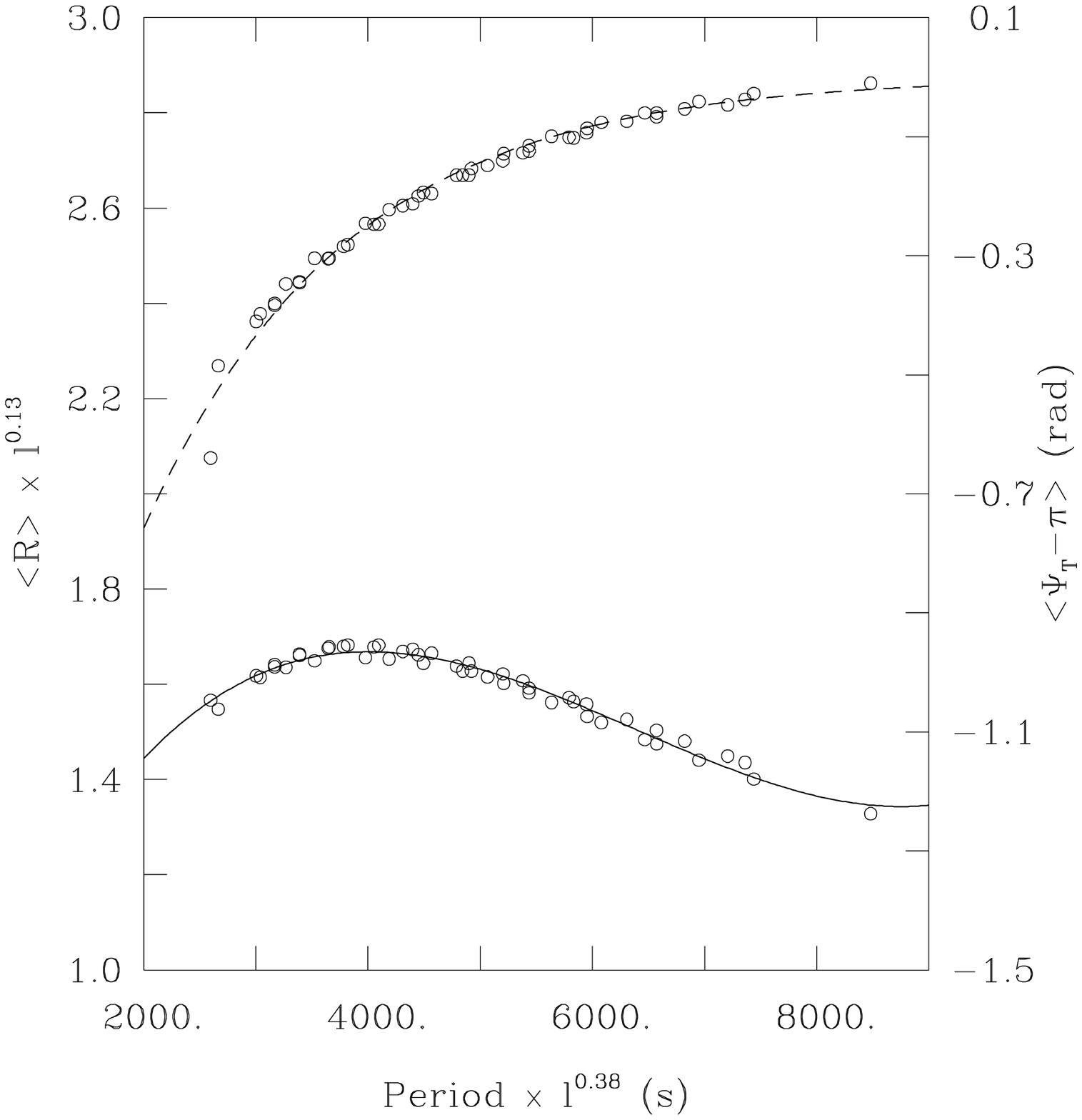}
\begin{flushright}
Figure 11
\end{flushright}
\end{figure}

\clearpage
\begin{figure}[p]
\plotone{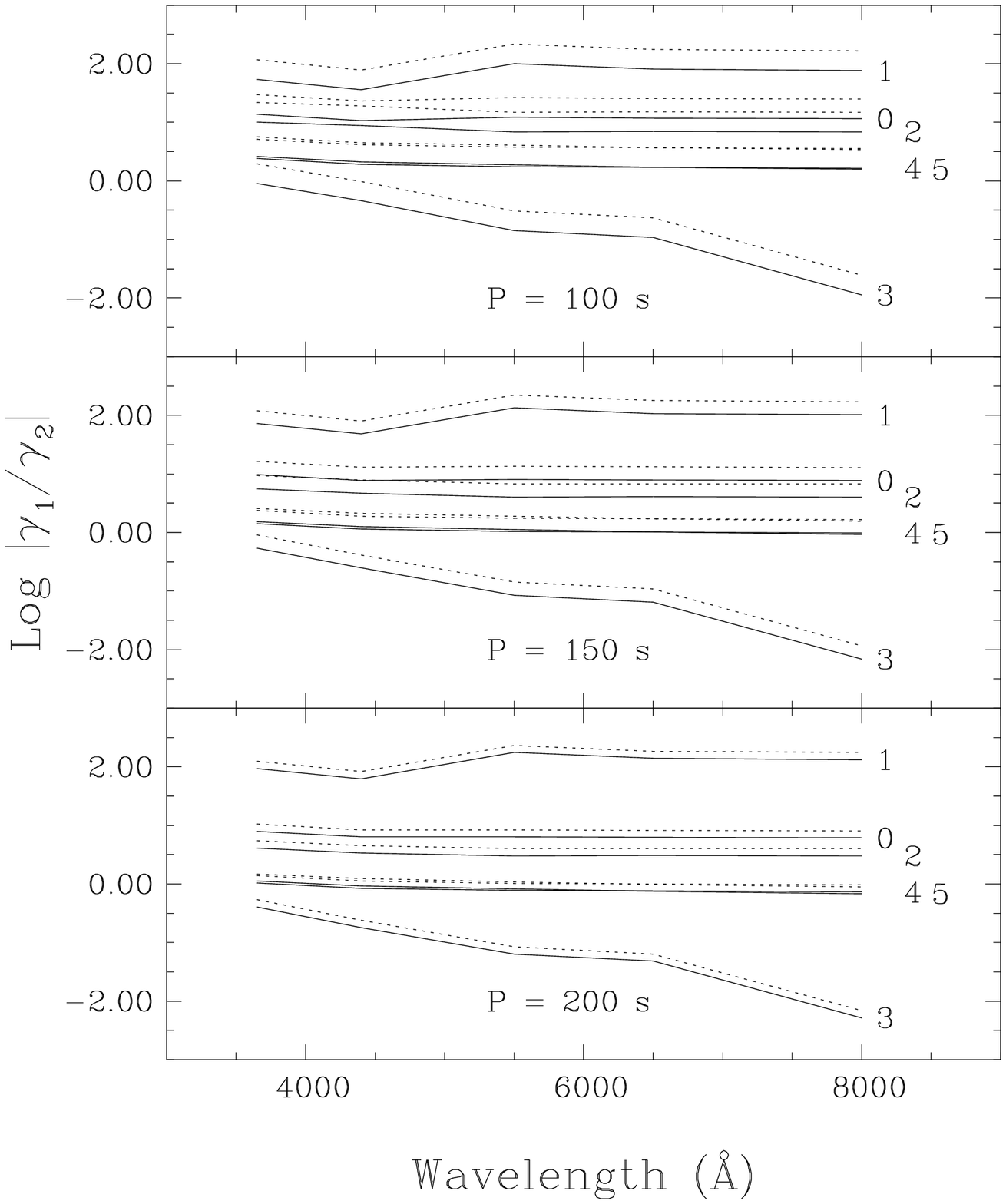}
\begin{flushright}
Figure 12
\end{flushright}
\end{figure}

\clearpage
\begin{figure}[p]
\plotone{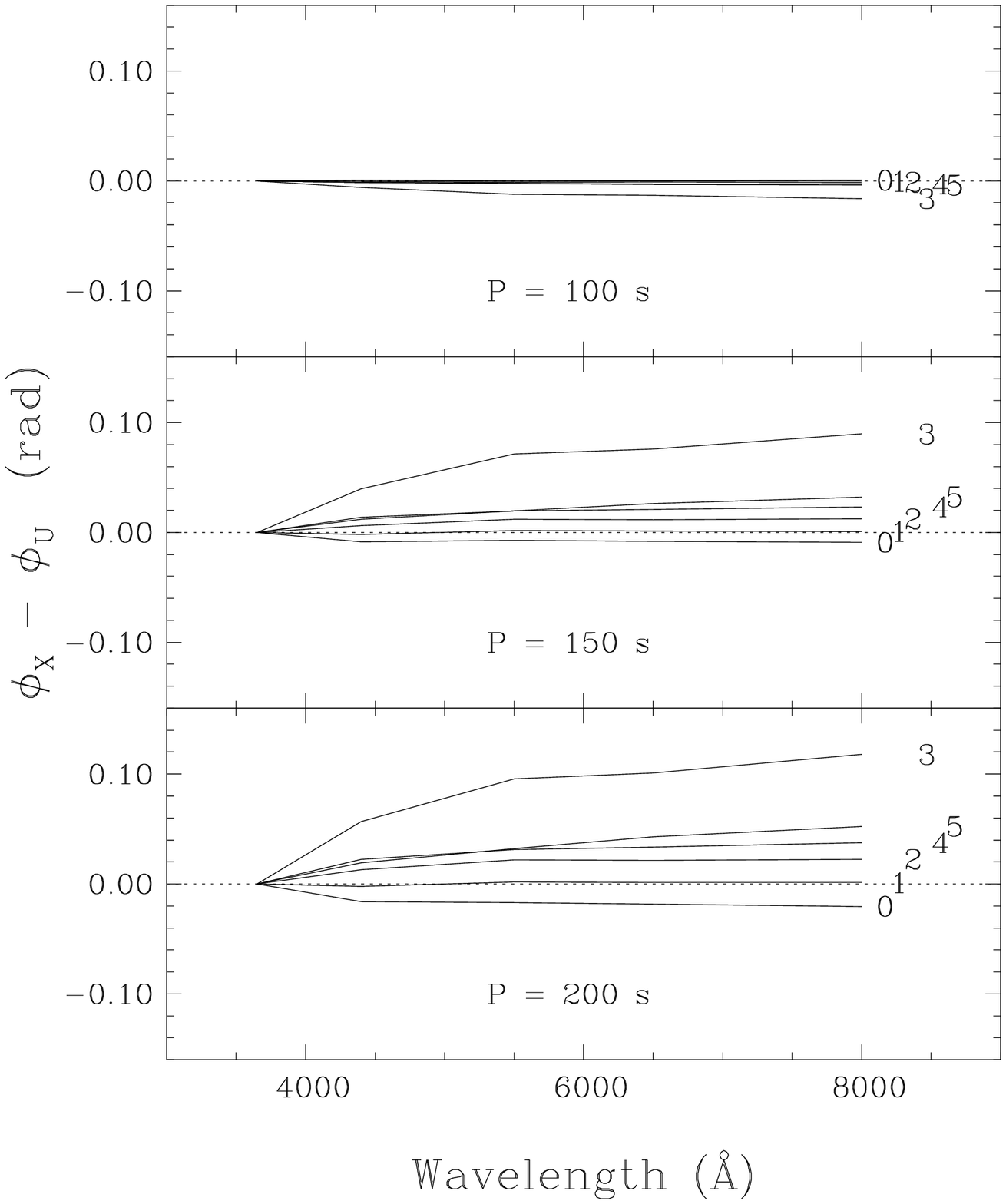}
\begin{flushright}
Figure 13
\end{flushright}
\end{figure}

\clearpage
\begin{figure}[p]
\plotone{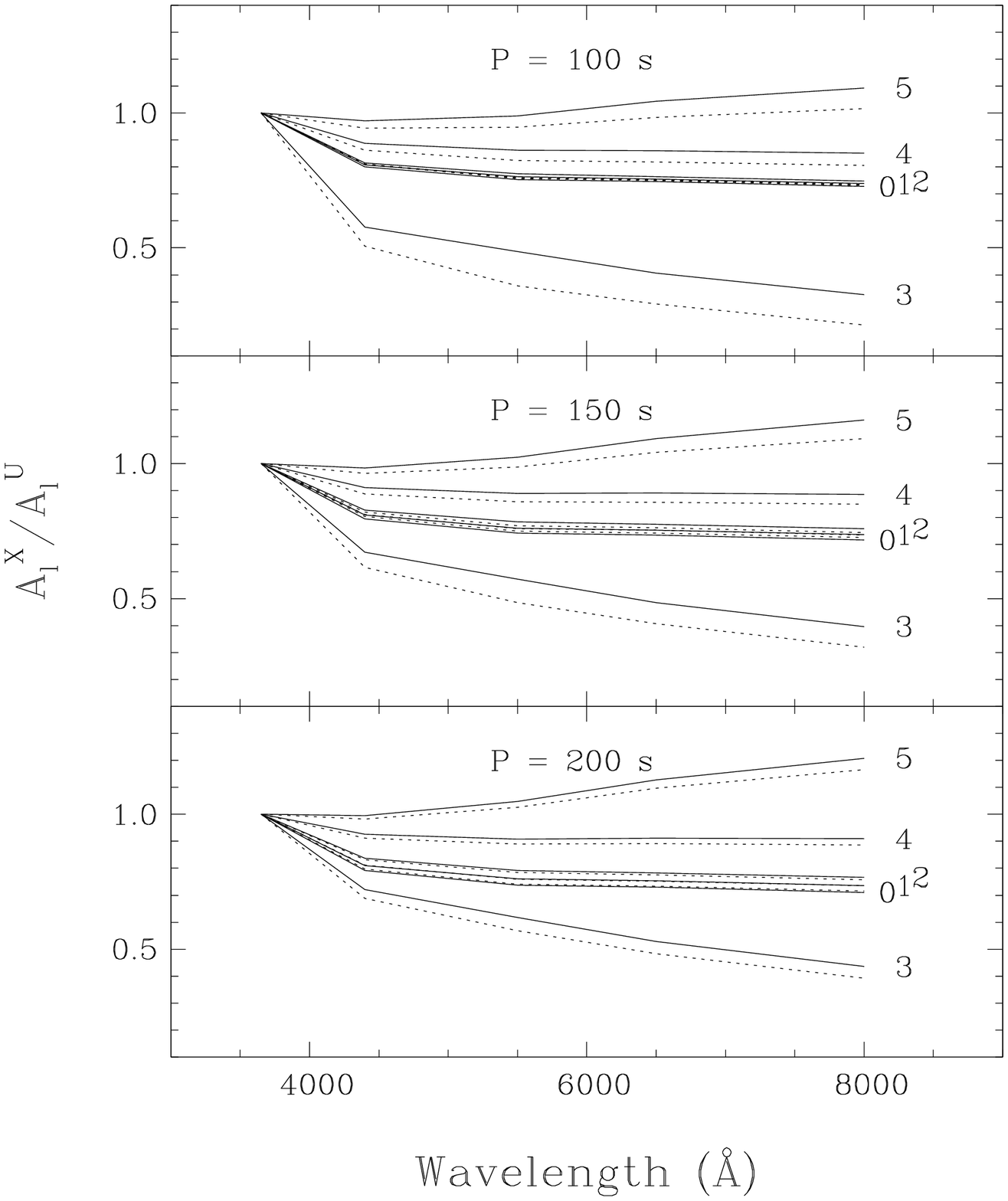}
\begin{flushright}
Figure 14
\end{flushright}
\end{figure}

\clearpage
\begin{figure}[p]
\plotone{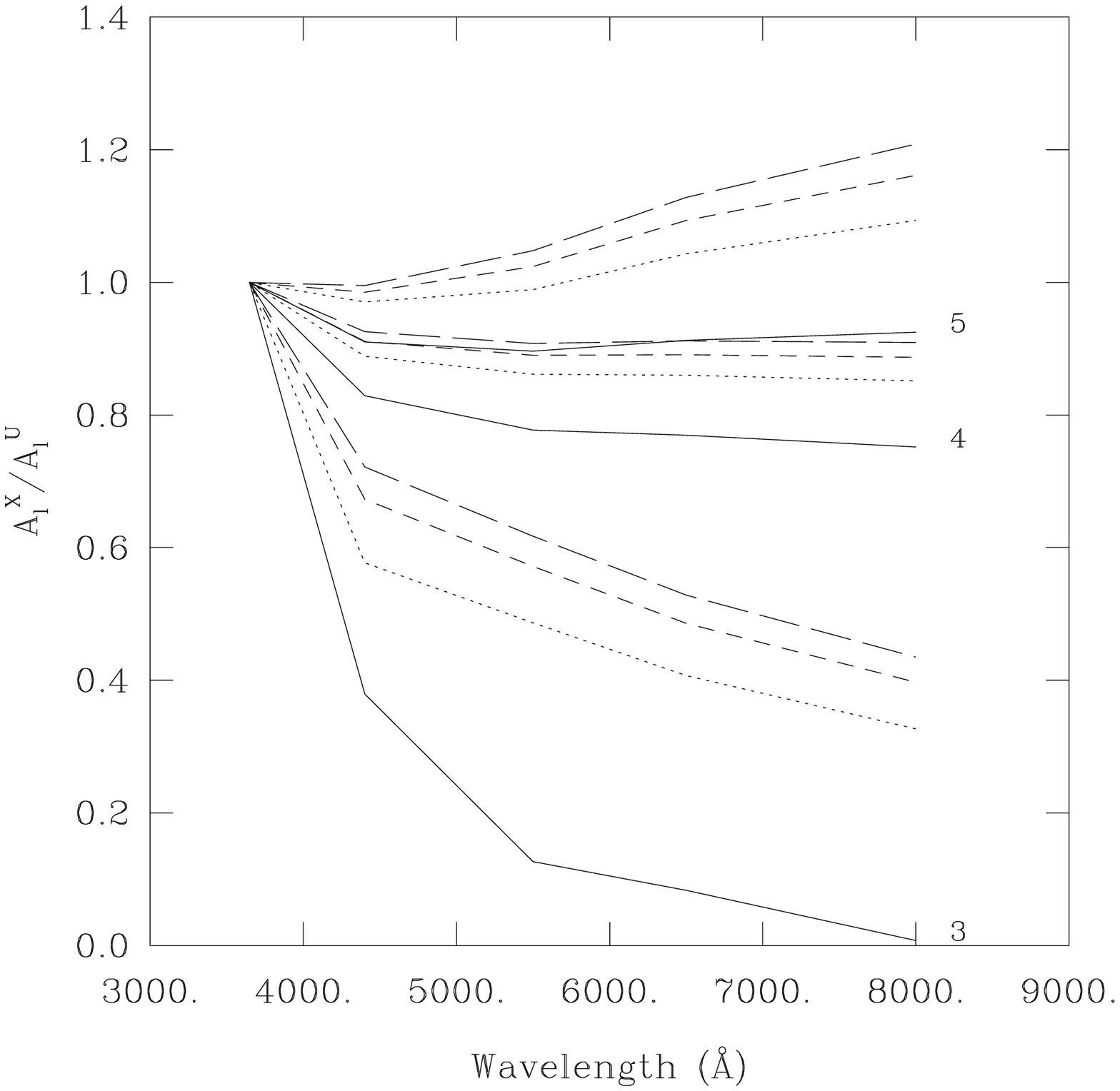}
\begin{flushright}
Figure 15a
\end{flushright}
\end{figure}

\clearpage
\begin{figure}[p]
\plotone{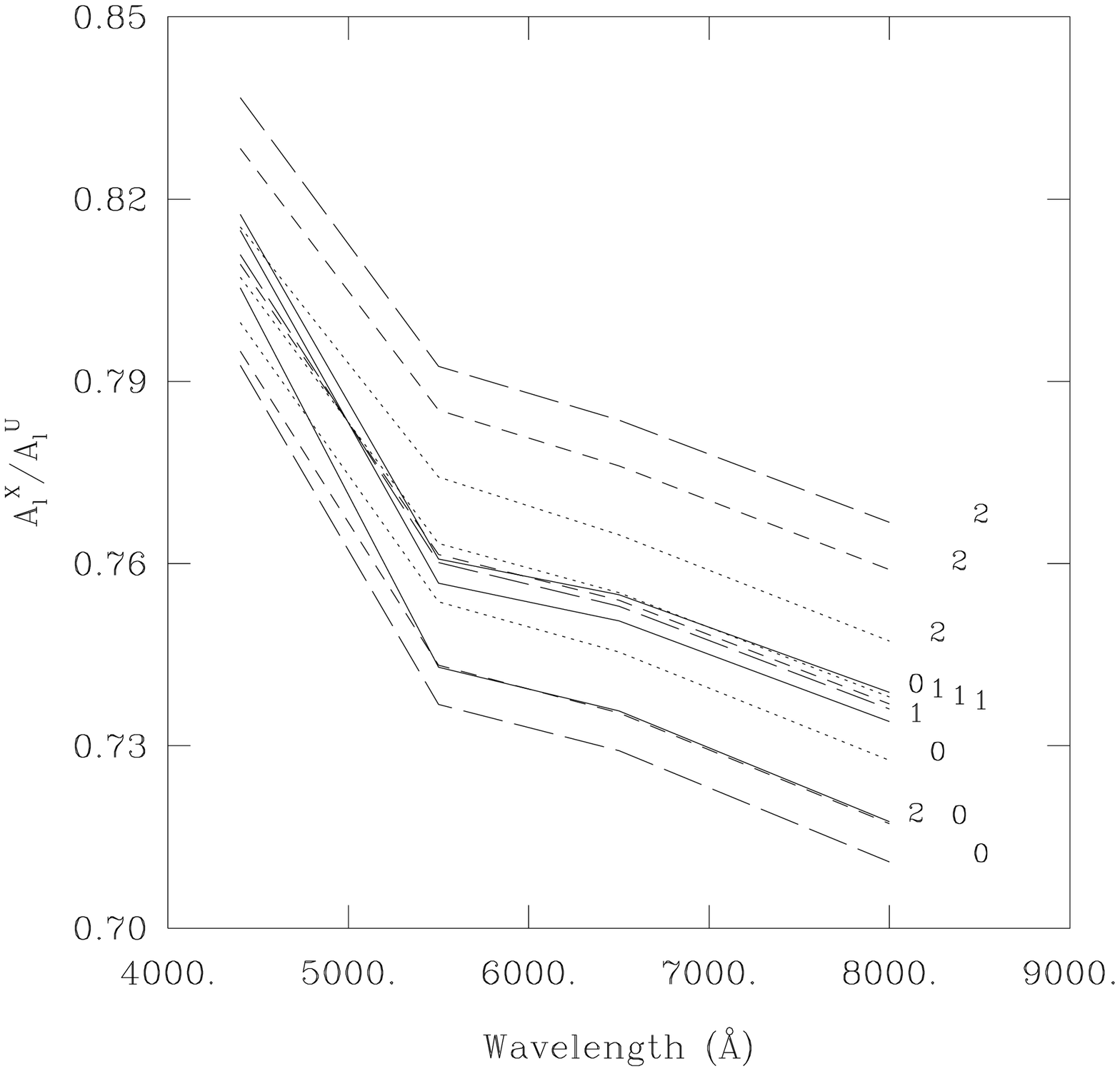}
\begin{flushright}
Figure 15b
\end{flushright}
\end{figure}

\clearpage
\begin{figure}[p]
\plotone{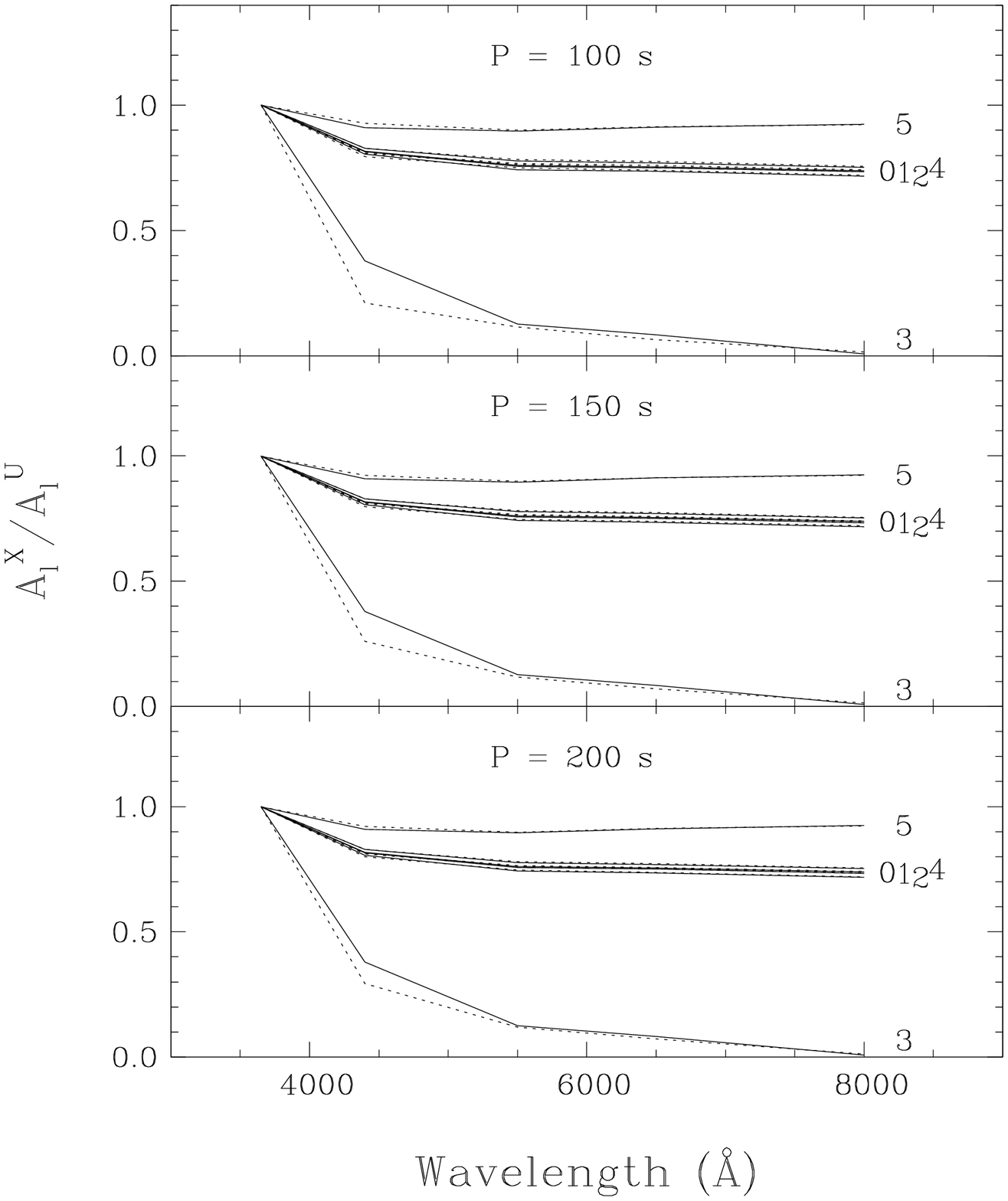}
\begin{flushright}
Figure 16
\end{flushright}
\end{figure}

\clearpage
\begin{figure}[p]
\plotone{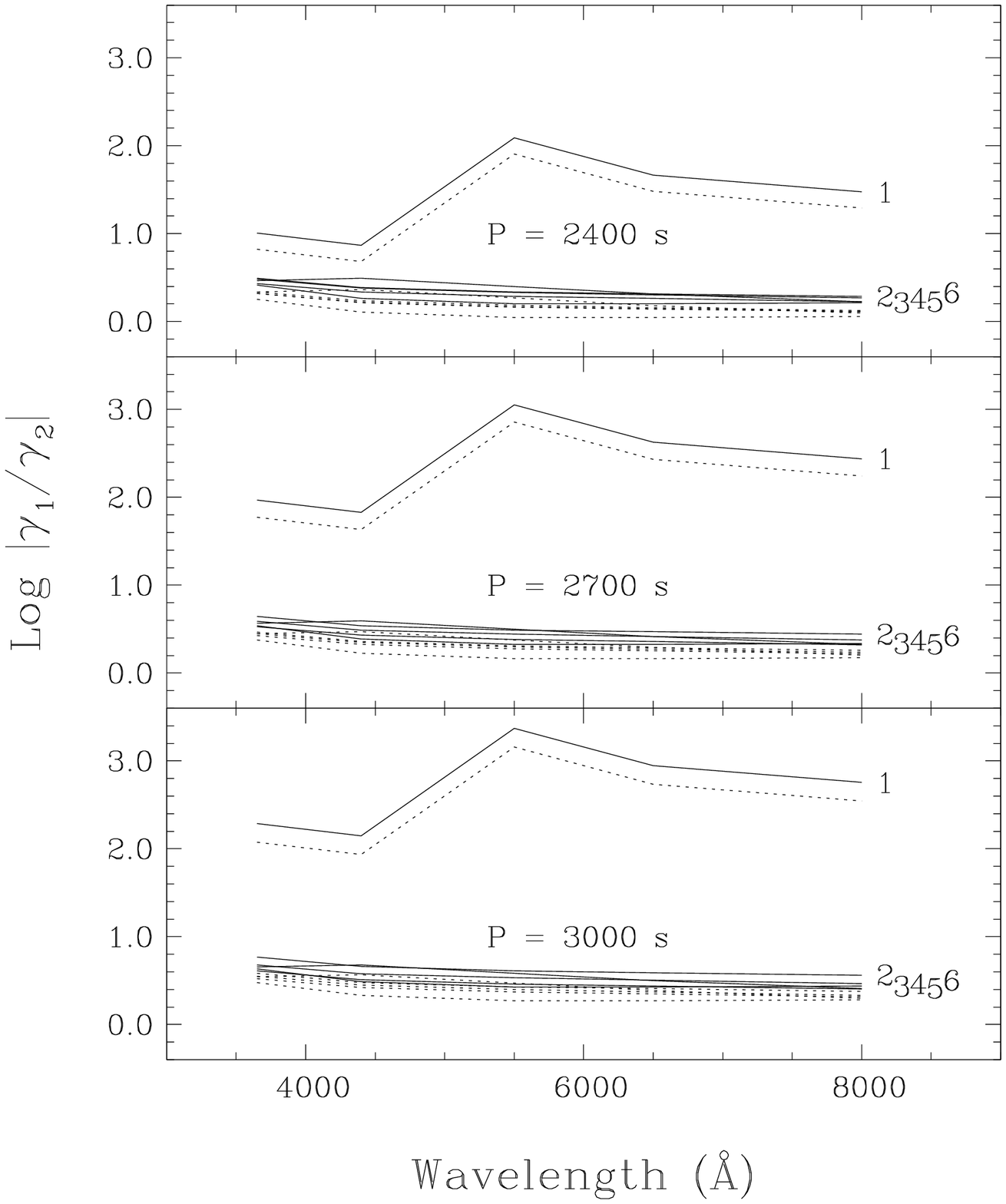}
\begin{flushright}
Figure 17a
\end{flushright}
\end{figure}

\clearpage
\begin{figure}[p]
\plotone{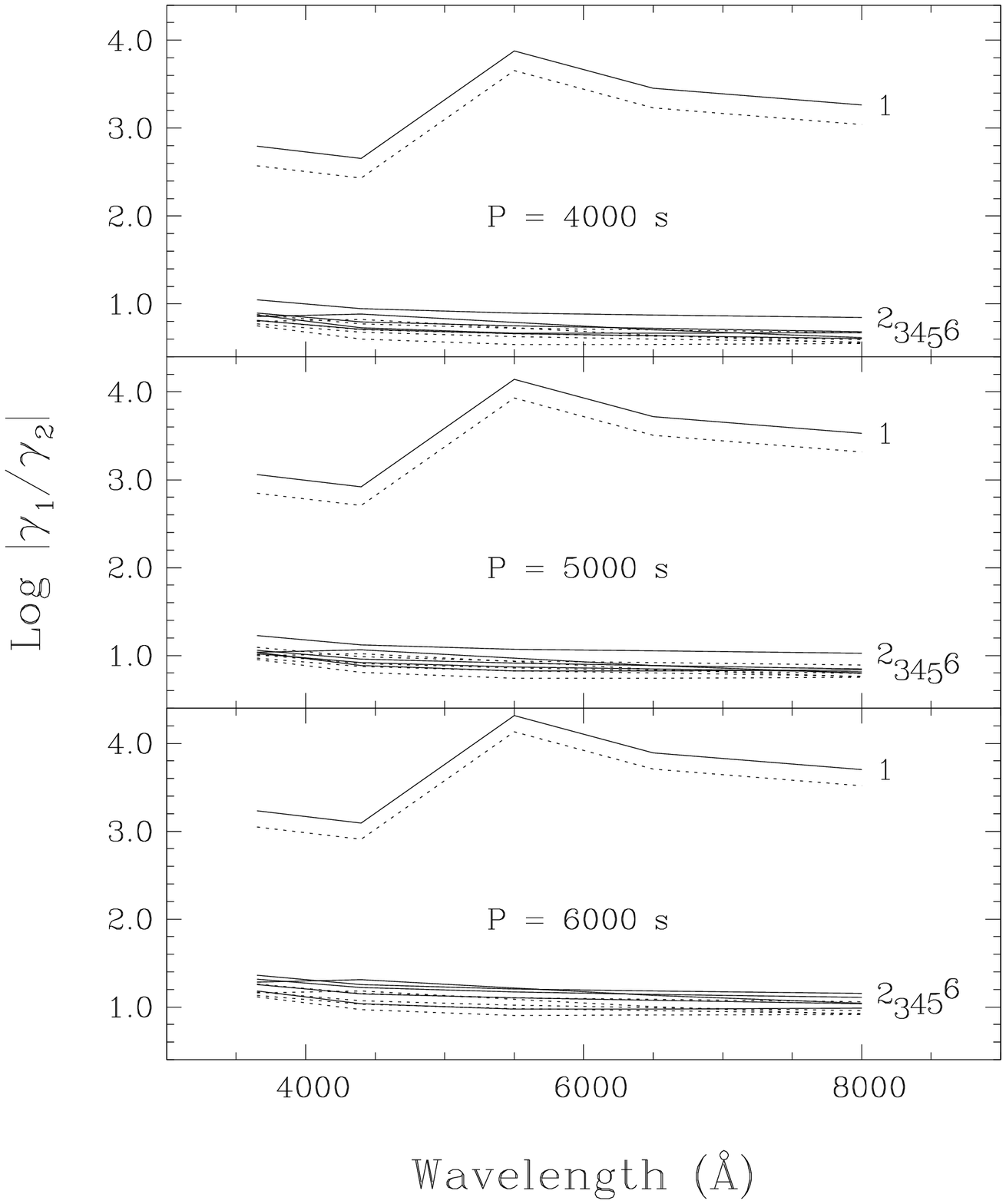}
\begin{flushright}
Figure 17b
\end{flushright}
\end{figure}

\clearpage
\begin{figure}[p]
\plotone{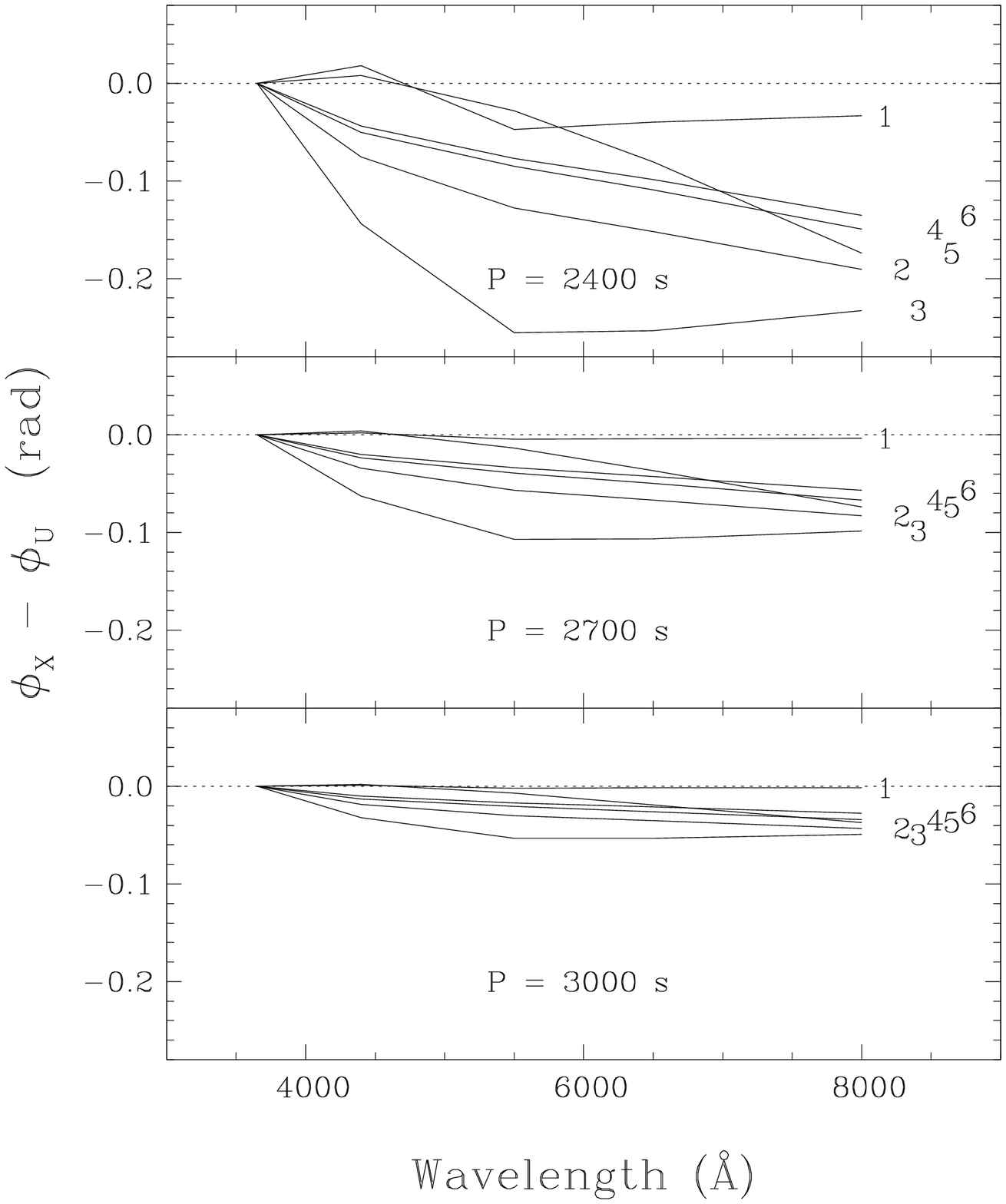}
\begin{flushright}
Figure 18a
\end{flushright}
\end{figure}

\clearpage
\begin{figure}[p]
\plotone{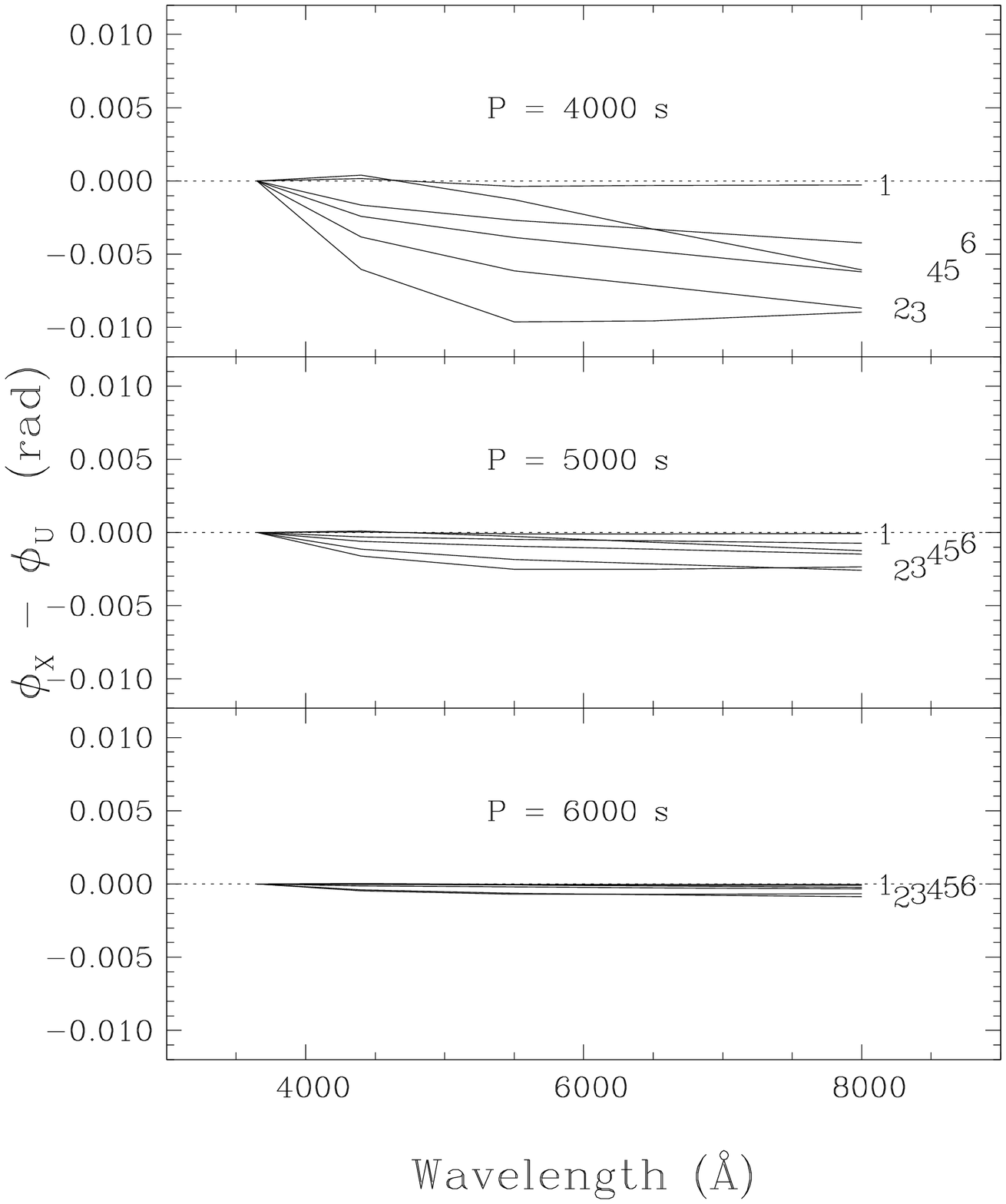}
\begin{flushright}
Figure 18b
\end{flushright}
\end{figure}

\clearpage
\begin{figure}[p]
\plotone{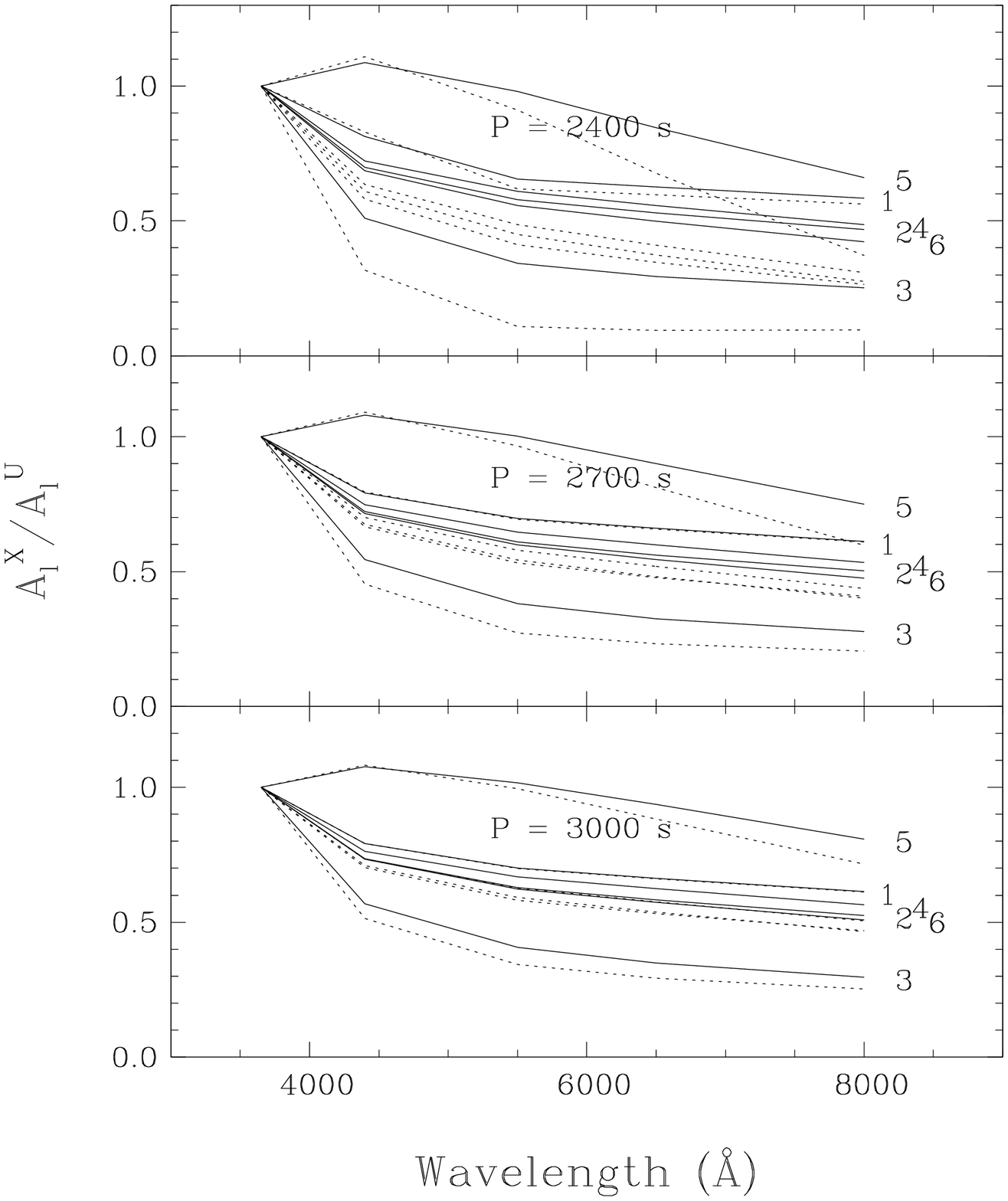}
\begin{flushright}
Figure 19a
\end{flushright}
\end{figure}

\clearpage
\begin{figure}[p]
\plotone{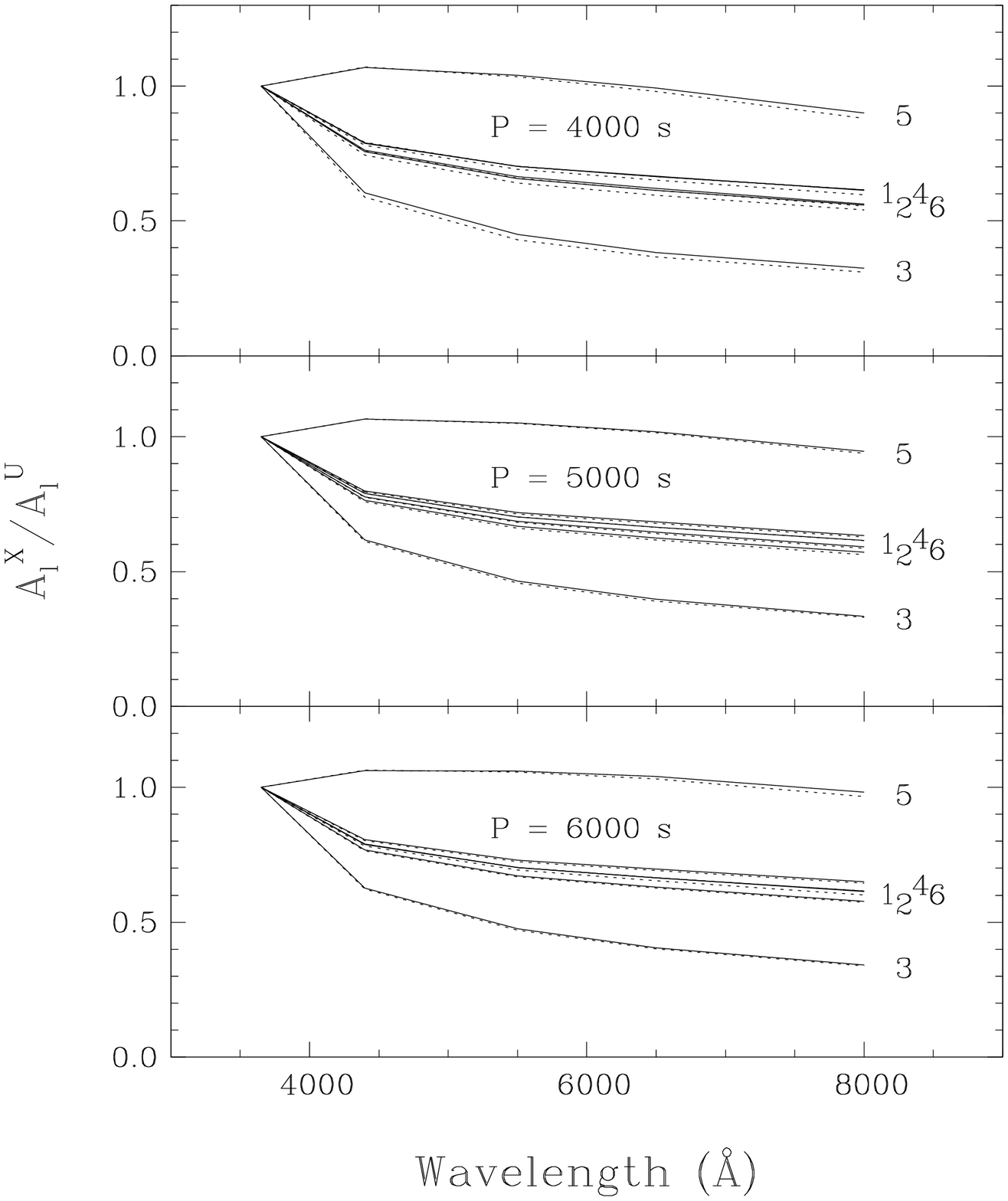}
\begin{flushright}
Figure 19b
\end{flushright}
\end{figure}

\clearpage
\begin{figure}[p]
\plotone{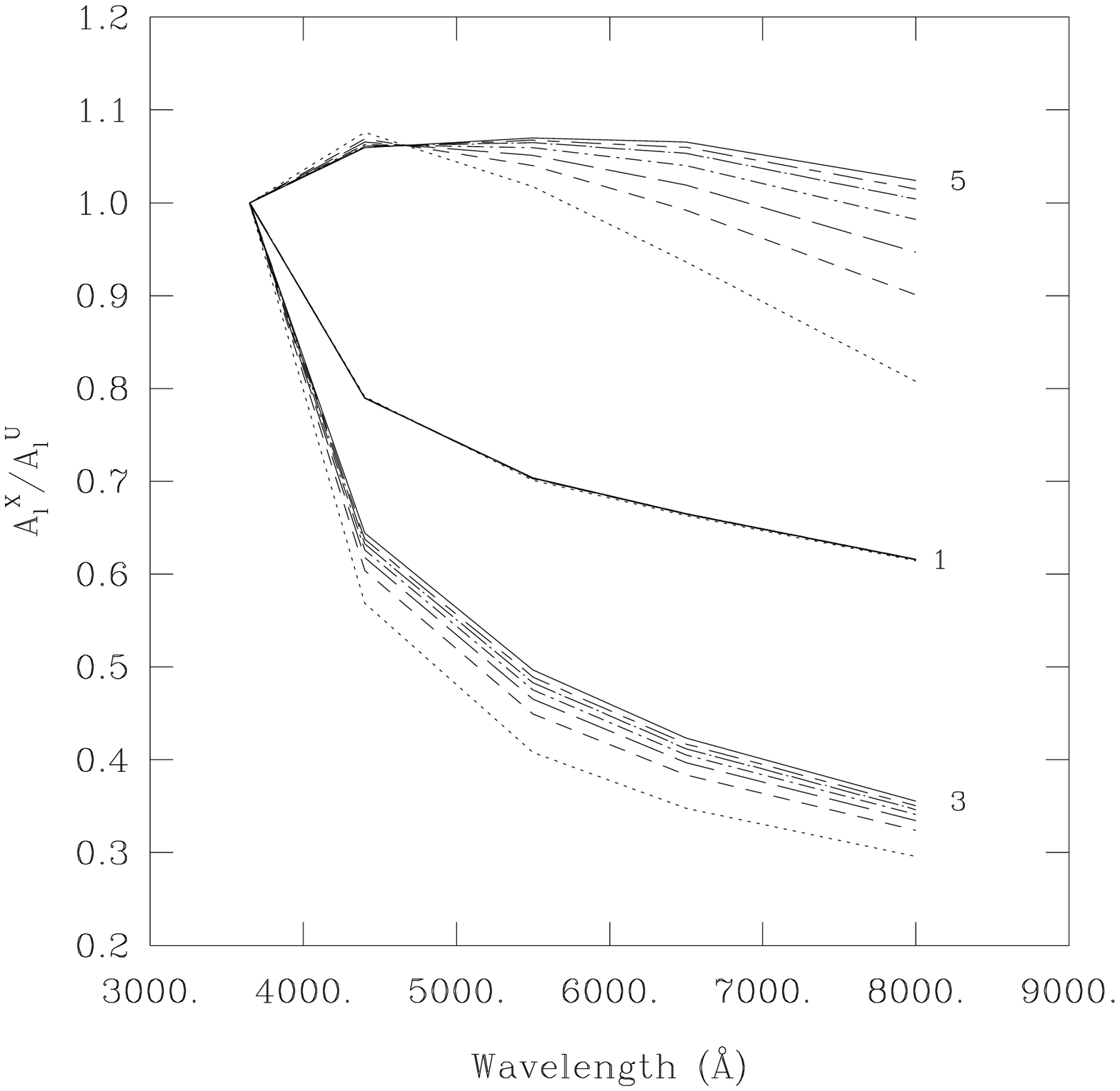}
\begin{flushright}
Figure 20a
\end{flushright}
\end{figure}

\clearpage
\begin{figure}[p]
\plotone{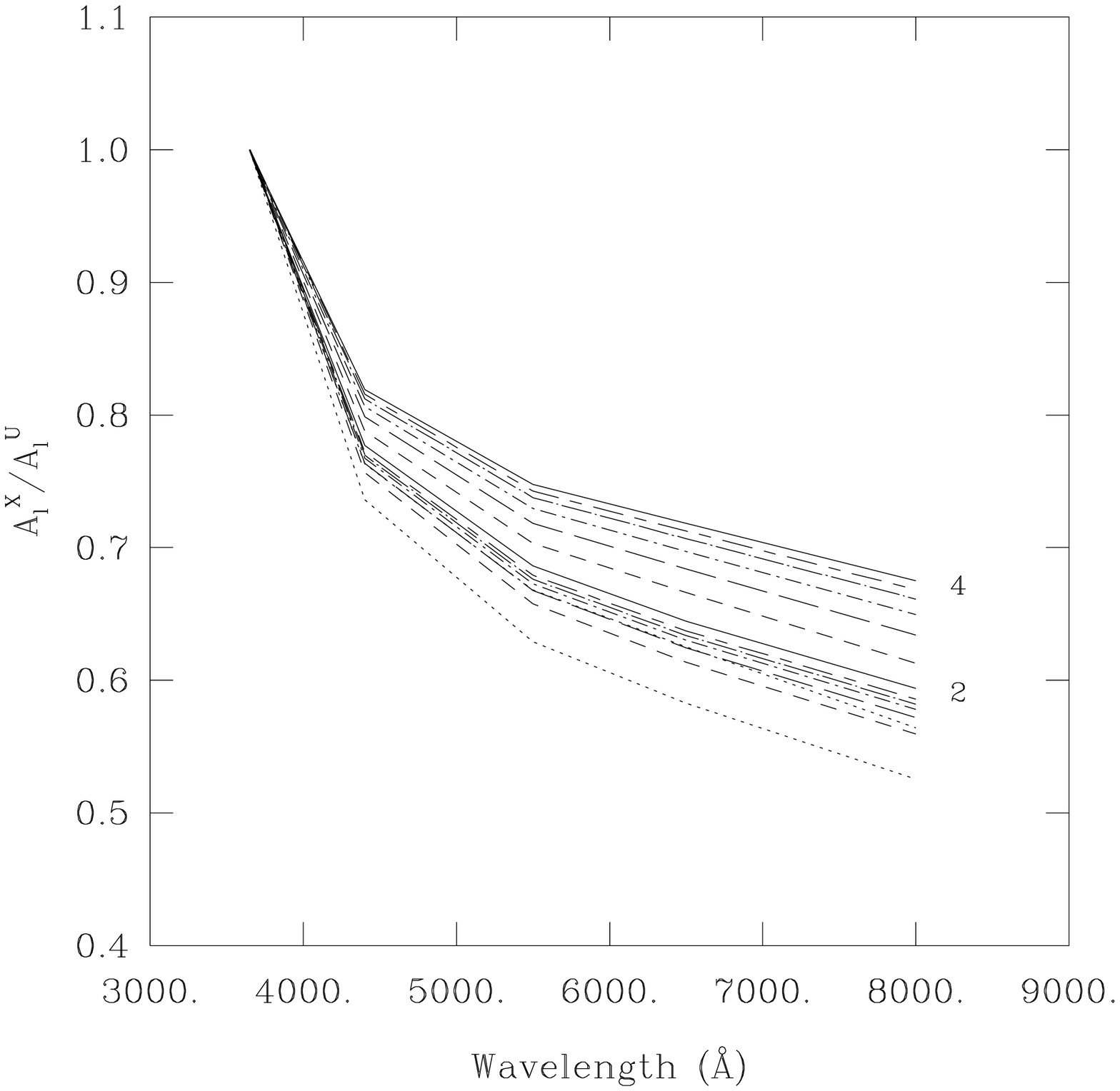}
\begin{flushright}
Figure 20b
\end{flushright}
\end{figure}

\clearpage
\begin{figure}[p]
\plotone{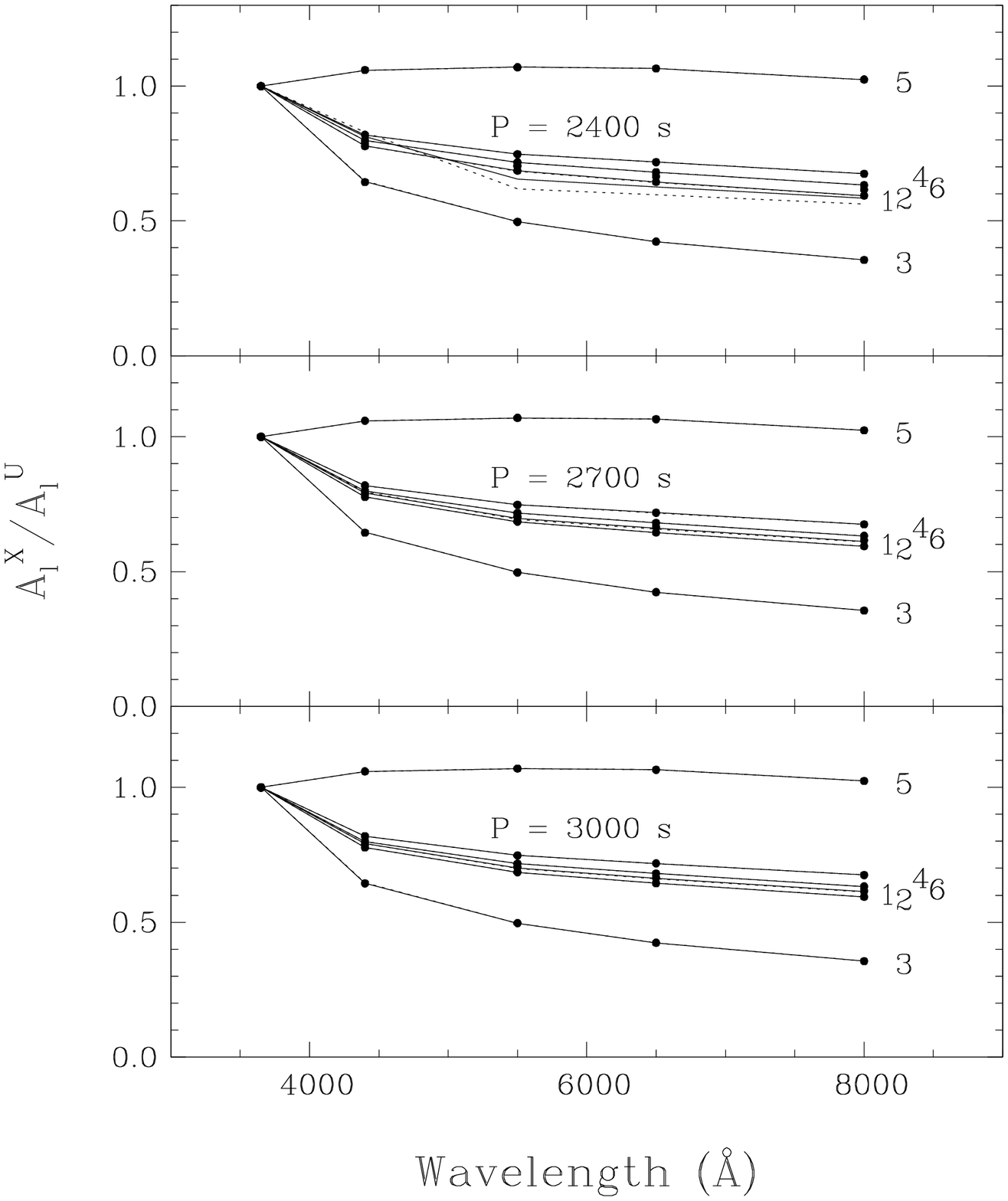}
\begin{flushright}
Figure 21
\end{flushright}
\end{figure}

\clearpage
\begin{figure}[p]
\plotone{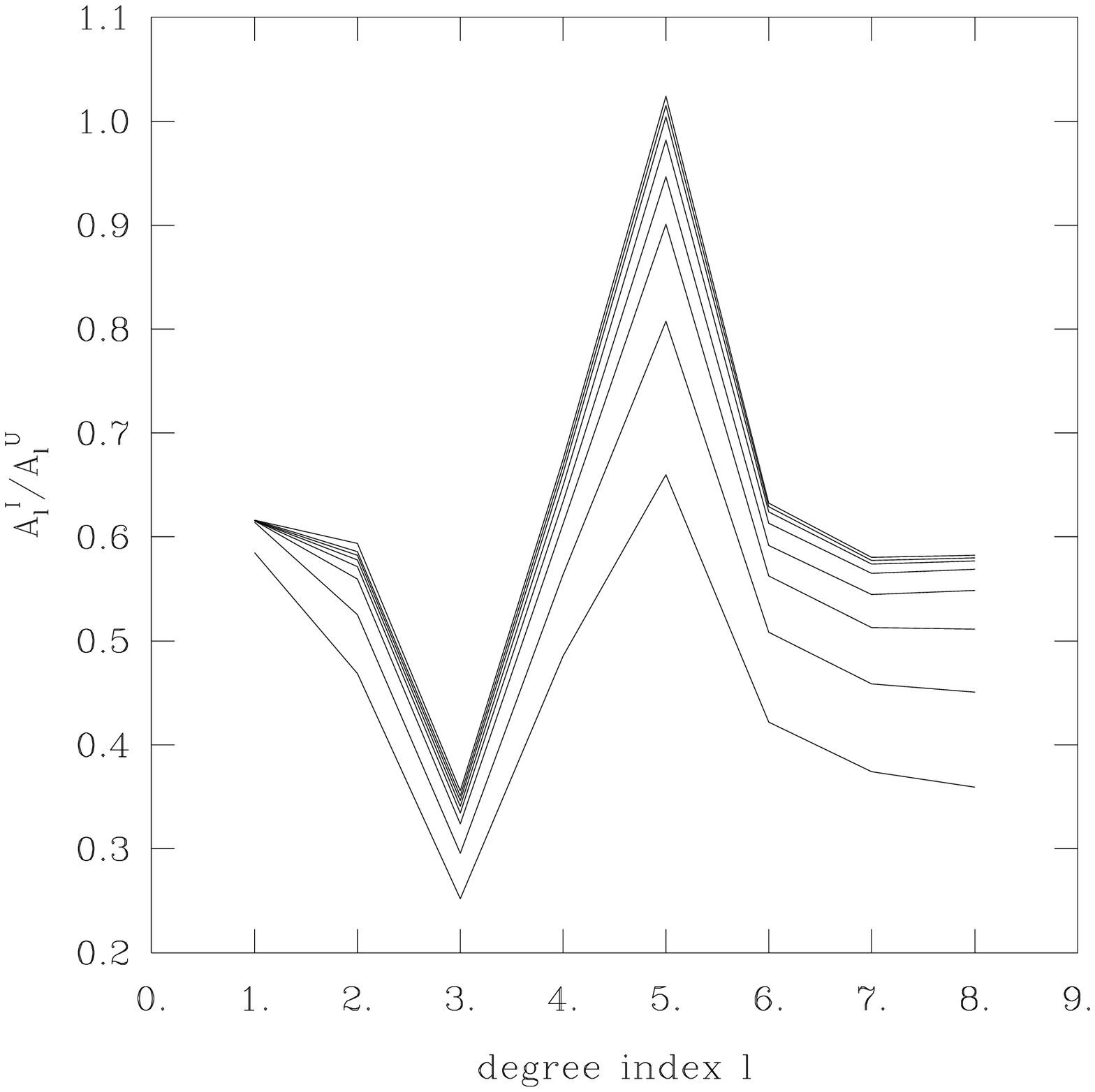}
\begin{flushright}
Figure 22
\end{flushright}
\end{figure}

\clearpage
\begin{figure}[p]
\plotone{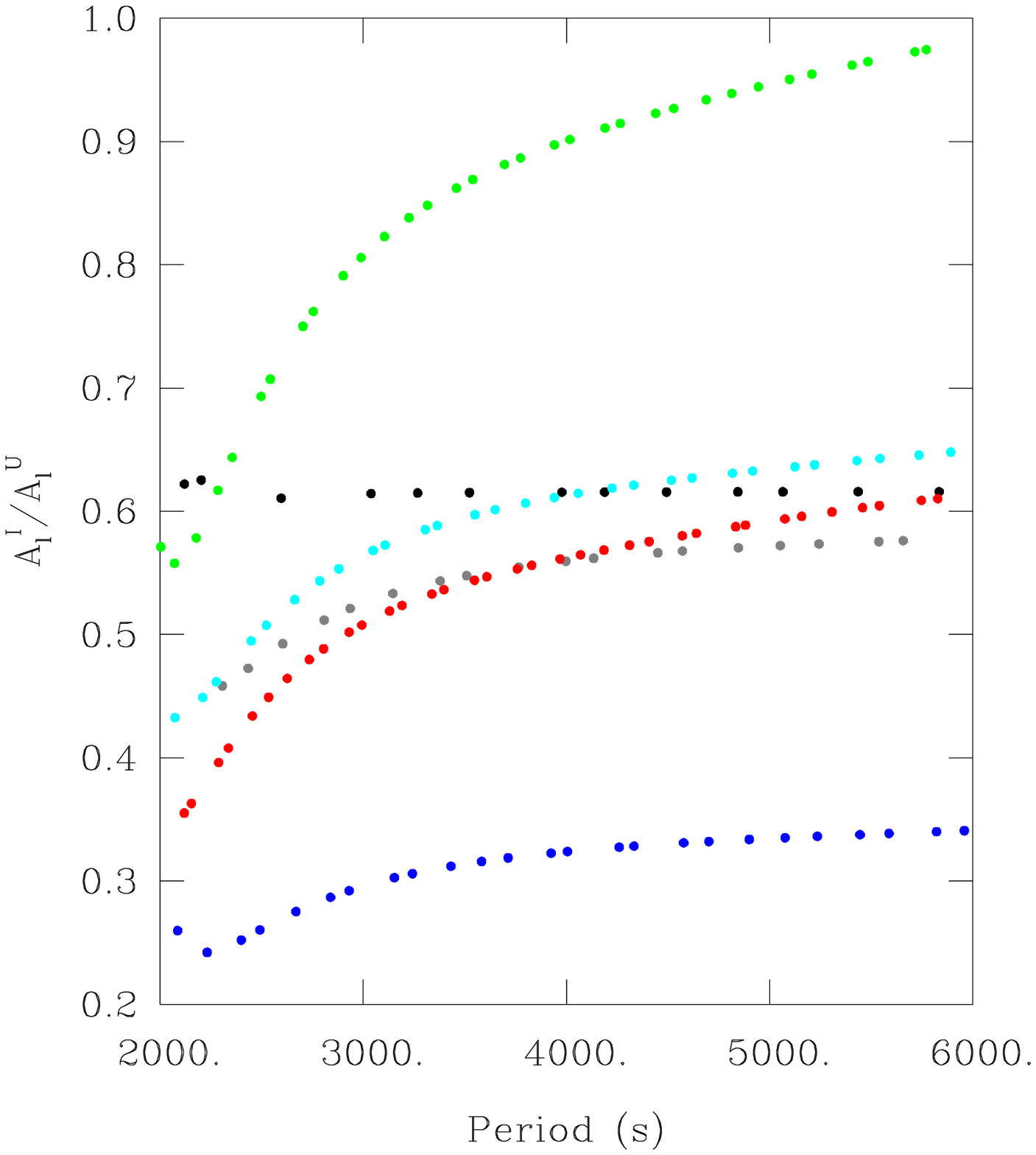}
\begin{flushright}
Figure 23
\end{flushright}
\end{figure}

\clearpage
\begin{figure}[p]
\plotone{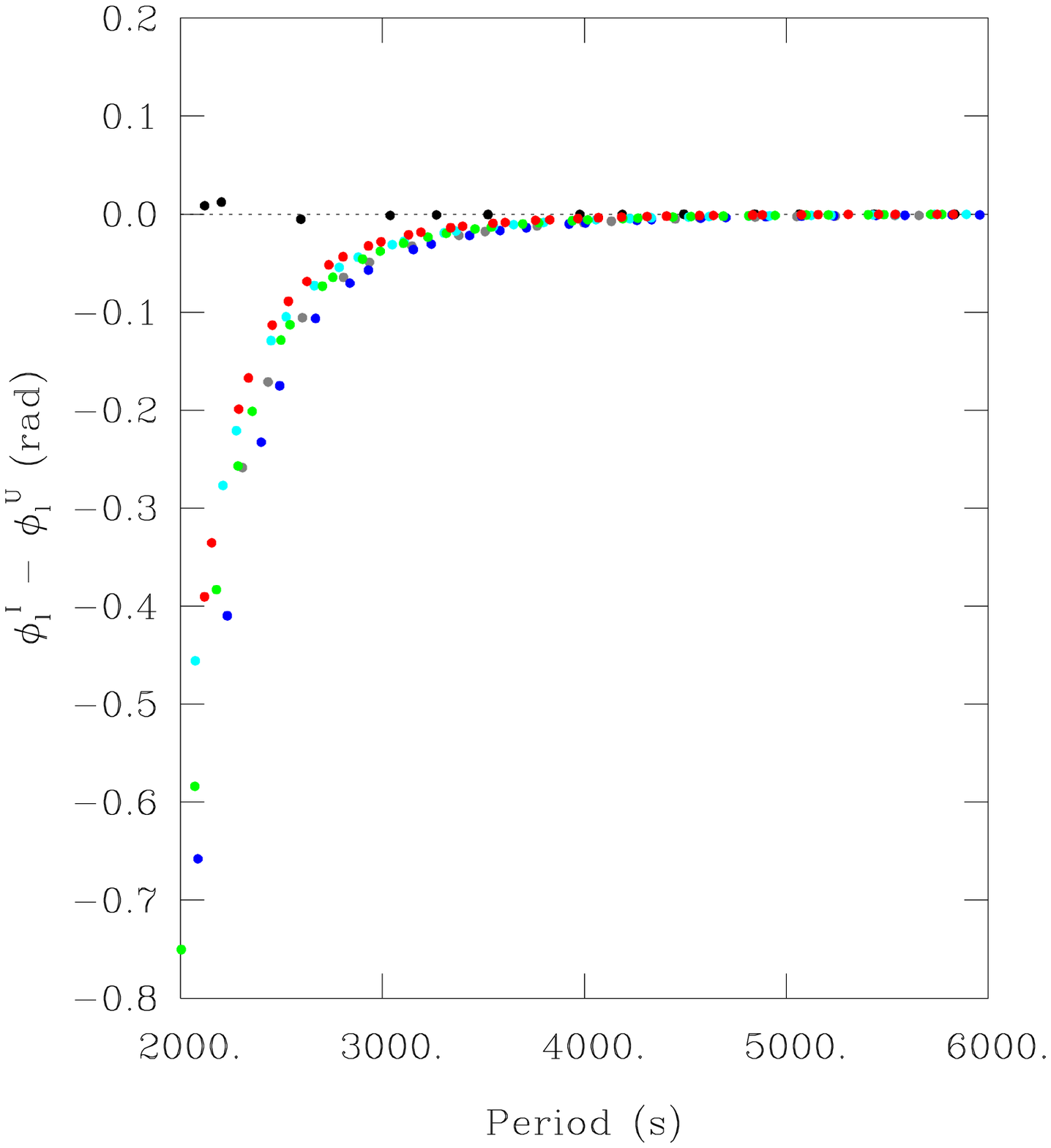}
\begin{flushright}
Figure 24
\end{flushright}
\end{figure}

\clearpage
\begin{figure}[p]
\plotone{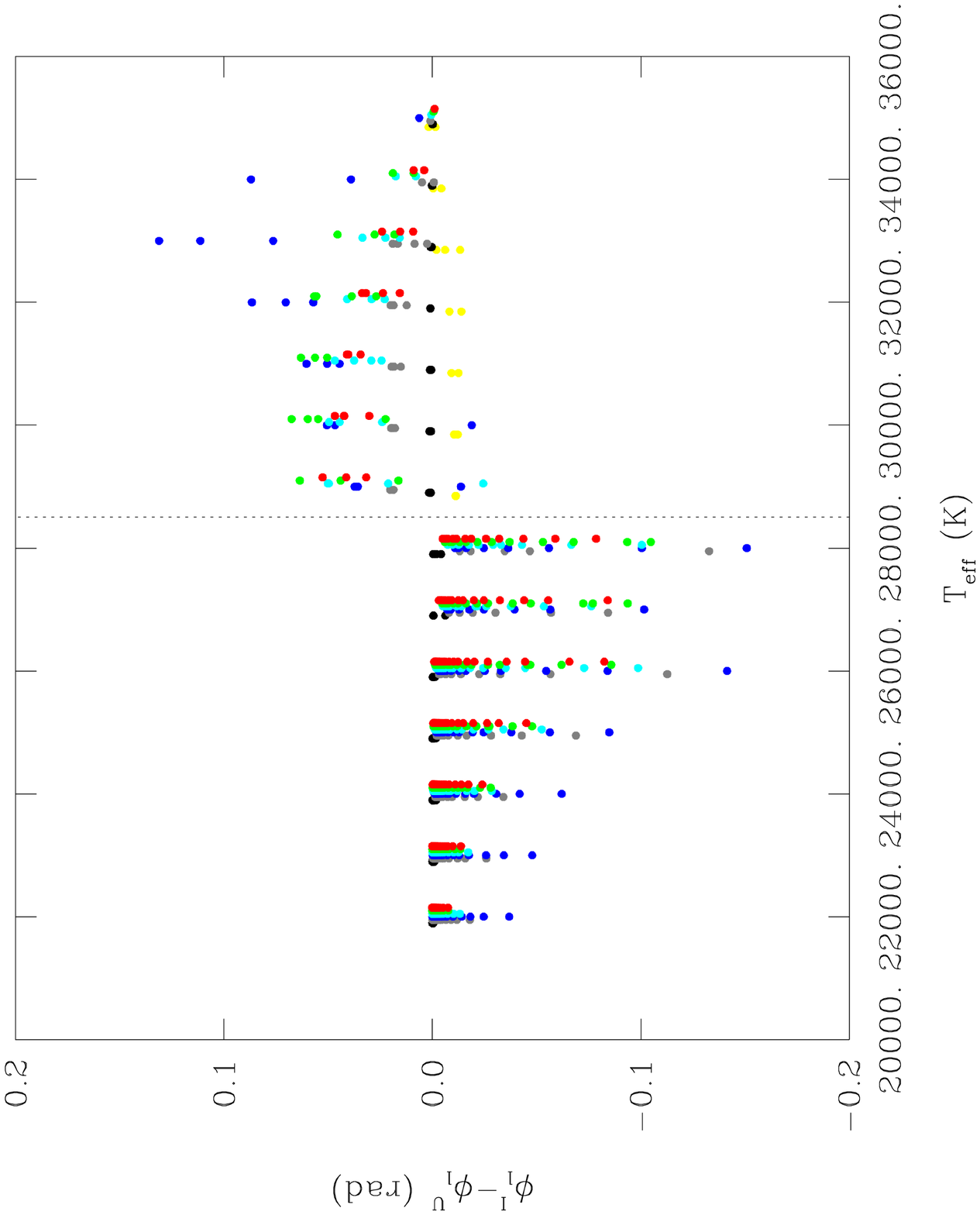}
\begin{flushright}
Figure 25
\end{flushright}
\end{figure}

\clearpage
\begin{figure}[p]
\plotone{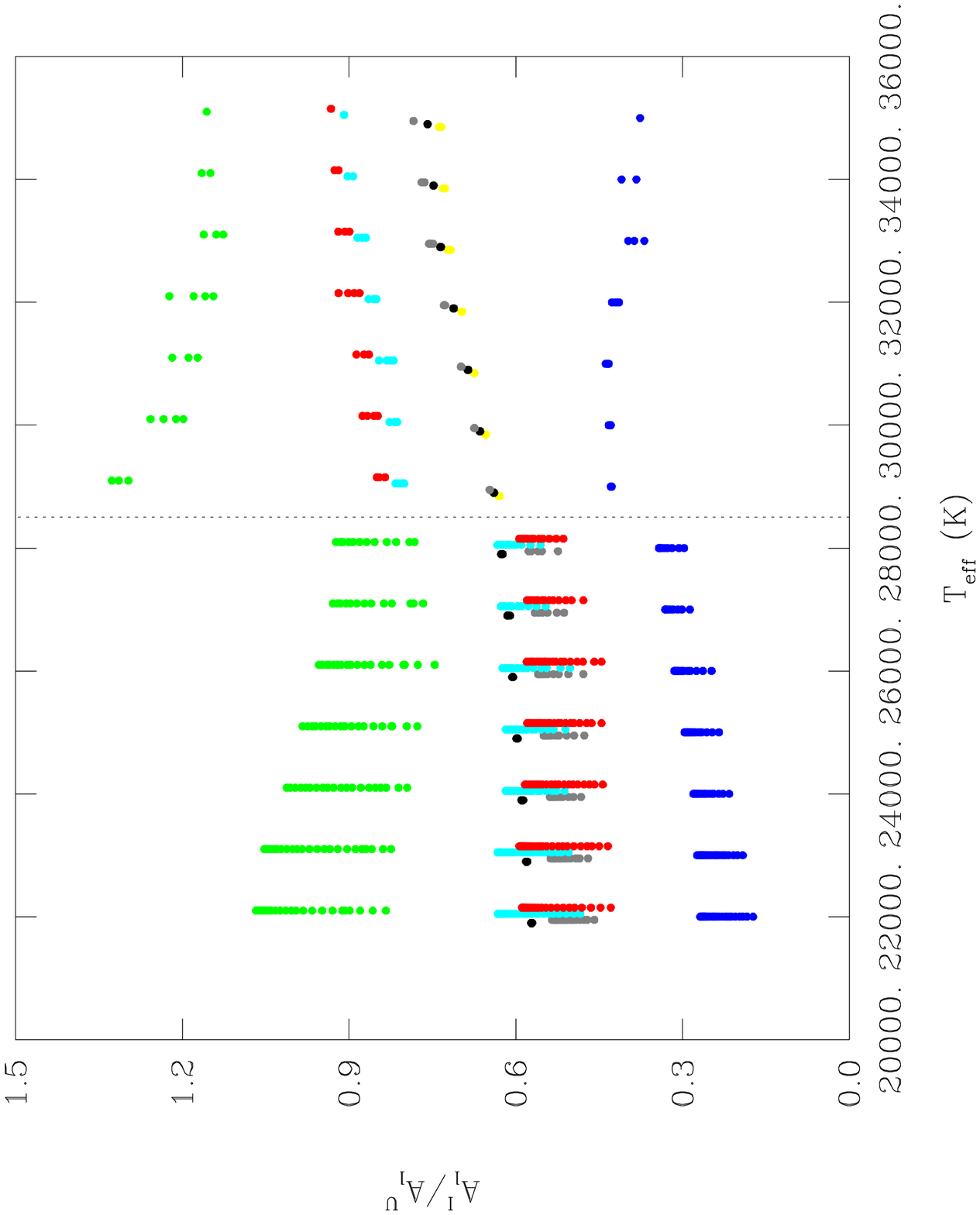}
\begin{flushright}
Figure 26
\end{flushright}
\end{figure}

\clearpage
\begin{figure}[p]
\plotone{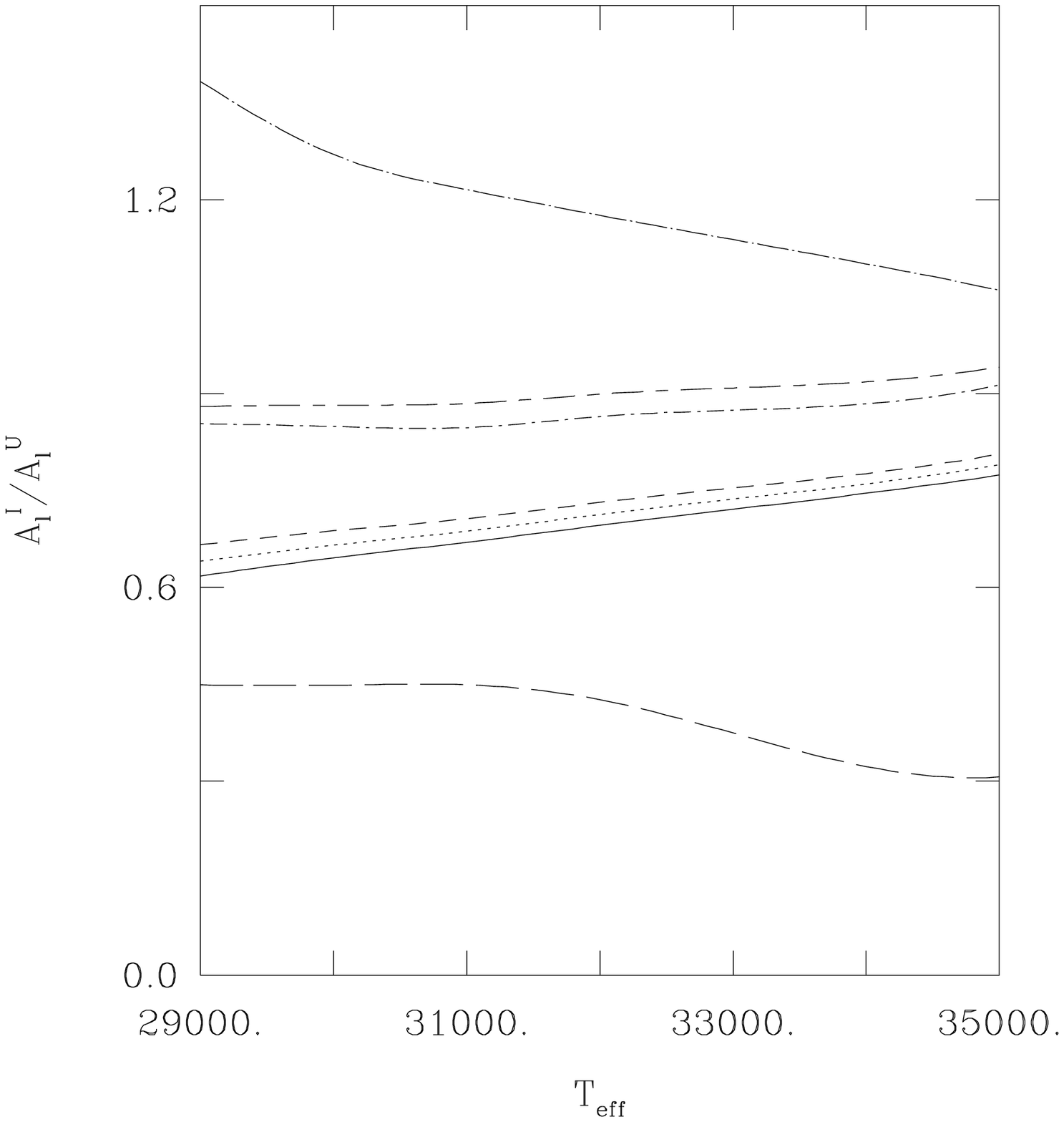}
\begin{flushright}
Figure 27a
\end{flushright}
\end{figure}

\clearpage
\begin{figure}[p]
\plotone{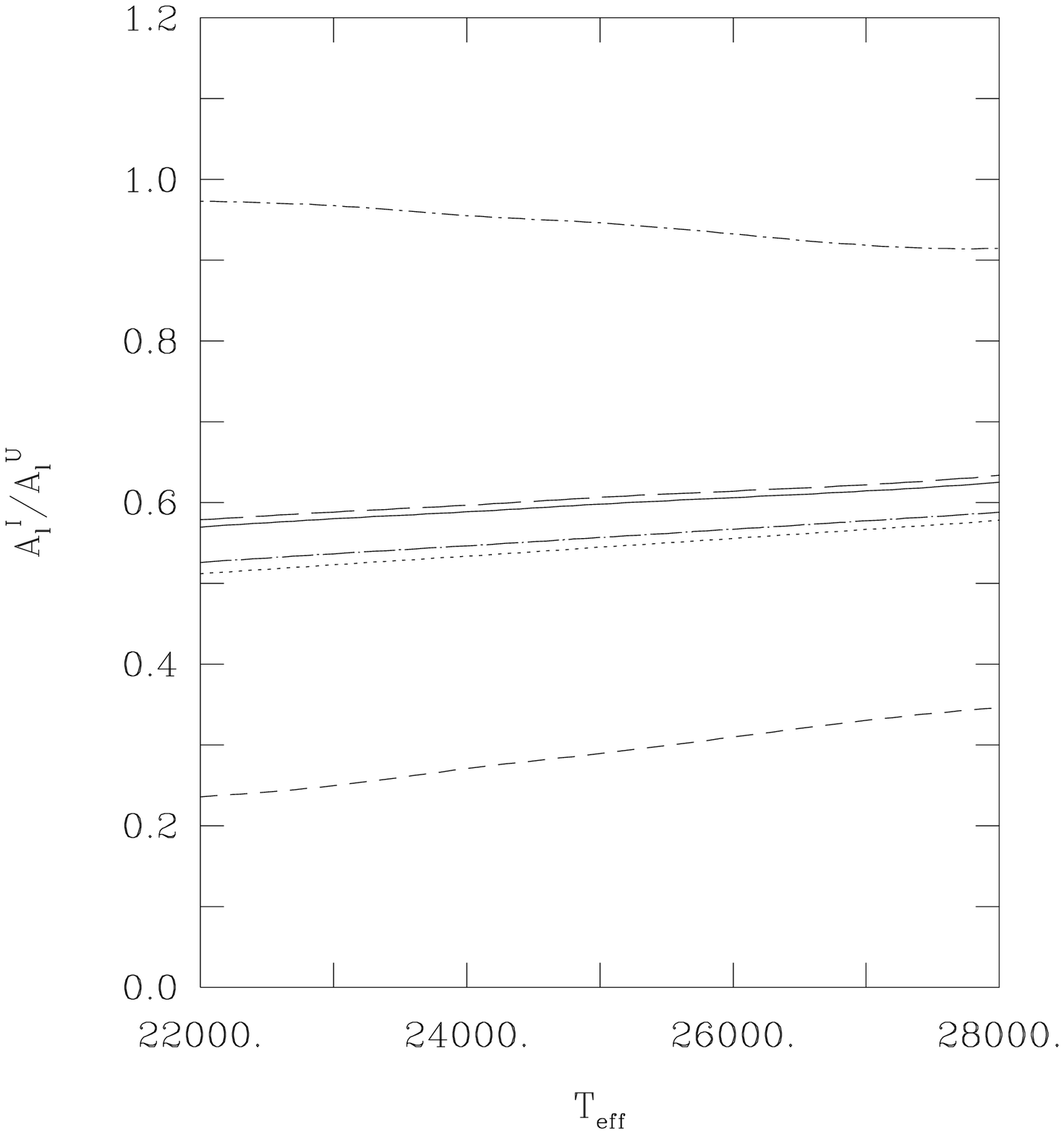}
\begin{flushright}
Figure 27b
\end{flushright}
\end{figure}

\clearpage
\begin{figure}[p]
\plotone{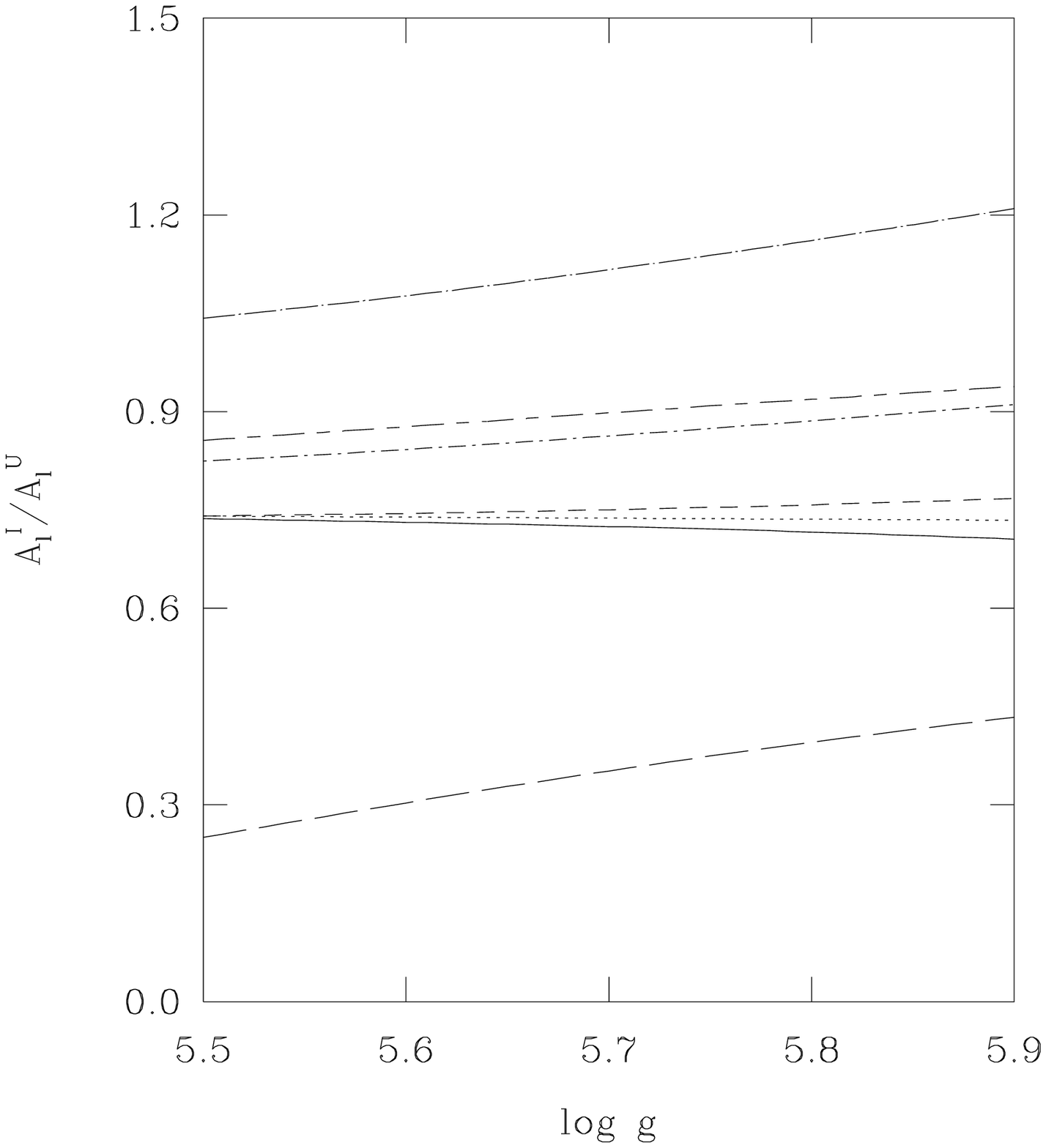}
\begin{flushright}
Figure 28a
\end{flushright}
\end{figure}

\clearpage
\begin{figure}[p]
\plotone{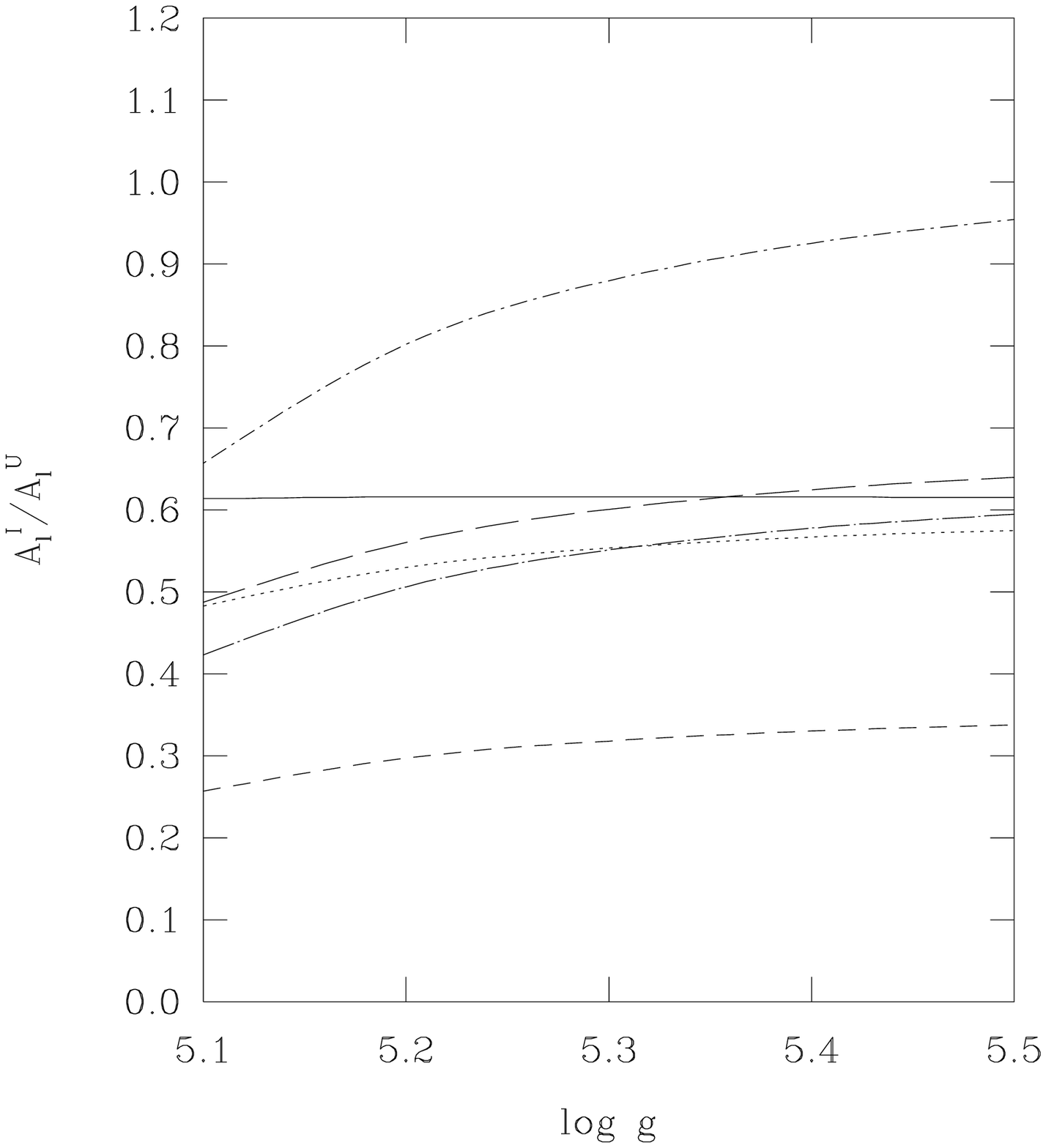}
\begin{flushright}
Figure 28b
\end{flushright}
\end{figure}

\clearpage
\begin{figure}[p]
\plotone{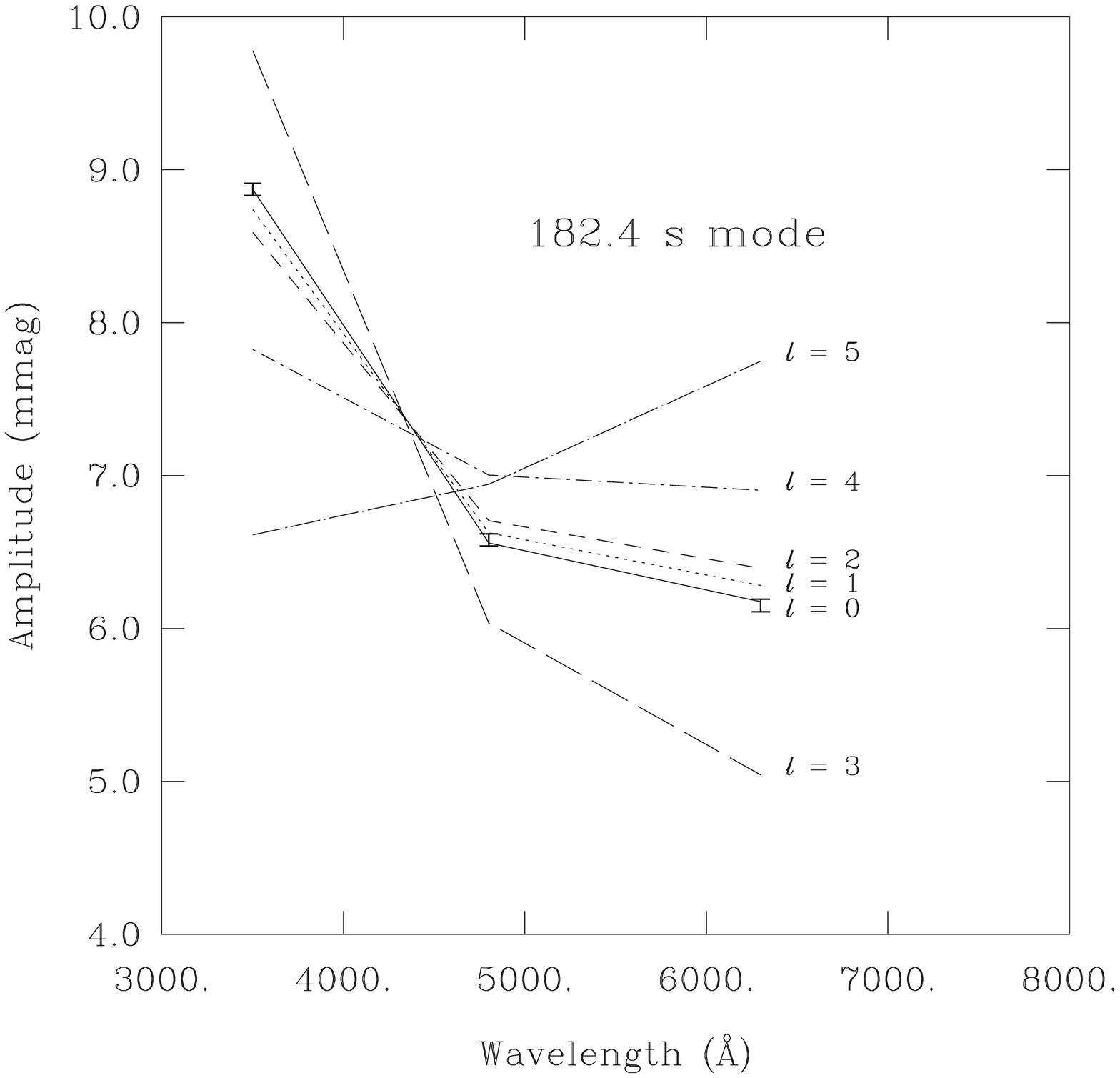}
\begin{flushright}
Figure 29
\end{flushright}
\end{figure}

\end{document}